\documentclass[%
 reprint,
 amsmath,amssymb,
 aps,
 pra, 
]{revtex4-1}

\usepackage{graphicx}
\usepackage{dcolumn}
\usepackage{bm}
\usepackage{color}
\usepackage{hyperref}

\newcommand{\dd}{\mathrm{d}}

\newcommand{\ee}{\mathrm{e}}
\newcommand{\ii}{\mathrm{i}\,}

\newcommand{\TT}{\mathcal{T}}

\newcommand{\VV}{\mathcal{V}}
\newcommand{\XX}{\mathcal{X}}
\newcommand{\FF}{\mathcal{F}}

\newcommand{\dg}{\dagger}

\newcommand{\phigood}{\tilde{\phi}}
\newcommand{\thtgood}{\tilde{\theta}}
\newcommand{\up}{\uparrow}
\newcommand{\dn}{\downarrow}

\begin{document}

\title{Ising ferromagnet to valence bond solid transition in a one-dimensional spin chain: Analogies to deconfined quantum critical points}

\author{Shenghan Jiang}
\email{jiangsh@caltech.edu}
\author{Olexei I. Motrunich}
\email{motrunch@caltech.edu}
\affiliation{Department of Physics and Institute for Quantum Information and Matter, \\
California Institute of Technology, Pasadena, California 91125, USA}

\date{\today}

\begin{abstract}
We study a one-dimensional (1d) system that shows many analogies to proposed two-dimensional (2d) deconfined quantum critical points~(DQCP).
Our system is a translationally invariant spin-1/2 chain with on-site $Z_2 \times Z_2$ symmetry and time reversal symmetry.
It undergoes a direct continuous transition from a ferromagnet (FM), where one of the $Z_2$ symmetries and the time reversal are broken, to a valence bond solid (VBS), where all on-site symmetries are restored while the translation symmetry is broken.
The other $Z_2$ symmetry remains unbroken throughout, but its presence is crucial for both the direct transition (via specific Berry phase effect on topological defects, also related to a Lieb-Schultz-Mattis-like theorem) and the precise characterization of the VBS phase (which has crystalline-SPT-like property).
The transition has a description in terms of either two domain wall species that ``fractionalize'' the VBS order parameter or in terms of two partons that ``fractionalize'' the FM order parameter, with each picture having its own $Z_2$ gauge theory structure.
The two descriptions are dual to each other and, at long wavelengths, take the form of a self-dual \emph{gauged} Ashkin-Teller model, reminiscent of the self-dual easy-plane non-compact CP$^1$ model that arises in the description of the 2d easy-plane DQCP.
We also find an exact reformulation of the transition that leads to a simple field theory description that explicitly unifies the FM and VBS order parameters; this reformulation can be interpreted as a new parton approach that does not attempt to fractionalize either of the FM and VBS order parameters but instead encodes them in instanton operators.
Besides providing explicit realizations of many ideas proposed in the context of the 2d DQCP, here in the simpler and fully tractable 1d setting with continuous transition, our study also suggests possible new line of approach to the 2d DQCP. 
\end{abstract}

\maketitle

\tableofcontents

\section{Introduction}
In the classical world, phases are classified according to symmetry properties, and are characterized by their order parameters.
Continuous phase transitions in classical systems, which describe critical phenomena between spontaneously symmetry breaking~(SSB) phases and symmetric phases, are captured by Landau-Ginzburg-Wilson~(LGW) theoretical framework.
Quantum phases, on the other hand, are more exotic, and cannot be fully characterized by SSB order parameters.
Examples include topological insulators of electrons, spin liquids in frustrated spin systems, as well as various quantum Hall phases.
Thus, it is natural to expect that quantum critical points involving exotic quantum phases are beyond the scope of the LGW framework.

Surprisingly, there is a special kind of exotic quantum criticality, named as deconfined quantum critical points~(DQCPs), where both sides of critical points are conventional SSB phases, with different symmetry breaking patterns.
The first example of such a phase transition was proposed to occur in a quantum spin-1/2 system on the two-dimensional square lattice~\cite{DQCP_science, DQCP_prb}.
By changing interactions, one can obtain an antiferromagnetic Neel order which breaks spin rotation symmetry, or a valence bond solid~(VBS) order which breaks lattice symmetries but preserves spin rotation symmetry.
Both theoretical and numerical studies~\cite{Sandvik2007, MelkoKaul2008, LuoSandvikKawashima2009, BanerjeeDamleAlet2010, Sandvik2010, HaradaSuzukiOkuboMatsuoLuoWatanabeTodoKawashima2013, JiangNyfelerChandrasekharanWises2008, ChenHuangDengKuklovProkofevSvistunov2013, NahumChalkerSernaOrtunoSomoza2015, NahumSernaChalkerOrtunoSomoza2015II, MotrunichVishwanath2008, KuklovMatsumotoProkofevSvistunovTroyer2008, Bartosch2013, CharrierAletPujol2008, ChenGukelbergerTrebstFabienBalents2009, CharrierAlet2010, SreejithPowell2015, ShaoGuoSandvik2016, SernaNahum2018} show that there is a second order (or weakly first order) phase transition between these two phases.
One may wonder if there is anything special about the spin-1/2 system on the square lattice and why the DQCP occurs in this system.
Answers to these questions are far from obvious. 
Before moving on, let us discuss a seemingly unrelated subject: the (generalized) Lieb-Schultz-Mattis~(LSM) theorem.

The original LSM theorem~\cite{LiebSchultzMattis1961} deals with spin-1/2 chains with $SO(3)$ symmetric and translationally invariant interactions.
The theorem implies that the ground state must either break translational symmetry, forming VBS order, or remain gapless.
Oshikawa~\cite{Oshikawa2000} and Hastings~\cite{Hastings2004} generalized the LSM theorem to (2+1)D.
They argued that for translationally symmetric systems with half-integer spin per unit cell, the ground states must be gapless, or if gapped, must have nontrivial degeneracy when systems are put on a torus.
In particular, gapped phases must be either topologically ordered, or break (discrete) symmetries.

The LSM theorem has been further generalized to various contexts~\cite{WatanabePoVishwanathZaletel2015, ChengZaletelBarkeshliVishwanathBonderson2016, PoWatanbeJianZaletel2017, QiFangFu2017, Huang2017,LuRanOshikawa2017, Lu2017,YangJiangVishwanathRan2017, KomargodskiSharonThorngrenZhou2017, KomargodskiSulejmanpasicUnsal2018, MetlitskiThorngren2017, SulejmanpasicTanizaki2018, TanizakiSulejmanpasic2018arXiv}.
Here, we focus on a particular generalization to translationally symmetric systems with an on-site symmetry $SG_\text{onsite}$ acting projectively on each unit cell.
The generalized LSM theorem tells us that it is impossible to construct a local Hamiltonian whose ground state is symmetric and gapped, and has no degeneracy on the torus.
In what follows, we call this kind of systems as LSM-systems.

How does the generalized LSM theorem help for the occurrence of DQCP? 
To see this, we first point out a simple fact: a conventional phase transition between an SSB phase and a trivial symmetric phase can never happen in the LSM-systems, since the latter phase does not exist in such systems!
So, the LSM-systems become a great platform to find exotic critical points.
In the 2d Neel-VBS DQCP context, this connection has been particularly elucidated in recent works Refs.~\cite{KomargodskiSharonThorngrenZhou2017, KomargodskiSulejmanpasicUnsal2018, MetlitskiThorngren2017}.

Now, let us turn our eyes away from 2d systems and instead focus on LSM-systems defined in 1d.
There are many examples in 1d.
For instance, the original LSM paper~\cite{LiebSchultzMattis1961} considered a translationally symmetric spin-1/2 chain with isotropic Heisenberg interactions preserving the full spin-rotation symmetry.
As already mentioned, in this case the ground state must be either gapless, or break translational symmetry~\cite{MajumdarGhosh1969, MajumdarGhosh1969II, Haldane1982}.
Similar results can be obtained for spin-1/2 models with anisotropic interactions that only preserve continuous spin-rotation symmetry about the $z$-axis and $\pi$ rotation about the $x$-axis and/or time reversal.
In this case, they are equivalent to hard-core boson systems at half-filling.
Using bosonization technique, one can find a phase transition from a quasi-long-range superfluid order to a translational symmetry breaking insulator.
This phase transition is the quantum version of the famous Kosterlitz--Thouless~(KT) transition~\cite{KosterlitzThouless1973, Kosterlitz1974, Haldane1982}.
The reason why we get the translational symmetry breaking insulator is related to properties of a $U(1)$ vortex: in half-filled systems, a spacetime vortex carries momentum $\pi$.
So, when proliferating vortices, one kills the quasi-long-range order while developing VBS order at the same time.

In this paper, we consider a translationally symmetric spin-1/2 chain with anisotropic spin-spin interactions which break the $SO(3)$ spin-rotation symmetry but preserve $\pi$-rotation symmetries about three orthogonal axes.
In other words, the global on-site symmetry is $Z_2^x \times Z_2^z$, which acts projectively on one unit cell.
Besides, we require that our system also hosts time reversal symmetry $\TT$.
We will propose and analyze a possible direct transition in such a spin-1/2 chain between a ferromagnetic (FM) phase (say, with magnetization in the $z$-direction) and a valence bond solid phase, and will draw interesting parallels with the 2d easy-plane DQCP (EP-DQCP).
To have a concrete system in mind, we will consider a model with ferromagnetic nearest-neighbor interactions and anti-ferromagnetic second-neighbor interactions.

The paper is organized as follows:
In Section~\ref{sec:summary}, we briefly mention various approaches to this problem and summarize our main results.
In Section~\ref{sec:model}, we present a concrete spin model to ground our analysis and give some rough idea about the phase diagram. 
In Sections~\ref{sec:bosonization}, \ref{sec:domain_wall}, and \ref{sec:dw_parton}, we use various techniques, including bosonization, duality, and parton constructions, to build up analysis of this model.
In Section~\ref{sec:goodvars}, we arrive at ``good variables'' to provide complete description of the FM to VBS transition; this is our key section in the paper.
In Section~\ref{sec:fermionic_parton}, we use the fermionic parton approach to describe the criticality.
Finally, in Section~\ref{sec:conclusion}, we discuss possible generalizations and future directions.
Several appendices contain some of the more technical details.
Particularly noteworthy are Appendix~\ref{app:ising_dual} presenting an interesting exact formulation of the Ising duality in 1d used throughout the paper; Appendix~\ref{app:deriv_good_vars} presenting a non-parton perspective on the good variables; and Appendix~\ref{app:qn_fermionic_parton} presenting a general algorithm to extract quantum numbers for Gutzwiller-projected wavefunctions.

\section{Summary of results}~\label{sec:summary}
From the outset, we should say that analytical tools available in 1d are very powerful, and, with many known results on related problems, one can get to our main results in many different ways.
Thus, we can start with a bosonized description of a $U(1)$-symmetric XZ spin chain and add spin anisotropy, and realize that the transition is likely described by a strongly-coupled field theory with precisely balanced competing cosines of standard conjugate phase and density variables, see Eq.~(\ref{eq:direct_bosonization_action}).
The structure resembles $Z_4$ clock ordering transition~\cite{JoseKadanoffKirkpatrickNelson1977, LecheminantGogolinNersesyan2002},
but with different ``periodicity'' conditions on the field variables, related to the fact that in the present case we have two-fold ground state degeneracy on both sides of the transition, while in the $Z_4$ clock model the transition is from a non-degenerate to four-fold degenerate ground states.
At this point, we can appeal to the precisely balanced structure of our theory and known properties of the $Z_4$ clock transition and already guess some properties of our ferromagnet to VBS transition, but with nagging questions about the different periodicity structure and physical observables in our case~\cite{LecheminantGogolinNersesyan2002}.

We will get to assuredly right results by a more circuitous route that will closely resemble developments in the 2d EP-DQCP theory.
The specific 2d setting for drawing such parallels has spin-1/2's on the square lattice with easy-plane ferromagnetic interactions plus additional interactions that can drive transition from the EP-ferromagnet to the VBS phase; important symmetries are $U(1)$ symmetry of spin rotations in the easy plane, symmetry of $\pi$-rotations around an in-plane axis, and time reversal symmetry, plus lattice symmetries.
This is qualitatively equivalent to a half-filled bosonic system with unfrustrated hopping, with specific symmetries (in particular, guaranteeing the half-filling), and interactions that drive superfluid to Mott insulator transition, where the insulator has valence-bond character.

We will start by thinking in terms of topological defects in the ordered phase.
In 2d EP systems, these are vortices, which are quantum particles coupled to a non-compact gauge field [i.e., $U(1)$ gauge field with no monopoles], and the superfluid order is destroyed by proliferating the vortices~\cite{Peskin1978, DasguptaHalperin1981, FisherLee1989}.
In the 2d EP-DQCP setting, the half-filling of the bosonic system leads to the presence of two low-energy vortex fields with non-trivial symmetry transformation properties, and simultaneous condensation of these fields produces the VBS Mott insulator~\cite{LannertFisherSenthil2001}.

In 1d Ising systems, topological defects are domain walls, which are also quantum particles; this description is typically obtained using Ising duality transformation that involves string operators, with subtleties arising when treating finite systems with periodic boundary conditions, which are understood but often ignored.
We will advocate an exact statement of this duality that does not use string operators but instead has the dual Ising field coupled to a $Z_2$ gauge field with no instanton dynamics, resembling the absence of monopole dynamics in the dual vortex theory for bosons in 2d.
The original Ising symmetry is encoded in the flux conservation of the dual $Z_2$ gauge field, paralleling how the $U(1)$ symmetry of bosons in 2d is encoded in the flux conservation of the dual gauge field in the dual vortex theory.
While this interpretation of the Ising duality does not give new results for the thoroughly understood quantum Ising chain, it will prove very useful in more complex situations that we will encounter, in particular with richer field content and/or where instanton operators are allowed in the dynamics.
For the 1d systems with $Z_2^x \times Z_2^z$ symmetry, the second $Z_2$ symmetry plays a role similar to the particle-hole symmetry in the 2d EP-DQCP systems, giving rise to two low-energy domain wall fields with non-trivial transformation properties, in particular, whose momenta differ by $\pi$.
These domain wall fields are coupled to the $Z_2$ gauge field with no instanton dynamics.
Simultaneous condensation of both such domain wall species gives the VBS order in this system.
Thus, the VBS order parameter is simply expressed in the domain wall variables.
On the other hand, the ferromagnetic order parameter is encoded in the instanton operator.

Microscopically, these two domain wall species can convert from one to another, but this inter-conversion may be suppressed on long length scales.
If we suppress the inter-conversion by hand, the theory has the structure of a \emph{gauged} version of the celebrated Ashkin-Teller (AT) model~\cite{AshkinTeller1943}.
At this point, one can appeal to known results for the criticality in the AT model~\cite{KadanoffWegner1971, JoseKadanoffKirkpatrickNelson1977, Kadanoff1979, KohmotoNijsKadanoff, Delfino1999, LecheminantGogolinNersesyan2002, RamolaDamleDhar2015, ChewMrossAlicea2018}.
However, extreme care is needed when identifying how symmetries and local observables are represented in the actual fully controlled theory of the transition, which is not in terms of the Ising fields but instead involves two rather special duality transformations~\cite{KohmotoNijsKadanoff}.

To proceed systematically in our problem, we will first dualize the effective theory of the two domain walls and will obtain a new theory with two Ising variables that are coupled to a new $Z_2$ gauge field.
Under this duality, the tunneling between the two domain wall species maps precisely to allowing instanton dynamics in the new gauge field, see Eqs.~(\ref{eq:model_mu_pm}) and (\ref{eq:model_bosonic_parton}).
The dual theory can in fact be viewed as an effective theory for a ``parton'' approach where one tries to ``fractionalize'' the ferromagnetic order parameter.
It is well known that parton approaches lead to gauge theories with allowed instantons, and our specific parton construction has $Z_2$ gauge structure.
As far as the FM and VBS order parameters are concerned, the situation is reversed compared to the domain wall theory: the FM order parameter is now readily represented using the parton fields, while the VBS order parameter is encoded in the instantons of the corresponding $Z_2$ gauge field.

The above description parallels the original development of the 2d EP-DQCP~\cite{DQCP_science, DQCP_prb}, where the theory of two vortices with inter-conversion between the two species on the lattice scale is dual to a theory of partons with $U(1)$ gauge structure and with allowed monopoles carrying specific Berry phases.
Also, the VBS order parameter is easily expressed in the vortex variables, while the superfluid order parameter is easily represented in the parton variables.
In the 2d EP-DQCP theory on the square lattice, going from the lattice to the continuum theory in the vortex fields, only quadrupled inter-conversions between vortex species survive due to lattice $C_4$ rotation symmetry (in the parton language, only quadrupled monopoles survive), and the theory~\cite{DQCP_science, DQCP_prb} conjectures that these processes are irrelevant at the transition, leading to so-called easy-plane non-compact CP$^1$ (EP-NCCP$^1$) field theory
(see Sec.~\ref{subsec:parallelsNCCP1} below for a brief recap).
In our 1d case, the inter-species tunneling completely disappears in the continuum limit in the domain wall fields, since a ``doubled'' domain wall is not distinguishable from no domain wall on long length scales (in the parton language, doubled spacetime vison is not distinguishable from no vison).
One should still worry if one can use such a continuum limit, since these Ising-like fields are not the best variables to describe the AT criticality, but we will later see controlled treatments and precise meaning for this phenomenon.

If we ignore the domain wall inter-conversion (instanton dynamics in the parton language), the emergent self-dual structure of the gauged AT model is reminiscent of the self-duality in the EP-NCCP$^1$ model~\cite{MotrunichVishwanath2004}; in particular, one can already see that the ferromagnetic and VBS order parameters should behave similarly at the transition.
However, just like the EP-NCCP$^1$ field theory is not tractable in its field variables, the gauged AT model is also not tractable in its variables.

We then embark on finding precise analogs of ``good variables'' that help understand the AT criticality as a Gaussian theory with only one relevant cosine operator, where the transition is obtained by tuning this coupling through zero.
We succeed in finding an exact such reformulation for our effective domain wall theory (parton theory), with the final theory, Eq.~(\ref{eq:model_at_dual_two}), having the structure of an XY chain on quarter-integer sites (where the physical spins reside on integer sites), with staggered bond couplings.
The transition corresponds to changing the sign of the staggering, or more precisely, to moving the pattern of entangled pairs in the ground state from one sublattice of bonds to the other in this quarter-integer chain.
The full structure in our problem is that this XY chain is gauged by a new $Z_2$ gauge field with allowed instanton dynamics.
Here we can truly apply the power of the bosonization in 1d and obtain complete description of the transition, including unification of the FM and VBS order parameters
and precise understanding of all observables.
This is the main result of the paper.

A possible parallel to our good variables in the 2d EP-DQCP problem is an attempt to describe this transition by ``unifying'' the $U(1)$ superfluid order parameter and the emergent $U(1)$ VBS order parameter into an $O(4)$ vector, mapping to an anisotropic $O(4)$ non-linear sigma model with a topological term at $\theta = \pi$~\cite{TanakaHu2005, SenthilFisher2006}.
Furthermore, the self-duality of the 2d EP-NCCP$^1$ model at criticality suggests a discrete ``symmetry'' rotating the superfluid to VBS order parameter, thus reducing the number of allowed terms that break the $O(4)$ symmetry at criticality.
A very interesting (but not established) possibility is that all such symmetry-allowed terms that break the $O(4)$ symmetry are irrelevant, and the $O(4)$ symmetry emerges at criticality; the transition is then driven by a single relevant perturbation that breaks the ``symmetry'' between the superfluid and VBS order parameters.
While this scenario is only conjectured in the 2d EP-DQCP problem, something like this does appear to happen in our 1d Ising DQCP problem.
However, we emphasize that our approach does not start from the FM and VBS order parameters but is more microscopic and directly attacks the good-variable reformulation using abelian bosonization.
The unification of the two order parameters does happen, but these are encoded in instantons of the good-variable model.
Our critical point indeed has emergent symmetry and allows only a single relevant perturbation that drives the transition between the ferromagnet and VBS phases.
Interestingly, in the 1d Ising DQCP, we actually have a line of fixed points with continuously varying critical indices.

While our good variables provide essentially complete description of the 1d Ising DQCP, we will continue with more perspective on this problem, with the hope of learning useful lessons for the 2d DQCP theories.
First, we will provide an interesting perspective on the good variables as a different parton theory where we do not try to fractionalize the FM or VBS order parameter, but instead ``fractionalize'' some other order parameters that are not present on either side of the transition.
In this new parton theory, the partons remain gapped (i.e., are not condensed) on either side of the transition; instead, one can think of the partons as being in distinct fully symmetric parton phases, i.e., ``parton SPT'' phases.
Both the FM and VBS order parameters crucially contain $Z_2$ instanton operators, which are nevertheless tractable in bosonization.
We do not know of a similar picture in the 2d EP-DQCP problem.

This somewhat unexpected ``good parton'' solution will motivate yet another perspective on the transition, now without using partons.
We can in fact relate the FM to VBS transition to the following problem.
Starting with an XY system with $U(1)$ symmetry, with ferromagnetic nearest-neighbor interactions, it is well known that second-neighbor antiferromagnetic interactions can drive it to the VBS order, while nearest-neighbor antiferromagnetic $\sigma^z$-$\sigma^z$ interactions can drive it to an Ising antiferromagnet~\cite{Haldane1982, SachdevBook, FurukawaSatoFurusaki2010, FurukawaSatoOnodaFurusaki2012}.
One also understands the transition from the Ising ferromagnet to the VBS order, which is described by a Gaussian theory with one relevant cosine, whose coupling changes sign across the transition.
Breaking the $U(1) \times Z_2^x$ symmetry down to $Z_2^z \times Z_2^x$, in fact, gives only irrelevant perturbations over a large window of the phase transition line; thus, we know how to describe the Ising antiferromagnet to VBS transition in such an anisotropic spin model with nearest-neighbor and second-neighbor interactions.
Now, we can simply rotate the spins on every-other site by $\pi$ around the $\hat{x}$ axis, thus obtaining a model with nearest-neighbor ferromagnetic $\sigma^x \sigma^x$ and $\sigma^z \sigma^z$ interactions and antiferromagnetic second-neighbor interactions that undergoes a transition from the ferromagnet in the $\sigma^z$ direction to a VBS phase.

For completeness, we will also explore fermionic parton approaches, looking for parallels with fermionic $N_f = 2$ QED$_3$ and fermionic parton descriptions of the 2d EP-DQCP problem~\cite{SenthilFisher2006, XuYou2015, KarchTong2016, SeibergSenthilWangWitten2016, WangNahumMetlitskiXuSenthil2017, MrossAliceaMotrunich2017}.
One such fermionic parton construction is motivated by a naive attempt to Jordan-Wigner fermionize the domain wall/bosonic parton theory (this is loosely in analogy of how the $N_f = 2$ QED$_3$ description is obtained by fermionizing bosonic partons in the EP-NCCP$^1$ description in Refs.~\cite{SenthilFisher2006, XuYou2015, KarchTong2016, SeibergSenthilWangWitten2016, WangNahumMetlitskiXuSenthil2017, MrossAliceaMotrunich2017}, although one does not use Jordan-Wigner fermionization there).
This naive approach fails to produce theory that could be consistently interpreted within a fermionic parton approach for the FM to VBS transition.
Nevertheless, we are able to guess one mean field ansatz (more precisely, one projective symmetry group or PSG) within the motivated parton decomposition that reproduces both the FM and VBS phases, which are represented as distinct (in the SPT sense) paired phases of the fermionic partons.
Interestingly, when we try a Jordan-Wigner fermionization of the good parton variables instead, this can be more readily interpreted as a consistent fermionic parton approach.
This gives a different parton decomposition and ansatz where the two phases are represented as topologically distinct ``bipartite hopping'' gapped phases of the new fermionic partons.
However, we find that both fermionic parton approaches have the same PSG equations, and in fact can be exactly mapped to each other, so we can focus on just one.
One lesson from the fermionic partons is that as long as we can access the desired phases and transition within the same PSG, the fermionic partons appear to immediately provide ``good variables'' for describing the criticality.
We do not know of similar fermionic parton approaches to the 2d DQCP problems, where both sides of the transition are gapped SPT-like phases of partons within the same PSG.

\section{Model and symmetries} \label{sec:model}

\begin{figure}
    \centering
    \includegraphics{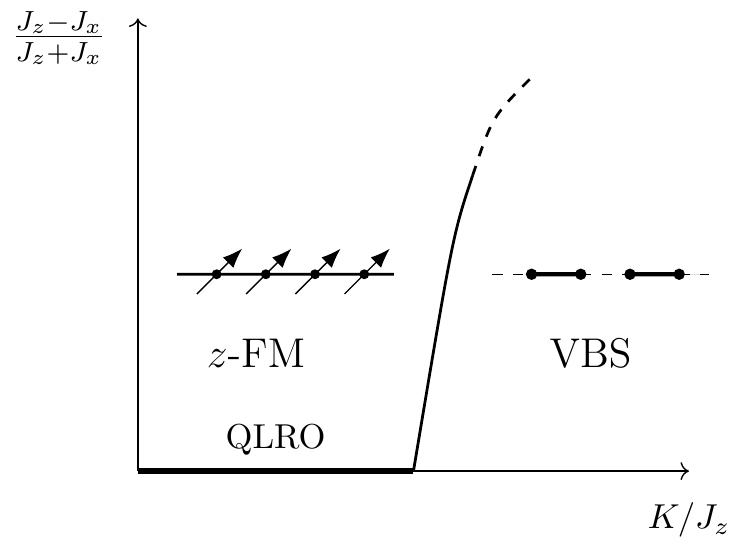}
    \caption{A schematic phase diagram for the spin model defined in Eq.~(\ref{eq:spin_model}). 
    Here, we set $K \equiv K_{2x} = K_{2z}$.
    When $J_z = J_x$, we are on the horizontal axis, and the Hamiltonian has $U(1)$ symmetry; the corresponding phase diagram was studied in Refs.~\cite{FurukawaSatoFurusaki2010, FurukawaSatoOnodaFurusaki2012}, where increasing $K$ leads to a direct Kosterlitz-Thouless-like transition from the Luttinger liquid phase with quasi-long-range order to the dimerized~(VBS) phase.
    Adding anisotropy between the $J_z$ and $J_x$ couplings kills the Luttinger liquid phase, and helps to develop ferromagnetic order, here $z$-FM for $J_z > J_x$.
    According to our analysis, there is a critical line between the $z$-FM and VBS phases, with varying critical exponents that depend on a single parameter $\tilde{g}$ defined in Eq.~(\ref{eq:gauged_bosonization_action}): the correlation length exponent is given in Eq.~(\ref{eq:nu}) and the scaling dimension for both the $z$-FM and VBS order parameters in Eq.~(\ref{eq:scalingdims_zFM_VBS}).
    The dashed line in this figure denotes undetermined behaviors.}
    \label{fig:phase_diagram}
\end{figure}

To be concrete, let us consider a spin-1/2 chain, with anisotropic spin-spin interactions.
The Hamiltonian reads
\begin{align}
  H = & \sum_j \left( -J_x \, \sigma^x_j \sigma^x_{j+1} - J_z \, \sigma^z_j \sigma^z_{j+1} \right) \notag \\ 
  & + \sum_j \left( K_{2x} \, \sigma^x_j \sigma^x_{j+2} + K_{2z} \, \sigma^z_j \sigma^z_{j+2} \right) + \cdots ~.
  \label{eq:spin_model}
\end{align}
This model is invariant under translational symmetry $T_x$.
For general coupling constants $J_\alpha$, the full spin rotation symmetry $SO(3)$ is broken down to $Z_2^x \times Z_2^z$ symmetry, which is generated by $g_{x,z}$ defined below.
Due to the absence of spin interaction terms containing an odd number of spin operators, this model also hosts time reversal symmetry $\TT$.
To summarize, generators of the global symmetry group are
\begin{align}
  T_x &:\ \sigma^\alpha_j \rightarrow \sigma^\alpha_{j+1} ~; \notag \\
  g_x \equiv \prod_j \sigma^x_j &:\ \sigma^x_j \rightarrow \sigma^x_j ~, \; \sigma^{y,z}_j \rightarrow -\sigma^{y,z}_j ~; \notag \\
  g_z \equiv \prod_j \sigma^z_j &:\ \sigma^{x,y}_j \rightarrow -\sigma^{x,y}_j ~, \; \sigma^z_j \rightarrow \sigma^z_j ~; \notag \\
  \TT \equiv \left( \prod_j \ii \sigma^y_j \right) \! \mathcal{K} &:\ \sigma^\alpha_j \rightarrow -\sigma^\alpha_j ~, \; \ii \rightarrow -\ii ~.  
\end{align}
The specific model also has lattice inversion symmetry $I: \sigma^\alpha_j \to \sigma^\alpha_{-j}$; for most of the discussion, this symmetry will not play any role.
The ``$\cdots$'' terms in Eq.~(\ref{eq:spin_model}) represent other terms respecting the global symmetry group, such as $\sum_j -J_y \sigma^y_j \sigma^y_{j+1}$, etc.

We point out that single-unit-cell parts of $g_x$ and $g_z$ actually anti-commute, $\sigma_j^x \sigma_j^z = -\sigma_j^z \sigma_j^x$.
Although excitations of this Hamiltonian are linear representations of the $Z_2^x \times Z_2^z$ symmetry,
this symmetry acts projectively on each unit cell.
Hence, the generalized LSM theorem can be applied here.
Namely, it is impossible to construct any local Hamiltonian with a gapped and fully symmetric ground state (i.e., respecting all the internal and translation symmetries).

Turning to a more concrete setting such as the above Hamiltonian, we will focus on the regime where all couplings in Eq.~(\ref{eq:spin_model}) are positive; that is, we have ferromagnetic nearest-neighbor interactions and antiferromagnetic second-neighbor interactions.
It is easy to identify simple phases in this model.
When the $J_z$ term dominates, we have ferromagnetic ordering of spins along the $\sigma^z$ direction, which breaks $g_x$ and $\TT$ but preserves $g_z$ and $T_x$; we will call this phase ``$z$-FM.''
On the other hand, when the $J_x$ term dominates, we have ferromagnetic ordering of spins along the $\sigma^x$ direction, which instead breaks $g_z$ but preserves $g_x$; we will label this phase ``$x$-FM.''
For $J_x = J_z$ and $K_x = K_z$, this model has $U(1)$ symmetry and was studied in Refs.~\cite{FurukawaSatoFurusaki2010, FurukawaSatoOnodaFurusaki2012}
Moderate second-neighbor couplings drive the system into a phase which is closely related to the celebrated dimerized phase in the $J_1 - J_2$ antiferromagnetic Heisenberg model~\cite{MajumdarGhosh1969, MajumdarGhosh1969II, Haldane1982, SachdevBook}; it preserves all internal symmetries but breaks the translation symmetry, and we will refer to this phase as ``valence bond solid'' (VBS).
With the $U(1)$ symmetry and for dominant $J_x = J_z$, we have Luttinger liquid phase with quasi-long-range order (QLRO), and there is a direct transition from this phase to the VBS phase, which is in the Kosterlitz-Thouless universality.
We are interested in the case with only discrete spin symmetries $g_x$ and $g_z$, in which case the QLRO is replaced by long range orders such as in the $z$-FM and $x$-FM phases.
The corresponding LSM theorem still guarantees that there is no fully symmetric phase which can intervene between such magnetically ordered phases and the VBS phase (although other phases, e.g., with coexisting magnetic and VBS orders may still intervene).
Our main focus is on the possibility of a direct continuous transition, say, between the $z$-FM phase and the VBS phase, and how to describe such criticality.
A schematic phase diagram is presented in Fig.~\ref{fig:phase_diagram}.

\section{Direct bosonization method} \label{sec:bosonization}
Spin-1/2 chains can be studied using bosonization techniques~\cite{Haldane1981, Shankar_Acta, GiamarchiBook, SachdevBook}.
To achieve this, we map each spin-1/2 to a hard-core boson, which is then approximated by a $U(1)$ quantum rotor as follows:
\begin{align}
& \sigma^y_j \sim 2 \left(n_j - \frac{1}{2} \right) ~, \notag \\
& \sigma^z_j \sim \cos(\phi_j) ~, \quad \sigma^x_j \sim -\sin(\phi_j) ~, 
\label{eq:rotorvars}
\end{align}
where the rotor number and phase variables satisfy $[n_j, \phi_{j'}] = \ii \delta_{jj'}$. 

To obtain a hydrodynamic description, we define new variables
\begin{align}
\theta_{j+1/2} = \sum_{j' \le j} \pi \, n_{j'} - \frac{\pi}{2} j ~,
\end{align}
such that
\begin{align}
\sigma^y_j \sim 2 (\theta_{j+1/2} - \theta_{j-1/2})/\pi ~.
\end{align}
The commutator between the $\theta$ and $\phi$ variables reads $[\theta_{j+1/2}, \phi_{j'}] = \ii \pi\, \Theta(j + 1/2 - j')$, where $\Theta(x)$ is a Heaviside step function.

To get an effective field theory description, let us work with long-wavelength fields defined in the continuum space.
The field $\theta(x)$ is a real-valued variable with periodicity $\pi$ (more precisely, a global shift by $\pi$ corresponds to the same physical state), while $\phi(x)$ has periodicity $2\pi$.
Their commutation relation reads
\begin{align}
\left[ \frac{\partial_x \theta(x)}{\pi}, \phi(x') \right] = \ii \delta(x - x') ~.
\end{align}
The $\sigma^{z,x}$ spin components have dominant contributions at zero momentum, which can be obtained by simply replacing $\phi_j$ in Eq.~(\ref{eq:rotorvars}) with the long-wavelength field $\phi(x)$.
On the other hand, the $\sigma^y$ component also obtains an important contribution at wavevector $\pi$:
\begin{align}
& \sigma^y_j \sim \frac{2\partial_x \theta}{\pi} + A (-1)^j \sin(2\theta) ~; \notag \\
& B_{j+1/2} \sim C (-1)^j \cos(2\theta) ~.
\end{align}
In the last line, we have also shown a similar important contribution at wavevector $\pi$ to a bond energy $B_{j+1/2}$ [which can be essentially any symmetric term associated with a bond $(j, j+1)$].

We can readily identify how the symmetries of our model act on the continuum fields:
\begin{align}
& T_x:\ \phi \rightarrow \phi ~, \quad \theta \rightarrow \theta + \frac{\pi}{2} ~; \notag \\
& g_x:\ \phi \rightarrow -\phi + \pi ~, \quad \theta \rightarrow -\theta ~; \notag \\
& g_z:\ \phi \rightarrow -\phi ~, \quad \theta \rightarrow -\theta ~; \notag \\
& \TT:\ \phi \rightarrow \phi + \pi ~, \quad \theta \rightarrow -\theta ~, \quad \ii \rightarrow -\ii ~.
\label{eq:sym_bosonization_cont}
\end{align}
Notice that for these continuum fields, $T_x$ acts as an internal symmetry.
Due to the $T_x$ symmetry, terms like $\cos(2\theta)$ are not allowed.

The symmetry-preserving action in the Euclidean space-time reads
\begin{align}
S[\phi, \theta] \!=\! &\int\! \dd\tau \, \dd x \left[ \frac{\ii}{\pi} \partial_\tau \phi \, \partial_x \theta + \frac{v}{2\pi} \left( \frac{1}{g} (\partial_x \theta)^2 + g (\partial_x \phi)^2 \right) \right] \notag \\
& + \int\! \dd\tau \, \dd x \left[ \lambda_u \cos(4\theta) + \lambda_a \cos(2\phi) \right] + \cdots ~,
\label{eq:direct_bosonization_action}
\end{align}
where ``$\cdots$'' terms include less important symmetry-allowed terms.
\footnote{We point out that $\sum_j (-1)^j \cos(2\theta_{j+1/2})$ is allowed on the lattice.
However, due to the staggered phase, this sum has rapid oscillations (assuming small coupling and hence slowly-varying field $\theta$) and hence disappears in the continuum.}

Now, let us identify gapped phases that can be described by this field theory.
We consider the following four limits:
\begin{enumerate}
  \item $\lambda_u \gg 0, \, \lambda_a \sim 0$:
    To minimize the action, $\theta$ is pinned to $\theta = \pi/4$ or $3\pi/4$, corresponding to two degenerate ground states.
    According to Eq.~(\ref{eq:sym_bosonization_cont}), we identify this phase as an antiferromagnetic phase with spins pointing in the $\sigma^y$ direction.
    Indeed, either such ground state breaks $T_x$, $g_x$, $g_z$, and $\TT$, and preserves $g_y \equiv \prod_j \sigma^y_j \sim g_x g_z$.
    We label this phase as $y$-AFM.

  \item $\lambda_u \ll 0, \, \lambda_a \sim 0$:
    To minimize the action, $\theta$ is pinned to $\theta = 0$ or $\pi/2$.
    This phase is identified as VBS phase, which only breaks the translational symmetry and preserves all on-site symmetries (remember that a global shift of $\theta$ by $\pi$ produces the same physical state).

  \item $\lambda_u \sim 0, \, \lambda_a \gg 0$:
    To minimize the action, $\phi$ is pinned to $\phi = \pi/2$ or $3\pi/2$.
    This phase is ferromagnetic phase with spins pointing in the $\sigma^x$ direction, labeled as $x$-FM.

  \item $\lambda_u \sim 0, \, \lambda_a \ll 0$:
    To minimize the action, $\phi$ is pinned to $\phi = 0$ or $\pi$.
    This phase is ferromagnetic phase along the $\sigma^z$ direction, labeled as $z$-FM.
\end{enumerate}

After identifying the phases, let us discuss possible phase transitions between these phases described by the field theory in Eq.~(\ref{eq:sym_bosonization_cont}) with small $\lambda_{u,a}$.
To see the nature of phase transitions, we calculate scaling dimensions for various small perturbations near the Gaussian fixed point.
This calculation is standard textbook problem, and results are
\begin{align}
  \text{dim}[\cos(2n\theta)] = n^2 g ~, \quad \text{dim}[\cos(m\phi)] = \frac{m^2}{4g} ~.
\end{align}

We first consider the free fermion limit, which corresponds to $g = 1$.
In this case, $\text{dim}[\cos(2\phi)] = 1$ and $\text{dim}[\cos(4\theta)] = 4$.
Thus, $\cos(2\phi)$ is relevant, and $\cos(4\theta)$ is irrelevant.
Hence a continuous phase transition happens when $\lambda_a$ changes sign, which describes critical theory between the $x$-FM and $z$-FM phases.
This phase transition point is realized by the lattice Hamiltonian in Eq.~(\ref{eq:spin_model}) with $J_x = J_z$ and vanishing other terms.
By Jordan-Wigner transformation, one can see that it is equivalent to a free fermion Hamiltonian.
It is worth noting, however, that the same field theory describes the $x$-FM to $z$-FM transition also when the spin model never passes through such a special point with the additional $U(1)$ symmetry, e.g., in the presence of the second-neighbor anisotropic spin interactions.
[In some sense, the $x$-FM to $z$-FM transition is also a ``Landau-forbidden'' continuous transition, ``protected'' in our spin chain by the discrete symmetries.]

Actually, when $g > 1/2$, we always have $\cos(2\phi)$ relevant and $\cos(4\theta)$ irrelevant.
In this parameter range, a continuous phase transition between the $x$-FM and $z$-FM is obtained by changing sign of $\lambda_a$.
However, the critical indices vary for different values of $g$.
The correlation length exponent is given by $\nu = 1/(2 - \text{dim}[\cos(2\phi)]) = g/(2g - 1)$, while the FM order parameters have power-law decay $\sim x^{-1/(2g)}$.

Similarly, when $0 < g < 1/2$, $\cos(2\phi)$ is irrelevant, and $\cos(4\theta)$ is relevant.  
Thus, in this parameter range, the field theory in Eq.~(\ref{eq:direct_bosonization_action}) describes a continuous phase transition between the VBS and $y$-AFM when $\lambda_u$ changes sign, with varying critical indices that depend on $g$.

An interesting phenomenon happens when $g = 1/2$.
In this case, $\text{dim}[\cos(2\phi)] = \text{dim}[\cos(4\theta)] = 2$.
Thus, these two allowed perturbations are both marginal.
For our purposes, let us consider the case $\lambda_u, \lambda_a < 0$.
For very small (but comparable) $\lambda_u, \lambda_a$, by varying $g$ away from $1/2$, we obtain either the $z$-FM phase or the VBS phase depending on the sign of $g - 1/2$.
When $g = 1/2$ with small $\lambda_u, \lambda_a$ and $|\lambda_u| = |\lambda_a|$, these two perturbations are competing with each other and neither can win.
We expect that such a theory describes a continuous transition between the $z$-FM and VBS phases, {which is of main interest in this paper.
Note that the proposed critical theory has finely balanced competing cosine terms, which is a non-perturbative situation in these variables.
Nevertheless, from the $\phi \leftrightarrow 2\theta$ ``symmetry'', we can already guess that the $z$-FM order parameter represented by $\cos(\phi)$ and the VBS order parameter represented by $\cos(2\theta)$ will have the same scaling dimensions.
Similarly, the $x$-FM and $y$-AFM order parameters represented respectively by $\sin(\phi)$ and $\sin(2\theta)$ will have the same scaling dimensions.
Note also that since the strictly marginal term $\cos(2\phi) + \cos(4\theta)$ has cosines rather than sines, the scaling dimensions of $\cos(\phi)$ and $\sin(\phi)$ are different, and similarly for $\cos(2\theta)$ vs $\sin(2\theta)$.

The structure of the critical theory is very similar to the $Z_4$ clock transition~\cite{JoseKadanoffKirkpatrickNelson1977, LecheminantGogolinNersesyan2002},
except that in the latter the competing cosines would be $\cos(4\phi)$ and $\cos(2\theta)$, and since the fields $\phi$ and $\theta$ have different periodicities, the ground state degeneracies on the two sides of the clock ordering transition (1 and 4) would be different from the $z$-FM to VBS case (2 and 2).
We can still use this similarity and known properties of the $Z_4$ criticality, which is also related to the Ashkin-Teller criticality, to guess further properties of our $z$-FM to VBS transition, such as the fact that one has continuously varying critical indices and relations among them.
However, such an understanding of the $Z_4$ clock and Ashkin-Teller criticality~\cite{KadanoffWegner1971, JoseKadanoffKirkpatrickNelson1977, Kadanoff1979, KohmotoNijsKadanoff, Delfino1999, LecheminantGogolinNersesyan2002, RamolaDamleDhar2015, ChewMrossAlicea2018}
is not obtained by thinking in direct variables, and instead involves a highly non-local transformation to new ``good variables,'' in which the critical theory has the structure of a gaussian field theory with a \emph{single} relevant cosine.
It is a non-trivial task to make precise connections between our physical system and such a convenient field theory.
Rather than guessing here, in subsequent sections we will derive good variables appropriate for our model, where we will correctly capture physical observables and all global aspects such as the ground state degeneracy in each phase.}

\section{Domain wall description}\label{sec:domain_wall} 
In this section, we will consider a dual description of the original spin model using domain wall variables by performing Ising duality.
It turns out that the nature of phase transitions becomes more clear in these dual variables.
A detailed discussion of the Ising duality can be found in Appendix~\ref{app:ising_dual}.

\subsection{Model in dual variables}
We define dual variables as operator correspondences
\begin{align}
& \mu^x_{j+1/2} = \sigma^z_j \sigma^z_{j+1} ~, \notag \\ 
& \mu^z_{j-1/2} \, \rho^z_j \, \mu^z_{j+1/2} = \sigma^x_j ~, \notag \\
& \rho^x_j = \sigma^z_j ~, \notag \\
& \text{gauge constraint\ :}\  \rho_j^x \rho_{j+1}^x = \mu^x_{j+1/2} ~.
\label{eq:dw_dual_map}
\end{align}
Here, $\mu$ degrees of freedom reside on the dual lattice labeled by half-integers and can be roughly thought as describing domain walls in the order parameter of the $z$-FM phase.
As explained in Appendix~\ref{app:ising_dual}, we keep track of the global symmetries by introducing $Z_2$ gauge fields $\rho$ on the links of the dual lattice; these can be labeled either as $\rho_{j-1/2, j+1/2}$ on the link between $j-1/2$ and $j+1/2$, or more compactly as $\rho_j$.
The physical Hilbert space is defined by the gauge constraint (``Gauss law'') on each site of the dual lattice.
In this language, the $\mu$ variables can also be thought as describing matter field that carries $Z_2^\rho$ gauge charge of the gauge field $\rho$.

In these dual variables, the Hamiltonian in Eq.~(\ref{eq:spin_model}) becomes
\begin{align}
H = & -J_x \sum_j \mu^z_{j-1/2} \rho^z_j \rho^z_{j+1} \mu^z_{j+3/2} - J_z \sum_j \mu_{j+1/2}^x \notag \\
& + K_{2x} \sum_j \mu^z_{j-3/2} \rho^z_{j-1} \mu^z_{j-1/2} \, \mu^z_{j+1/2} \rho^z_{j+1} \mu^z_{j+3/2} \notag \\
& + K_{2z} \sum_j \mu_{j-1/2}^x \mu_{j+1/2}^x + \cdots ~,
\label{eq:model_dw}
\end{align}
with the gauge constraint in Eq.~(\ref{eq:dw_dual_map}).
We emphasize that on a chain with periodic boundary conditions, this is an exact rewriting of the original spin Hamiltonian, see Appendix~\ref{app:ising_dual}.

\subsection{Symmetry analysis in dual variables}
Now, let us analyze how the symmetries act on the dual variables $\mu$ and $\rho$.
According to Eq.~(\ref{eq:dw_dual_map}), the symmetry generators can be expressed in the dual variables as
\begin{align}
& g_x = \prod_j \rho^z_j ~, \notag \\
& g_z = \prod_\ell \mu^x_{2\ell-1/2} ~, \notag \\
& \TT = \left( \prod_\ell \mu^x_{2\ell-1/2} \, \prod_j \rho^z_j \right) \! \mathcal{K} ~.
\label{eq:sym_in_dual_variables}
\end{align}
The complex conjugation $\mathcal{K}$ is in the standard $\mu^x, \rho^x$ eigenbasis and coincides with the complex conjugation in the $\mu^z, \rho^z$ eigenbasis.
When writing $g_z$ and $\TT$, we have assumed periodic boundary conditions and even length of the chain, which we will assume throughout the paper.
It is straightforward to obtain how the symmetries act on the dual variables:
\begin{align}
T_x:\ & \rho^\alpha_j \rightarrow \rho^\alpha_{j+1} ~, \quad \mu^\alpha_{j-1/2} \rightarrow \mu^\alpha_{j+1/2} ~; \notag \\
g_x:\ & \rho^{x,y}_j \rightarrow -\rho^{x,y}_j ~, \quad \rho^z_j \rightarrow \rho^z_j ~, \notag \\
& \mu^\alpha_{j+1/2} \rightarrow \mu^\alpha_{j+1/2} ~; \notag \\
g_z:\ & \rho^\alpha_j \rightarrow \rho^\alpha_j ~, \notag \\
& \mu^x_{j+1/2} \rightarrow \mu^x_{j+1/2} ~, \quad \mu^{y,z}_{j+1/2} \rightarrow (-1)^j \mu^{y,z}_{j+1/2} ~; \notag \\
\TT:\ & \rho^x_j \rightarrow -\rho^x_j ~, \quad \rho^{y,z}_j \rightarrow \rho^{y,z}_j ~, \notag \\
& \mu^x_{j+1/2} \rightarrow \mu^x_{j+1/2} ~, \quad \mu^y_{j+1/2} \rightarrow (-1)^{j+1} \mu^y_{j+1/2} ~, \notag \\
& \mu^z_{j+1/2} \rightarrow (-1)^j \mu^z_{j+1/2} ~, \quad \ii \to -\ii ~.
\label{eq:dw_sym}
\end{align}

Notice that the form of the symmetry actions is far from unique due to the $Z_2^\rho$ gauge constraint: any $\mu^x_{j+1/2}$ in Eq.~(\ref{eq:sym_in_dual_variables}) can be replaced by $\rho^x_j \rho^x_{j+1}$ and vice versa.
For example, for symmetry $g_z$, we have 
\begin{align}
g_z = \prod_\ell \mu^x_{2\ell-1/2} = \prod_\ell \mu^x_{2\ell+1/2} = \prod_j \rho^x_j ~,
\label{eq:dw_sym_different_gauges}
\end{align}
as well as other forms.
All these forms act equivalently in the constrained Hilbert space.

It is convenient to choose a special gauge, such that the gauge connection $\rho^z_j$ is invariant under the symmetry actions.
Within this gauge choice, it is straightforward to extract quantum numbers of gauge invariant objects formed by $\mu^z$ and $\rho^z$: one can just neglect the $\rho^z$ part and focus on the symmetry action on $\mu^z$.
In Eq.~(\ref{eq:dw_sym}), on-site symmetries act trivially on $\rho^z$.
Furthermore, for spin chains with infinite length, we can choose gauge $\rho^z_j = 1$, which is also invariant under $T_x$.
\footnote{For finite-length chains, there are two gauge-inequivalent sectors: $\prod_j \rho^z_j = 1$ and $-1$.
For the nontrivial gauge flux sector, it is impossible to find a uniform $\rho^z_j$ configuration, and thus $T_x$ should be defined differently from Eq.~(\ref{eq:dw_sym}).}

Now, let us focus on the symmetry action on $\mu^z$.
Since $\mu^z$ carries $Z_2^\rho$ gauge charge and is not a local object, symmetries can act projectively on $\mu^z$~\cite{WenPSG}.
In particular, any gauge invariant object is formed by even number of $\mu^z$.
Thus, symmetry action on $\mu^z$ has a $Z_2$ phase ambiguity.
Consequently, $\mu^z$ forms a projective representation of the original symmetry group.
In our case, from Eq.~(\ref{eq:dw_sym}), we list the nontrivial group generator relations as following
\begin{align}
& T_x g_z \circ \mu^z_{j+1/2} = -g_z T_x \circ \mu^z_{j+1/2} ~, \notag \\
& T_x \TT \circ \mu^z_{j+1/2} = -\TT T_x \circ \mu^z_{j+1/2} ~.
\label{eq:dw_proj_rep}
\end{align}
Minus signs in the above two equations indicate nontrivial projective representation of the gauge charges $\mu^z$ under the symmetry actions.
We claim that due to these minus signs, it is impossible to obtain a symmetric phase by condensing the gauge charges.

To get a better understanding of the nontrivial projective representation, let us ignore the gauge field $\rho$ for a moment and focus on the Hamiltonian $H[\{\mu_{j+1/2}\}]$ of the gauge charges $\mu_{j+1/2}$.
$H[\{\mu_{j+1/2}\}]$ hosts an additional global $Z_2^\rho$ symmetry---called $Z_2^\rho$ invariant gauge group or $IGG$---whose generator acts as $\mu^z_j \to -\mu^z_j$.
The appearance of the $Z_2^\rho$ $IGG$ is related to the fact that $\mu^z$'s carry $Z_2^\rho$ gauge charge, and thus must appear in pairs in any gauge invariant local operator.
Notice that the minus sign in Eq.~(\ref{eq:dw_proj_rep}) is exactly the $Z_2^\rho$ $IGG$ action.

Furthermore, the formal symmetry group of $H[\{\mu_{j+1/2}\}]$, which is called projective symmetry group or $PSG$, is an $IGG$ extension of the original global symmetry group $SG$~\cite{WenPSG}:
\begin{align}
PSG/IGG = SG ~.
\end{align}
Such a group extension is far from unique, and different extensions can describe different sets of phases and phase transitions.
A trivial extension is defined as $PSG_\text{triv} = IGG \times SG$.

In our case, the extension is ``nontrivial'' according to Eq.~(\ref{eq:dw_proj_rep}). 
By condensing gauge charge $\mu^z$, the $Z_2^\rho$ $IGG$ is broken.
Since here the $PSG$ is a nontrivial extension, the remaining symmetry group $SH$ after condensing $\mu^z$ can only be a proper subgroup of $SG$: $SH \neq SG$ and $SH < SG$.

It is legitimate to ignore gauge field fluctuations when one studies phases obtained by condensing gauge charges, since the gauge field is Higgsed in the condensed phase.
One may wonder what happens when the gauge charge $\mu^z$ is gapped.
In this case, one should consider dynamics of the gauge flux.
In particular, for $Z_2^\rho$ gauge theory in (1+1)D, properties of the spacetime vison~(instanton) determine the resulting phases when gauge charges are gapped and instantons ``proliferate.''
In our case, the instanton operator $\rho^x_j$ is odd under $g_x$ and $\TT$, as shown in Eq.~(\ref{eq:dw_sym}), which would lead to SSB phase when $\mu^z$ is trivially gapped.
We will discuss the resulting phases in more detail in the next part.

Now, let us look more carefully at the dual model in Eq.~(\ref{eq:model_dw}).
If we replace $\rho^z$ with our uniform gauge choice, $\rho^z_j = 1$, then the Hamiltonian in Eq.~(\ref{eq:model_dw}) resembles qualitatively the celebrated
Ashkin-Teller\,(AT) model:
$\mu$'s living on the even links and on the odd links form two Ising chains, and they both have their own $Z_2$ symmetry generated by $\widehat{g}_e = \prod_\ell \mu^x_{2\ell+1/2}$ and $\widehat{g}_o = \prod_\ell \mu^x_{2\ell-1/2}$ respectively.
These two quantum Ising chains are coupled by symmetry-preserving energy-energy couplings.
In other words, $PSG$ of $H(\mu)$ is identified as the symmetry group for AT model.

Notice that $\widehat{g}_e$ and $\widehat{g}_o$ are not directly related to the original $Z_2^x \times Z_2^z$ symmetry.
By including the gauge field $\rho$, our discussion of Eqs.~(\ref{eq:dw_sym}) and (\ref{eq:dw_sym_different_gauges}) shows that $\widehat{g}_e$ and $\widehat{g}_o$ actually both correspond to representations of the original $g_z$ symmetry, and $\widehat{g}_e\cdot\widehat{g}_o$ is actually the generator for $Z_2^\rho$ $IGG$.
We conclude that our system is more properly thought as a ``gauged'' Ashkin-Teller model, albeit with no vison dynamics (i.e., no vison creation/annihilation terms since these are prohibited by the $g_x$ symmetry).

\subsection{Identification of phases}
For convenience, let us work in the continuum limit.
The continuum variables are
\begin{align}
m_1(a \cdot \ell) \sim \mu^z_{2\ell-1/2} ~, \quad 
m_2(a \cdot \ell) \sim \mu^z_{2\ell+1/2} ~,
\end{align}
where $a$ denotes some lattice constant (here covering one even and one odd dual lattice sites).
In these continuum variables, we have schematic energy density
\begin{align}
\epsilon \sim t\, (m_1^2 + m_2^2) + u\, (m_1^4 + m_2^4) + w\, m_1^2 m_2^2 ~.
\label{eq:model_dw_continuum}
\end{align}
One way to think about this continuum theory is as follows.
We can develop Euclidean path integral for the spin system in Eq.~(\ref{eq:model_dw}) in the $\mu^z$ basis, obtaining two Ising systems corresponding to the even and odd sublattices, with specific energy-energy coupling between the two.
The above field theory then arises naturally when studying ordering in these Ising systems (for simplicity, we did not show gradient terms).
One can also think of this as a field theory Hamiltonian for real-valued quantum fields $m_1$ and $m_2$, where again we did not show gradient terms and did not show conjugate field variables.
When $w = 0$, we have two decoupled Ising systems which both undergo ordering transition when $t$ changes sign from positive to negative.
In Eq.~(\ref{eq:model_dw}), this corresponds to $K_{2x} = K_{2z} = 0$, and the transition occurs at $J_x = J_z$.
The $w$ term represents energy-energy coupling between the two Ising systems and roughly corresponds to combined effects of the $K_{2x}$ and $K_{2z}$ terms.

In these variables, the symmetries act as
\begin{align}
& T_x:\; m_1 \rightarrow m_2 ~, \quad m_2 \rightarrow m_1 ~; \notag \\
& g_z:\; m_1 \rightarrow m_1 ~, \quad m_2 \rightarrow -m_2 ~; \notag \\
& \TT:\; m_1 \rightarrow m_1 ~, \quad m_2 \rightarrow -m_2, \quad \ii \rightarrow -\ii ~.
\end{align}
Notice that $g_x$ acts trivially on $m_{1,2}$ and is encoded instead in its action on the gauge field $\rho$.
Remember also that $m_{1,2}$ carry gauge charge with respect to $\rho$, and  only combinations containing even number of $m$'s correspond to local physical observables.
For example, $m_1 m_2$ is odd under $g_z$ and $\TT$ and even under $T_x$, and thus can be identified as order parameter for breaking the $g_z$ and $\TT$ symmetries, i.e., order parameter for ferromagnetic order with spins aligned in the $\sigma^x$ direction. 
Similarly, $m_1^2 - m_2^2$ is odd under $T_x$ and even under all on-site symmetries, which serves as order parameter for breaking $T_x$, i.e., VBS order parameter.

Now we are able to analyze possible phases for this model in the dual variables.
\begin{itemize}
\item $J_z$ dominant, or $t > 0$.
We have $\langle m_{1,2} \rangle = 0$.
To identify this phase in the original variables, we need to include the gauge field $\rho$.
As shown in Eq.~(\ref{eq:dw_sym}), the $Z_2^\rho$ instanton operator $\rho^x$ transforms non-trivially under $g_x$ and $\TT$.
We can then argue that we obtain a ferromagnetic phase with magnetization pointing in the $\sigma^z$ direction.
Indeed, we can integrate out the trivially gapped matter fields $\mu$ and obtain a pure gauge theory Hamiltonian.
We can loosely say that the $J_z$ term aligns $\mu$'s in the $\mu^x$ direction, which via the constraint induces $\langle \rho^x \rangle \neq 0$, thus breaking both the $g_x$ and $\TT$ symmetries.
More precisely, in the present setup arising from the duality for the spin model with the $g_x$ symmetry, this pure gauge theory does not allow local terms that can mix or distinguish the two flux sectors $\prod_j \rho_j^z = \pm 1$; hence, the ground state is two-fold degenerate corresponding to spontaneous breaking of the $g_x$ symmetry.

\item $J_x$ dominant, or $t < 0$ and $w < 2u$.
In this case, both $m_1$ and $m_2$ obtain a nonzero expectation value, with $\langle m_1 \rangle = \pm \langle m_2 \rangle \neq 0$.
Due to the gauge charge condensation, we get Higgs phase and can ignore the gauge field.
For this condensation pattern, $\langle m_1 m_2 \rangle \neq 0$ while $\langle m_1^2 - m_2^2 \rangle = 0$.
We conclude that $g_z$ and $\TT$ are broken while $T_x$ is preserved.
Furthermore, $g_x$ is also preserved due to finite energy splitting between the even and odd flux sectors; this can be argued by noting that the two flux sectors correspond to periodic vs antiperiodic boundary conditions on the condensing matter fields, or by examining minimization of the $J_x$ terms in Eq.~(\ref{eq:model_dw}).
Thus, one obtains a ferromagnetic phase with the spins pointing in the $\sigma^x$ direction.

\item $K_{2x}, K_{2z}$ comparable with $J_x, J_z$, or $t < 0$ and $w > 2u$.
To minimize the energy here, the condensation pattern is chosen as $\langle m_1 \rangle \neq 0,\ \langle m_2 \rangle = 0$ or $\langle m_1 \rangle = 0,\ \langle m_2 \rangle \neq 0$.
In either case, we have $\langle m_1 m_2 \rangle = 0$ while $\langle m_1^2 - m_2^2 \rangle \neq 0$.
Thus, $g_z$ is not broken, while $T_x$ is broken.
As in the previous case, due to the gauge charge condensation, this phase also preserves the $g_x$ symmetry.
One can further check that $\TT$ is also preserved.
Thus, we conclude that the resulting phase is a VBS phase, with two-fold ground state degeneracy characterized by $\langle m_1^2 - m_2^2 \rangle > 0$ or $< 0$.
\end{itemize}

\subsection{New domain wall variables and the $z$-FM to VBS transition}
\label{subsec:new_dw}
We now turn to the phase transitions of interest to us.
The $z$-FM to $x$-FM transition in the dual language corresponds to simultaneous condensation of the $m_1$ and $m_2$ fields, which lands on the much studied line of continuously varying criticality in the Ashkin-Teller (AT) model~\cite{AshkinTeller1943, KohmotoNijsKadanoff} (which contains also a point corresponding to two decoupled Ising models).
However, we emphasize that here we are interested in the specific ``gauged'' Ashkin-Teller model, albeit with no $Z_2^\rho$ instanton dynamics in the Hamiltonian.
While such gauge field does not change thermodynamic critical properties, including it is important for correct identification of phases as well as physical observables in the theory.
The criticality in the AT model can be conveniently described by a two-step duality transformation of its lattice spins (here $\mu$'s) to new ``good variables,'' such that the Hamiltonian in the new variables looks like a perturbed XY chain and naturally leads to a Gaussian field theory with a single relevant cosine interaction~\cite{KohmotoNijsKadanoff}.
In our specific case of the gauged AT model for the $z$-FM to $x$-FM transition, finding such good variables corresponds to simply returning to the original spin chain with dominant ferromagnetic nearest-neighbor $J_x \sim J_z$ interactions, and the field theory for the transition is the one described in Sec.~\ref{sec:bosonization}.

On the other hand, the transition between the $z$-FM and VBS phases corresponds to condensation of either $m_1$ or $m_2$, but not both.
To describe this transition in more familiar terms, we perform a change on the continuum variables as follows:
\begin{align}
& m_+ = \frac{1}{\sqrt{2}} (m_1 + m_2) ~, \notag \\
& m_- = \frac{1}{\sqrt{2}} (m_1 - m_2) ~.
\end{align}
Then the symmetry actions on $m_\pm$ read
\begin{align}
& T_x:\ m_+ \rightarrow m_+ ~, \quad m_- \rightarrow -m_- ~; \notag \\
& g_z:\ m_+ \rightarrow m_- ~, \quad m_- \rightarrow m_+ ~; \notag \\
& \TT:\ m_+ \rightarrow m_- ~, \quad m_- \rightarrow m_+ ~, \quad \ii \rightarrow -\ii ~.
\label{eq:m_mp_symm}
\end{align}
The schematic energy density, Eq.~(\ref{eq:model_dw_continuum}), becomes in the $m_\pm$ variables
\begin{align}
\epsilon \sim t\, (m_+^2 + m_-^2) + u'\, (m_+^4 + m_-^4) + w'\, m_+^2 m_-^2 ~,
\label{eq:model_dw_continuum_+-}
\end{align}
where $u' = u/2 + w/4$ and $w' = 3u - w/2$.
It is easy to check that $w' - 2u' = -(w - 2u)$, and condensation patterns in the $m_+, m_-$ variables are reversed compared to the $m_1, m_2$ variables.
The $z$-FM to VBS transition becomes transition where both $m_+$ and $m_-$ fields condense simultaneously, so in these variables we land onto the familiar Ashkin-Teller-like criticality.

The VBS order parameter that breaks $T_x$ but preserves all internal symmetries is identified as 
\begin{align}
\Psi_\text{VBS} \sim m_+ m_- ~,
\end{align}
We can also identify the $x$-FM order parameter that breaks $g_z$ and $\TT$ but preserves $g_x$ and $T_x$:
\begin{align}
M_x^\text{FM} \sim m_+^2 - m_-^2 ~.
\end{align}
However, remember that this $M_x^\text{FM}$ does not order on either side of the $z$-FM to VBS transition.
Also remember that the $\sigma^z$ component of the physical spin is the instanton in the gauge field, $\sigma^z_j = \rho^x_j$, so the $z$-FM order parameter crucially requires including the gauge field and will be discussed later.

We can now obtain some predictions for the $z$-FM to VBS transition from known results for the AT model~\cite{KohmotoNijsKadanoff}.
Thus, we expect a second-order transition whose critical indices depend on the microscopics and can vary continuously but are parametrized by a single parameter.
For the case of $\Psi_\text{VBS}$ and $M_x^\text{FM}$ observables that have ``gauge-invariant'' expressions in terms of the matter fields, we are safe to ignore the gauge field and can use the AT model results to deduce the following relations between the corresponding scaling dimensions and the correlation length exponent $\nu$:
\begin{align}
\text{dim}[\Psi_\text{VBS}] = \frac{2 - 1/\nu}{4} ~, \\
\text{dim}[M_x^\text{FM}] = \frac{1}{2 - 1/\nu} ~.
\end{align}
``Self-duality'' structure of the AT model leads to predictions also for scaling dimensions of objects that involve ``disorder operators'' $\tau^z_+$ and $\tau^z_-$ that are ``dual'' to $m_+$ and $m_-$.
For example, such analyses predict that $\tau^z_+ \tau^z_-$ has the same scaling dimension as $m_+ m_-$.
However, $\tau^z_+$ and $\tau^z_-$, as well as $\tau^z_+ \tau^z_-$ are non-local in the AT model (and hence ``non-observable'' using local operators).
Handily for us, as we will see in the next section, the gauge structure in our ``gauged'' AT model and this non-locality in the ``non-gauged'' AT model conspire to turn the analog of $\tau^z_+ \tau^z_-$ into a local observable, whose meaning is precisely the $z$-FM order parameter,
\begin{align}
M_z^\text{FM} \sim \tau^z_+ \tau^z_- ~.
\label{eq:MzFM_AT}
\end{align}
We will give a careful derivation of this in the next Sec.~\ref{sec:dw_parton}, while in Sec.~\ref{sec:goodvars} we will present a derivation of ``good variables'' where the nature of the transition and critical properties of physical observables become particularly transparent.

We conclude with one last remark.
The AT criticality also predicts that the scaling dimensions for the $m_+$ and $m_-$ fields, as well as for the dual $\tau^z_+$ and $\tau^z_-$ fields, are always fixed at $1/8$ even though other exponents can vary continuously.
The $m_+$ and $m_-$ are local observables in the AT model and hence readily measurable; on the other hand, the $\tau_+$ and $\tau_-$ are non-local and cannot be measured using local operators.
However, in our gauged AT model, the $m_+$ and $m_-$ variables that describe the domain walls are not local observables, since they are gauge-charged with respect to $\rho$.
The corresponding dual variables $\tau^z_+$ and $\tau^z_-$ will also turn out to be non-local; in fact, they will turn our to be ``parton'' variables that appear when attempting to ``fractionalize'' the physical spins, and so they will also be gauge-charged, here with respect to a new gauge field that appears in such parton constructions.

\section{Connection between the domain wall and parton approaches} \label{sec:dw_parton}
In this section, we will study the transition between the $z$-FM and VBS phases in more detail.
By realizing Eq.~(\ref{eq:model_dw_continuum_+-}) on the lattice and performing Ising duality, we will find that this theory actually has interesting ``self-duality'' features. 
Furthermore, the dual theory can be viewed as an effective theory for bosonic partons.
The domain wall variables provide simple description of the VBS order parameter, while they require including instanton effects to describe the magnetic order, and the situation is reversed in the parton variables.

To describe the critical properties, in a subsequent Sec.~\ref{sec:goodvars} we will perform a different ``two-step duality,'' which is an analog of the approach introduced in Ref.~\cite{KohmotoNijsKadanoff} to analyze the quantum Ashkin-Teller model.
In the final theory, the $z$-FM order and the VBS order appear ``democratically;'' thus, this theory explicitly ``unifies'' the two order parameters.
The critical theory turns out to be described by a standard Luttinger-liquid-like theory (i.e., a Gaussian theory), where we can easily calculate critical exponents.
Our treatment of dualities keeps track of all symmetries and corresponding global aspects, which allows us to see how these appear in the final theory, and in particular allows us to unambiguously identify all physical observables.

\subsection{Lattice realization of the new domain wall variables}
To perform further analysis, in particular to capture the important physics that the domain wall fields $m_+, m_-$ see the $Z_2^\rho$ gauge field and to capture all global aspects, let us try to realize the continuum Hamiltonian Eq.~(\ref{eq:model_dw_continuum_+-}) with symmetries Eq.~(\ref{eq:m_mp_symm}) in a lattice system.
Note that the change of variables $(m_1, m_2) \to (m_+, m_-)$, while physically reasonable and convenient on the coarse-grained fields in the continuum, cannot be done exactly for the Ising variables on the lattice.
This is why we will introduce a new lattice model in terms of new quantum Ising variables $\mu_+, \mu_-$, whose continuum limit will coincide with the above model in terms of $m_+, m_-$, including matching the symmetries.

We define discrete variables $\mu^z_{\pm, j+1/2} \sim m_\pm(a \cdot j)$ that reside on the same dual lattice sites as the original domain wall variables.
The $\mu_\pm^z$ variables carry gauge charge with respect to the same gauge field $\rho$ that appeared under the original duality in Sec.~\ref{sec:domain_wall}.
$\mu^x_\pm$'s are conjugate variables to $\mu^z_\pm$, which are roughly identified as $\ii\partial_t m_{\pm}$ in the continuum theory.
We can now write a lattice Hamiltonian that, as we will argue, captures the desired physics:
\begin{align}
\tilde{H} = & \sum_j \sum_{\sigma = \pm} \left( -J\,\mu^x_{\sigma, j+1/2} - h\,\mu^z_{\sigma, j-1/2} \, \rho^z_j \, \mu^z_{\sigma, j + 1/2} \right) \notag \\
& - \sum_j \Gamma \, (-1)^j\, \mu^z_{+, j+1/2} \mu^z_{-, j+1/2} ~.
\label{eq:model_mu_pm}
\end{align}
The gauge constraint is
\begin{align}
\rho^x_j \rho^x_{j+1} = \mu^x_{+, j+1/2} \mu^x_{-, j+1/2} ~,
\label{eq:mu_pm_gauge_constraint}
\end{align}
which reflects the fact that the re-latticized model allows the two species to reside on the same site.
The physics of the different terms in the Hamiltonian will become clear below.

The symmetry action on $\rho$ is the same as in Eq.~(\ref{eq:dw_sym}), while action on $\mu_\pm$ reads
\begin{align}
  T_x:\ & \mu^\alpha_{+, j-1/2} \rightarrow \mu^\alpha_{+, j+1/2} ~, \notag \\
  & \mu^x_{-, j-1/2} \rightarrow \mu^x_{-, j+1/2} ~, \quad \mu^{y,z}_{-, j-1/2} \rightarrow -\mu^{y,z}_{-, j+1/2} ~; \notag \\ 
  g_x:\ & \mu^\alpha_{\pm, j+1/2} \rightarrow \mu^\alpha_{\pm, j+1/2} ~; \notag \\
  g_z:\ & \mu^\alpha_{\pm, j+1/2} \rightarrow \mu^\alpha_{\mp, j+1/2} ~; \notag \\
  \TT:\ & \mu^{x,z}_{\pm, j+1/2} \rightarrow \mu^{x,z}_{\mp, j+1/2} ~, \quad
\mu^y_{\pm, j+1/2} \rightarrow -\mu^y_{\mp, j+1/2} ~, \notag \\
& \ii \rightarrow -\ii ~.
\label{eq:mu_pm_sym}
\end{align}
In particular, this captures that $m_+$ and $m_-$ carry physical momenta $0$ and $\pi$ respectively, and that they are interchanged under $g_z$ and $\TT$.
The re-latticized version of the physical translation symmetry allows the $\mu_+$ and $\mu_-$ species to inter-convert via the $\Gamma$ term in Eq.~(\ref{eq:model_mu_pm}).
One cannot write such a term in the continuum limit with only slowly varying $m_+$ and $m_-$, but it is important on the lattice scale.
We think that it captures the fact that we cannot go from the $m_1, m_2$ variables to the $m_+, m_-$ variables exactly for Ising degrees of freedom on the lattice, and it will also nicely encode staggered bond energy density when both $m_+$ and $m_-$ condense.
While this model, which has twice as many domain wall variables as under the original duality mapping, may seem somewhat ad hoc, its physical appropriateness will be further supported also by an exact connection to a parton approach below.
The $J$ term represents energy cost of the domain walls and roughly corresponds to the $J_z$ term in Eq.~(\ref{eq:model_dw}), while the $h$ term represents hopping of the domain walls and roughly corresponds to the combined effects of the $J_x$ and $K_{2x}$ terms.

Using the symmetry transformation properties, we can readily identify phases in this model, in agreement with our discussion in Sec.~\ref{sec:domain_wall}.
Thus, for $J \gg h, \Gamma$, we get the $z$-FM phase due to the instanton event ``condensation,'' schematically, $\langle \rho^x \rangle \neq 0$. 
On the other hand, for $h \gg J, \Gamma$, we get the VBS phase due to condensation of both domain walls, $\langle \mu_+^z \rangle = \pm \langle \mu_-^z \rangle \neq 0$; the $\Gamma$-term energy explicitly shows staggering in the static bond energy in this case.
Thinking now about the $z$-FM to VBS transition in terms of long-wavelength components of $\mu^z_\pm$, for small $\Gamma$ the corresponding term in Eq.~(\ref{eq:model_mu_pm}) washes out due to the rapidly oscillating factor $(-1)^j$, and one can think about it as being irrelevant at the transition.
One may then conclude that the phase transition happens at $J \approx h$, and is described by two-decoupled-Ising criticality.
However, additional symmetric terms, such as energy-energy coupling terms $\mu^x_{+, j+1/2} \mu^x_{-, j+1/2}$, $\mu^z_{+, j-1/2} \mu^z_{+, j+1/2} \mu^z_{-, j-1/2} \mu^z_{-, j+1/2}$, etc, are allowed in Eq.~(\ref{eq:model_mu_pm}), which will drive the universality away from the decoupled-Ising criticality to a more general Ashkin-Teller criticality.
Just as in Sec.~\ref{subsec:new_dw}, we can already deduce some properties of the transition from the known properties of the AT criticality, but again we are yet to learn, e.g., how the $z$-FM order parameter is represented, and how to find a complete and efficacious field-theoretic description.

\subsection{Bosonic parton approach}
\label{subsec:bosonic_parton}
As we have shown in the preceding section, some physics of the phase transition between the $z$-FM and VBS orders becomes more clear using the domain wall variables $\mu_\pm$.
In this section, we will make connection with a parton (a.k.a.~``slave particle'') method to study this phase transition, which will clarify some additional physics.

\subsubsection{Hard-core-boson parton representation for physical spins}
Let us first introduce a hard-core-boson parton construction for physical spins $\sigma$ (this can be viewed as a particular Schwinger boson construction where in addition we make the bosonic partons as hard-core, hence the name).
We enlarge the local spin Hilbert space to four dimensions, labeled by two qubits $\tau_+$ and $\tau_-$, which we refer to as (bosonic) partons.
Local spin states are identified as
\begin{align}
|\sigma^x = \pm 1 \rangle\ \leftrightarrow\ |\tau^x_+ = \pm 1 ~, \, \tau^x_- = \mp 1 \rangle ~.
\end{align}
Equivalently, we impose a local constraint as
\begin{align}
\tau^x_+ \tau^x_- = -1 ~, \quad \text{or} \quad \tau^x_+ + \tau^x_- = 0 ~.
\label{eq:bosonic_parton_constraint}
\end{align}
In these parton variables, the physical spin operators are represented as
\begin{align}
& \sigma^x = \frac{1}{2} (\tau^x_+ - \tau^x_-) ~, \notag \\
& \sigma^y = \frac{1}{2} (\tau^y_+ \tau^z_- - \tau^z_+ \tau^y_-) ~, \notag \\
& \sigma^z = \frac{1}{2} (\tau^z_+ \tau^z_- + \tau^y_+ \tau^y_-) ~.
\end{align}
One can readily check validity of this representation in the constrained Hilbert space.
The specific choice in some sense corresponds to ``fractionalizing'' the $\sigma^z$ order parameter, which can be written equivalently as $\sigma^z = \tau^z_+ \tau^z_-$.
(For example, such a parton writing of an Ising magnetic order parameter was used in Appendix D in Ref.~\cite{MotrunichSenthil2005} to describe fractionalization in a two-dimensional quantum Ising system, i.e., to describe Ising-symmetry-enriched topological order.)
We will see that this choice is convenient to provide connection with the preceding domain wall approach and to add to the discussion of the $z$-FM to VBS transition.

\subsubsection{Effective theory for bosonic partons}
To write down a general form of an effective Hamiltonian for the partons, one should figure out how they transform under symmetries.
Notice that a single parton field $\tau^z_\pm$ is not a local object.
Instead, partons should be viewed as gauge charges coupled to a $Z_2^\zeta$ gauge field $\zeta$~\cite{SenthilFisher2000},
\footnote{Strictly speaking, it is possible to have a $U(1)$ gauge group rather than $Z_2^\zeta$.
However, for our purposes here, we always allow Higgs terms to break the $U(1)$ gauge group to $Z_2^\zeta$.}.
Thus, similarly to the domain wall variables, $\tau^z_\pm$ transform projectively under symmetries, and an effective theory for partons should be invariant under some PSG~\cite{WenPSG}.

However, we point out that even for the same symmetry group, the choice of PSG is not unique.
In the presence of the $Z_2^\zeta$ gauge field, PSGs are classified by the second cohomology group $H^2(SG, Z_2^\zeta)$, where $SG$ denotes the whole symmetry group including both on-site and spatial symmetries.
Thus, a natural question arises: which PSG should we choose to describe the phase transition between the $z$-FM and VBS phases?

One specific proposal that we consider here is motivated by the domain wall description discussed in the last section:
Effective theory for the bosonic partons $\tau_\pm$ can be obtained by performing a duality transformation on the theory of the domain wall variables $\mu_\pm$.
We define duality by the following operator mappings:
\begin{align}
& \mu^x_{\sigma, j+1/2} = \tau^z_{\sigma, j} \, \zeta^z_{j+1/2} \, \tau^z_{\sigma, j+1} ~, \notag \\
& \mu^z_{\sigma, j-1/2} \, \rho^z_j \, \mu^z_{\sigma, j+1/2} = \tau^x_{\sigma, j} ~, \notag \\
& \rho^x_j = \tau^z_{+, j} \, \tau^z_{-, j} ~, \notag \\
& \mu^z_{+, j+1/2} \, \mu^z_{-, j+1/2} = \zeta^x_{j+1/2} ~.
\label{eq:dw_to_parton}
\end{align}
Notice that the constraint $\rho^x_j \rho^x_{j+1} = \mu^x_{+, j+1/2} \, \mu^x_{-, j+1/2}$ is automatically satisfied.
Also, a new gauge constraint (``Gauss law'') arises when imposing this mapping, which reads
\begin{align}
\zeta^x_{j-1/2} \, \zeta^x_{j+1/2} = \tau^x_{+, j} \, \tau^x_{-, j} ~.
\label{eq:bosparton_gauge_constraint}
\end{align}

Using this duality mapping, the Hamiltonian defined in Eq.~(\ref{eq:model_mu_pm}) becomes
\begin{align}
  \tilde{H} = & \sum_j \sum_{\sigma = \pm} \left( -J\, \tau^z_{\sigma, j} \zeta^z_{j+1/2} \tau^z_{\sigma, j+1} - h\, \tau^x_{\sigma, j} \right) \notag \\
  & - \sum_j \Gamma \, (-1)^j \, \zeta^x_{j + 1/2} ~,
  \label{eq:model_bosonic_parton}
\end{align}
with the gauge constraint given by Eq.~(\ref{eq:bosparton_gauge_constraint}).
We emphasize that this is an exact mapping between the two matter-gauge models, one with the fields $\mu_\pm, \rho$, and the other with $\tau_\pm, \zeta$; the mapping is exact on a chain with periodic connectedness.
One way to see this is to ``solve'' the constraints in each model (e.g., one can solve for eigenvalues of $\mu_+^x$ in terms of $\mu_-^x$ and $\rho^x$ in the first gauge theory and solve for eigenvalues of $\tau_-^x$ in terms of $\tau_+^x$ and $\zeta^x$ in the second theory), and then match the corresponding unconstrained Hamiltonians.
In Appendix~\ref{app:dw_parton_equiv}, we present a detailed proof of the equivalence between Eqs.~(\ref{eq:model_mu_pm}) and (\ref{eq:model_bosonic_parton}).

From Eqs.~(\ref{eq:dw_sym}), (\ref{eq:mu_pm_sym}), and (\ref{eq:dw_to_parton}), we can figure out symmetry actions on $\zeta$ and $\tau_\pm$ as
\begin{align}
  T_x:\ & \zeta^{x,y}_{j-1/2} \rightarrow -\zeta^{x,y}_{j+1/2} ~, \notag \\
  & \tau^\alpha_{\pm, j} \rightarrow \tau^\alpha_{\pm, j+1} ~; \notag \\
  g_x:\ & \zeta^{x,y}_{j+1/2} \rightarrow \zeta^{x,y}_{j+1/2} ~, \notag \\
  & \tau^\alpha_{+, j} \rightarrow \tau^\alpha_{+, j} ~, \quad \tau^x_{-, j} \rightarrow \tau^x_{-, j} ~, \quad \tau^{y,z}_{-, j} \rightarrow -\tau^{y,z}_{-, j} ~; \notag \\
  g_z:\ & \zeta^{x,y}_{j+1/2} \rightarrow \zeta^{x,y}_{j+1/2} ~, \notag \\
  & \tau^\alpha_{+, j} \rightarrow \tau^\alpha_{-, j} ~, \quad \tau^\alpha_{-, j} \rightarrow \tau^\alpha_{+,j } ~; \notag \\
  \TT:\ & \zeta^x_{j+1/2} \rightarrow \zeta^x_{j+1/2} ~, \quad \zeta^y_{j+1/2} \rightarrow -\zeta^y_{j+1/2} ~, \notag \\
  & \tau^x_{\pm, j} \rightarrow \tau^x_{\mp, j} ~, \quad \tau^y_{+, j} \rightarrow -\tau^y_{-, j} ~, \quad \tau^y_{-, j} \rightarrow \tau^y_{+, j} ~, \notag \\
  & \tau^z_{+, j} \rightarrow \tau^z_{-, j} ~, \quad \tau^z_{-, j} \rightarrow -\tau^z_{+, j} ~, \quad \ii \rightarrow -\ii ~.
  \label{eq:bosonic_parton_sym}
\end{align}
Again, we have fixed a special gauge such that $\zeta^z$ is invariant under all symmetry actions.
It is straightforward to check that the Hamiltonian in Eq.~(\ref{eq:model_bosonic_parton}) is invariant under the symmetry actions defined above.

Note that the symmetry actions on the $\tau_\pm$ variables are fixed (up to a global gauge transformation $\tau_{\pm, j}^{y,z} \to -\tau_{\pm, j}^{y,z}$ for all $j$) by the symmetry actions on the $\mu_\pm$ variables.
It is straightforward to check that $\tau^z_{\pm, j}$ have nontrivial projective transformations under the $g_x$, $g_z$, and $\TT$ symmetries:
\begin{align}
& g_x g_z \circ \tau^z_{\pm, j} = -g_z g_x \circ \tau^z_{\pm, j} ~, \notag \\
& g_x \TT \circ \tau^z_{\pm, j} = -\TT g_x \circ \tau^z_{\pm, j} ~, \notag \\
& \TT^2 \circ \tau^z_{\pm, j} = -\tau^z_{\pm, j} ~.
\end{align}
Remember that a parton decomposition itself does not fix a PSG; usually, an appropriate PSG is chosen by, e.g., energetics considerations for a given Hamiltonian, in an attempt to describe phases of interest.
Here, the PSG is fixed by the connection to the $\mu_\pm$ domain wall variables, and some such energetics considerations happened when motivating the domain wall theory that can access the $z$-FM to VBS transition.

We have claimed that Eq.~(\ref{eq:model_bosonic_parton}) can be viewed as an effective theory for bosonic partons~\cite{SenthilFisher2000}.
To see this, we take the gauge field coupling parameter $\Gamma$ to be very large. 
Then, to minimize the energy, $\zeta^x_{j+1/2} = (-1)^j$ [assuming $\Gamma > 0$ for concretness].
Using the Gauss law at each site, we have $\tau^x_{+, j} \tau^x_{-, j} = -1$, which is exactly the constraint for the microscopic (``bare'') bosonic parton approach in Eq.~(\ref{eq:bosonic_parton_constraint}).
Using Eqs.~(\ref{eq:dw_dual_map}) and (\ref{eq:dw_to_parton}), we have $\sigma^z_j = \rho^x_j = \tau^z_{+, j} \tau^z_{-, j}$.
Furthermore, $\frac{1}{2} (\tau^x_{+, j} - \tau^x_{-, j})$ is odd under $g_z$ and $\TT$ and even under $g_x$; hence, it has the same symmetry properties as $\sigma^x_j$, and can be identified as the $\sigma^x_j$ operator in the effective theory.
Thus, in the limit of very large $\Gamma$, we recover the lattice spin system with some Hamiltonian which can be derived perturbatively and resembles the ferromagnetic $J_z$ and $J_x$ terms in the original spin model, Eq.~(\ref{eq:spin_model}),
\footnote{A perturbative scheme where we treat the $\Gamma$ and $h$ terms in Eq.~(\ref{eq:model_bosonic_parton}) as an unperturbed Hamiltonian and the $J$ term as a perturbation, assuming $\Gamma > 2h$ [so that the unperturbed ground state satisfies the bare parton constraints in Eq.~(\ref{eq:bosonic_parton_constraint})] and $\Gamma - 2h \gg J$, gives the following spin Hamiltonian at second order in $J$:
\begin{align}
H_\text{spin} = \sum_j \frac{-J^2 \Gamma}{\Gamma^2 - 2 h^2 (1 + \sigma^x_j \sigma^x_{j+1})} (1 + \sigma^z_j \sigma^z_{j+1}) ~.
\end{align}
This Hamiltonian wants to have ferromagnetic nearest-neighbor $\sigma^z \sigma^z$ and $\sigma^x \sigma^x$ correlations, akin to the effect of the $J_z$ and $J_x$ terms in Eq.~(\ref{eq:spin_model}).}.

We now further assume that this theory gives qualitatively correct physics for arbitrary values of $\Gamma$, in the sense that it produces phases and transitions that can be realized in the original spin system.
Below, we provide further support for this assumption.

Let us analyze phases of the Hamiltonian in Eq.~(\ref{eq:model_bosonic_parton}) in the small $\Gamma$ limit.
\begin{itemize}
\item When $J > h$, we get ordered states for $\tau^z_\pm$, where $\langle \tau^z_+ \rangle = \pm \langle \tau^z_- \rangle \neq 0$.
By analyzing symmetries of gauge-invariant objects (e.g., $\sigma^z \sim \tau^z_+ \tau^z_-$), we conclude that $g_x$ and $\TT$ are broken, while $T_x$ and $g_z$ are preserved.
Thus, we get the ferromagnetic phase with magnetic order in the $\sigma^z$ direction.
\item When $J < h$, we get disordered states for the $\tau^z_\pm$ variables with $\langle \tau^z_\pm \rangle = 0$.
Hence, the internal symmetries $g_x$, $g_z$, and $\TT$ are preserved.
To identify the nature of the resulting phase, we should include the gauge field dynamics.
In the limit of very large $h$, we have $\tau^x_\pm = 1$; hence, two states labeled by $\zeta^x_{j+1/2} = 1$ for all $j$ or $\zeta^x_{j+1/2} = -1$ for all $j$ have degenerate energy.
Since $\zeta^x_{j+1/2}$ changes sign under $T_x$, we actually obtain a translational symmetry breaking phase, which is identified as the VBS ordered phase.
We can also see directly from the $\Gamma$ term in the Hamiltonian, Eq.~(\ref{eq:model_bosonic_parton}), that such states with uniform $\zeta^x$ have staggered bond energy density, as expected in the VBS states.
\end{itemize}

We can now discuss the transition between the $z$-FM and VBS phases from the parton perspective.
The partons are gapped in the VBS phase, and the transition occurs by simultaneous condensation of both parton species.
The $z$-FM order parameter is obtained by combining the parton fields, $\sigma^z \sim \tau^z_+ \tau^z-$ [thus proving Eq.~(\ref{eq:MzFM_AT}) claimed in Sec.~\ref{subsec:new_dw}], while the VBS order parameter is given by the instanton field $\zeta^x$.
This is to be compared with the domain wall theory, where the VBS order parameter is obtained by combining the domain wall fields, $\Psi_\text{VBS} \sim \mu^z_+ \mu^z_-$, while the $z$-FM order parameter is given by the instanton field $\rho^x$.
Note that the domain wall inter-conversion term [i.e., the $\Gamma$ term in Eq.~(\ref{eq:model_mu_pm})] mapped to the instanton creation term in the parton theory, and hence these two theories in general have qualitatively different structure.
However, if the $\Gamma$ term is small and is irrelevant in RG sense at the transition (which we argued is plausible in the domain wall theory with long-wavelength $\mu_\pm$ fields), then the two theories have similar structure, and we can anticipate ``self-duality'' property at the transition.
In particular, we can now appeal to known results for the AT model discussed in Sec.~\ref{subsec:new_dw} and argue that the $z$-FM and VBS order parameters should have the same scaling dimension at the transition.
The above properties bear close resemblance to the easy-plane deconfined criticality theory between the Neel and VBS phases on the 2d square lattice; we will now discuss such interesting similarities in more detail.

\subsection{Parallels with the easy-plane NCCP$^1$ description of the 2d easy-plane DQCP}
\label{subsec:parallelsNCCP1}
We begin with a brief recap of the 2d Neel-VBS EP-DQCP on the square lattice~\cite{DQCP_science, DQCP_prb}.
By performing a $\pi$ rotation around the $z$-axis on one sublattice, the resulting spin system can be mapped to a bosonic system at half-filling with unfrustrated nearest-neighbor hopping, undergoing a transition between a superfluid phase and a Mott insulator phase with VBS character.
Reference~\cite{LannertFisherSenthil2001} studied this system from the dual vortex perspective and found that there are two low-energy vortex fields $\psi_{1,2}$ that transform projectively under lattice symmetries.
One arrives at the following continuum Lagrangian in Euclidean space-time:
\begin{align}
\mathcal{L} = & \mathcal{L}_{\text{NCCP}^1} + \mathcal{L}_\text{tunn.} ~, \notag \\
\mathcal{L}_{\text{NCCP}^1} = & \sum_{a = 1, 2} \left[|(\nabla_\mu - \ii b_\mu) \psi_a|^2 + r_d |\psi_a|^2 + u_d (|\psi_a|^2)^2 \right] \notag \\
& + w_d |\psi_1|^2 |\psi_2|^2 + \kappa_d (\epsilon_{\mu\nu\rho} \partial_\nu b_\rho)^2 ~, \notag \\
\mathcal{L}_\text{tunn.} = & -v_8 [(\psi_1^*\psi_2)^4 + \text{c.c.}] ~.
\end{align}
As usual under the boson-vortex duality, the original $U(1)$ symmetry is encoded as flux conservation of the dual gauge field $b_\mu$ [equivalently, the monopole operator that creates $2\pi$ of the $b_\mu$ flux carries $U(1)$ charge].
When the vortex fields $\psi_{1,2}$ are gapped, one obtains the superfluid order.
The part $\mathcal{L}_{\text{NCCP}^1}$ has $\psi_1$ and $\psi_2$ separately conserved; however, physically, there is only one vortex species, so microscopically there is ``tunneling'' between the $\psi_1$ and $\psi_2$ fields.
Because of the nontrivial transformation properties of these fields under the lattice symmetries (especially the four-fold rotation symmetry around site center), only the quadrupled tunneling between them survives in the continuum limit, which gives the term $\mathcal{L}_\text{tunn.}$.
Condensing one vortex field but not the other gives a charge density wave insulator, while simultaneously condensing both vortex fields gives a VBS phase, whose details further depend on the sign of the $v_8$ coupling.

Under a boson-vortex duality applied separately to the $\psi_1$ and $\psi_2$ fields, the $\mathcal{L}_{\text{NCCP}^1}$ maps to a theory with the same structure~\cite{MotrunichVishwanath2004, DQCP_science, DQCP_prb}, while the quadrupled inter-species tunneling maps to allowing quadrupled monopoles.
Schematically:
\begin{align}
\mathcal{L}_{\text{NCCP}^1} \to\ & \sum_{a = 1, 2} \left[|(\partial_\mu - \ii a_\mu) z_a|^2 + r |z_a|^2 + u(|z_a|^2)^2 \right] \notag \\
& + w |z_1|^2 |z_2|^2 + \kappa (\epsilon_{\mu\nu\rho} \partial_\nu a_\rho)^2 ~, \notag \\
\mathcal{L}_\text{tunn.} \to\ & \text{allow quadrupled monopoles} ~.
\end{align}
Here, $z_1$ and $z_2$ can be viewed as spinon fields from a parton decomposition of the physical spin, while $a_\mu$ can viewed as a compact gauge field (i.e., with allowed monopole dynamics) that arises in such a parton approach.
The spinon fields transform projectively under the spin rotation symmetry, and condensing spinons leads to magnetically ordered phases.
On the other hand, the monopole operator for $a_\mu$ carries lattice quantum numbers, which is why only quadrupled monopoles survive in the continuum limit.
When the spinons are gapped, proliferation of the monopoles leads to a VBS phase.
A key conjecture of the EP-DQCP theory~\cite{DQCP_science, DQCP_prb} is that the $\mathcal{L}_\text{tunn.}$ term is irrelevant at the Neel-VBS transition.

Our domain wall theory in terms of the $m_+, m_-, \rho$ fields is analogous in spirit to the above vortex theory in terms of the $\psi_1, \psi_2, b_\mu$ fields (though not in specifics, since these are, of course, different problems).
Thus, the $Z_2^\rho$ instanton carries charge under the $g_x$ symmetry and hence has no dynamics in the theory.
If the $m_\pm$ fields are trivially gapped, one gets the $z$-FM order that breaks the $g_x$ symmetry.
The $m_\pm$ fields transform projectively under the $g_z$ and $T_x$ symmetries, and one obtains the $x$-FM phase or the VBS phase by condensing $m_\pm$ in different ways.
The tunneling between the two domain wall fields is manifest in the quantum lattice version in terms of the $\mu_+, \mu_-$ degrees of freedom in Eq.~(\ref{eq:model_mu_pm}).

Under the formal duality, the domain wall theory in the $\mu_+, \mu_-, \rho$ variables maps to the parton theory in the $\tau_+, \tau_-, \zeta$ variables, and the tunneling between the $\mu_+$ and $\mu_-$ fields maps to allowing instanton dynamics in the $Z_2^\zeta$ gauge field.
This is analogous in spirit to the described relation between the above vortex and spinon theories for the 2d EP-DQCP.
Continuing with analogies in the parton language, $\tau_\pm$ transform projectively under the $g_x$ and $g_z$ symmetries, and condensing $\tau_\pm$ leads to ferromagnetically ordered phases.
On the other hand, $Z_2^\zeta$ instanton operator $\zeta^x_j$ is odd under the translation, so one obtains the VBS phase if the $\tau_\pm$ are trivially gapped.

Unlike the 2d EP-DQCP problem where the quadrupled inter-vortex-species tunneling (quadrupled $a_\mu$ monopoles) survive in the continuum limit, doubled inter-domain-wall-species tunneling (doubled $Z_2^\zeta$ instantons) are indistinguishable from a trivial operator, so our lattice $\Gamma$ terms do not give non-trivial terms in the naive continuum limit.
Thus, we conclude that we do not have an analog of the $\mathcal{L}_\text{tunn.}$ term in the 1d Ising DQCP theory.
Dropping the $\Gamma$ term on the lattice scale gives us a gauged Ashkin-Teller model with no instanton dynamics.

Motivated by the observations so far, we follow with more parallels in slightly more abstract directions.
The NCCP$^1$ model as a classical statistical mechanics model in (2+1)D was introduced in Ref.~\cite{MotrunichVishwanath2004} as a description of a classical $O(3)$ spin model with complete suppression of hedgehog topological defects~\cite{KamalMurthy1993}.
Specifically, we can define a CP$^1$ variable as $(z_\up, z_\dn) \in \mathbb{C}^2$, $|z_\up|^2 + |z_\dn|^2 = 1$, where $(\ee^{\ii \gamma} z_\up, \ee^{\ii \gamma} z_\dn)$ is identified with $(z_\up, z_\dn)$; that is, the $(z_\up, z_\dn)$ representation has $U(1)$ gauge redundancy.
Such a CP$^1$ variable is equivalent to
an $O(3)$ spin $\vec{n} = (n_1, n_2, n_3) = z_\alpha^* \vec{\sigma}_{\alpha\beta} z_\beta$.
A generic CP$^1$ statistical mechanics model in (2+1)D formulated in terms of $(z_\up, z_\dn)$ fields, by definition, also has a dynamical compact $U(1)$ gauge field, and such a model is equivalent to a generic $O(3)$ spin model~\cite{SachdevPark2002}.
By examining a low-energy field configuration with a monopole in the CP$^1$ model, one can see that it corresponds to a hedgehog in the $O(3)$ spin model.
Hence, Ref.~\cite{MotrunichVishwanath2004} proposed that the $O(3)$ spin model where hedgehogs are completely suppressed is equivalent to a CP$^1$ model where monopoles in the $U(1)$ gauge field are completely suppressed; this is the origin of the name ``NCCP$^1$'' where ``NC'' stands for ``non-compactness'' of the $U(1)$ gauge field after the complete monopole suppression (perhaps another name could be ``no-monopole'' CP$^1$).
By starting with such $O(3)$ spins with the complete hedgehog suppression and introducing easy-plane spin anisotropy---e.g., allowing terms like $n_3^2 - n_1^2 - n_2^2$ that make the spins prefer to lie in the $(n_1, n_2)$ plane,---we obtain the easy-plane NCCP$^1$ model.

We can ask if there is an analog related to the 1d Ising DQCP.
It is easy to see, in either the domain wall language or the parton language, that we have a two-component field $(s_+, s_-) \in \mathbb{R}^2$ with a local $Z_2$ gauge redundancy, i.e., $(-s_+, -s_-)$ is identified with $(s_+, s_-)$.
Without changing qualitative properties, we can also require $s_+^2 + s_-^2 = 1$, in which case we can call it an RP$^1$ degree of freedom.
Such an RP$^1$ variable is actually equivalent to an $O(2)$ spin $\vec{n} = (n_1, n_2)$ via $n_1 + \ii n_2 \sim (s_+ + \ii s_-)^2$; for later convenience, we pick the overall phase such that $n_1 = 2 s_+ s_-$ and $n_2 = s_-^2 - s_+^2$.
By analogy with the CP$^1$ model in (2+1)D, we can define a generic RP$^1$ lattice statistical mechanics model in (1+1)D in terms of $(s_+, s_-)$ fields coupled to a dynamical $Z_2$ gauge field.
Such a model is equivalent to a generic $O(2)$ model in (1+1)D.
It is easy to see that in the RP$^1$ model, a low-energy configuration with a $Z_2$ instanton (``vison'') has a half-vortex in the two-component field $(s_+, s_-)$ and hence corresponds to a configuration with a full strength vortex in the $O(2)$ field $(n_1, n_2)$.
On the other hand, a full vortex in the two-component field $(s_+, s_-)$ does not require a $Z_2$ instanton and corresponds to a double-strength vortex in $(n_1, n_2)$.

Now, in analogy to how we obtain the NCCP$^1$ model by the complete monopole suppression in the CP$^1$ model, we can define a ``no-instanton'' RP$^1$ model by a complete suppression of $Z_2$ instantons in the RP$^1$ model; this is exactly what emerges from our domain wall/parton theory when we set $\Gamma = 0$, which is the presumable fate at long distances in our 1d Ising DQCP theory.
We also conclude that such a no-instanton RP$^1$ model is mathematically equivalent to an abstract $O(2)$ statistical mechanics model in (1+1)D with complete suppression of odd-strength vortices, while even-strength vortices are allowed.
To connect with the 1d Ising DQCP, we need to add appropriate anisotropy in the $(n_1, n_2)$ spin variables.
For concreteness, let the real-valued fields $(s_+, s_-)$ describe the parton Ising variables $(\tau^z_+, \tau^z_-)$; then $n_1$ can be roughly identified with the physical spin component $\sigma^z$, while $n_2$ can be identified with $\sigma^x$.
For our purposes, we want easy-axis anisotropy along the $\sigma^z$ direction, which can be realized by adding the following term to the energy: $\lambda_a (n_1^2 - n_2^2) = \lambda_a \cos(\phi)$ with $\lambda_a < 0$, where we have used the phase representation $n_1 + \ii n_2 \sim \ee^{\ii \phi}$ which matches with Eq.~(\ref{eq:rotorvars}).
In the familiar dual sine-Gordon description of the $O(2)$ model, allowing vortices corresponds to having terms $\cos(2\theta)$ and its multiples in the action, where the dual field $\theta$ is introduced identically to our direct bosonization treatment in Sec.~\ref{sec:bosonization}.
Then, prohibiting strength-one vortices by hand while allowing strength-two vortices corresponds to the leading allowed cosine being $\lambda_u \cos(4\theta)$.
We have thus recovered the direct bosonization description of the transition in Sec.~\ref{sec:bosonization}, which is not surprising since the $(n_1, n_2)$ give essentially the physical spin components $(\sigma^z, \sigma^x)$, and in that analysis the combination of translation and on-site symmetries effectively prohibited strength-one vortices on long distances, which in the abstract no-instanton RP$^1$ model we simply postulated by hand.

While we have arrived at the picture already described in Sec.~\ref{sec:bosonization}, we have also learned that the no-instanton model with the larger $O(2)$ symmetry relates to the easy-axis model of interest for the 1d Ising DQCP in the manner that resembles how the SU(2)-symmetric NCCP$^1$ model relates to the EP-NCCP$^1$.
In the present (1+1)D context, the higher-symmetry model corresponds to $\lambda_a = 0$ and has a QLRO phase when $\lambda_u$ is irrelevant, while the transition to the VBS phase is of Kosterlitz-Thouless type and is obtained when the $\lambda_u$ becomes relevant.
On the other hand, in the easy-axis model, the $z$-FM to VBS transition corresponds to $\lambda_a$ and $\lambda_u$ both non-zero and effectively, combining to a strictly marginal term.
While the (1+1)D physics is definitely important in these observations, it would be interesting to see if there may be some analogs in the (2+1)D DQCP theories.

\section{Formulation in ``good variables'' to describe criticality} \label{sec:goodvars}
In this section, we propose a new ``duality transformation'' to ``good variables'' which resembles a two-step duality transformation for the quantum Ashkin-Teller model in Appendix 2 in Ref.~\cite{KohmotoNijsKadanoff}.
In the new set of variables, the $z$-FM and VBS order parameters are treated more democratically, and at long wavelengths there emerges a $U(1)$ symmetry rotating these two order parameters into each other.
Furthermore, the phase transition between the $z$-FM and VBS orders is described by a Luttinger-liquid-like theory (i.e., a Gaussian theory) with only one relevant cosine operator, and critical exponents can be easily extracted.

\subsection{``Good variables'' as new ``duality'' on the domain wall/parton variables}
The derivation of the duality transformation is a two-step process described in Appendix~\ref{app:deriv_good_vars}.
Here we state the final result, whose nice structure can be appreciated already without the derivation.
The new degrees of freedom are ``matter'' fields $\nu$ (two-level systems) residing on ``quarter-integer'' lattice sites $j \pm 1/4$, as well as Ising ``gauge fields'' $\xi$ residing on links of the $\nu$-lattice, or, equivalently, at positions $j$ (original lattice site) and $j+1/2$ (original lattice link).
The physical Hilbert space is defined by gauge constraints
\begin{align}
  \xi^x_j\, \xi^x_{j+1/2} = \nu^x_{j+1/4} ~, \quad \xi^x_{j+1/2}\, \xi^x_{j+1} = \nu^x_{j+3/4} ~.
  \label{eq:at_dual_constraint}
\end{align}
The operator map between the domain wall variables $\mu_+, \mu_-, \rho$ [with constraint Eq.~(\ref{eq:mu_pm_gauge_constraint})] and these new variables (with the above constraint) is
\begin{align}
  & \mu^x_{+, j+1/2} = \tau^z_{+, j}\, \zeta^z_{j+1/2}\, \tau^z_{+, j+1} = \nu^y_{j+1/4}\, \xi^z_{j+1/2}\, \nu^y_{j+3/4} ~, \notag \\
  & \mu^x_{-, j+1/2} = \tau^z_{-, j}\, \zeta^z_{j+1/2}\, \tau^z_{-, j+1} = \nu^z_{j+1/4}\, \xi^z_{j+1/2}\, \nu^z_{j+3/4} ~, \notag \\
  & \mu^z_{+, j-1/2}\, \rho^z_j\, \mu^z_{+, j+1/2} = \tau^x_{+, j} = \nu^z_{j-1/4}\, \xi^z_j\, \nu^z_{j+1/4} ~, \notag \\
  & \mu^z_{-, j-1/2}\, \rho^z_j\, \mu^z_{-, j+1/2} = \tau^x_{-, j} = \nu^y_{j-1/4}\, \xi^z_j\, \nu^y_{j+1/4} ~, \notag \\
  & \rho^x_j = \tau^z_{+, j}\, \tau^z_{-, j} = (-1)^j \xi^x_j ~, \notag \\
  & \mu^z_{+, j+1/2}\, \mu^z_{-, j+1/2} = \zeta^x_{j+1/2} = (-1)^j \xi^x_{j+1/2} ~.
  \label{eq:good_vars}
\end{align}
To bring out the structure more clearly, we have also included map to the parton variables $\tau_+, \tau_-, \zeta$ [with constraint Eq.~(\ref{eq:bosparton_gauge_constraint})], which is just the duality map between the domain wall and parton variables, Eq.~(\ref{eq:dw_to_parton}).
From the above equations, we can loosely say that the new variables ``straddle'' (or ``unify'') the domain wall and parton variables.
We emphasize that the above operator map is an \emph{exact} relation between the domain wall theory and the new theory on a chain with periodic boundary conditions.
This can be also verified directly without going through the two-step procedure of Appendix~\ref{app:deriv_good_vars}, e.g., by ``solving'' the constrained theory $\rho_j, \mu_{+, j+1/2}, \mu_{-, j+1/2}$ in terms of unconstrained variables $\rho_j, \mu_{+, j+1/2}$, and by ``solving'' the constrained theory $\nu_{j \pm 1/4}, \xi_j, \xi_{j+1/2}$ in terms of unconstrained variables $\xi_j, \xi_{j+1/2}$, and carefully matching physical operators in the unconstrained theories.

Using these operator mappings, we can rewrite the Hamiltonian Eq.~(\ref{eq:model_mu_pm}) as
\begin{align}
  \tilde{H} = & -J \sum_j \xi^z_{j+1/2} \left( \nu^z_{j+1/4} \nu^z_{j+3/4} + \nu^y_{j+1/4} \nu^y_{j+3/4} \right) \notag \\
  & -h \sum_j \xi^z_j \left( \nu^z_{j-1/4} \nu^z_{j+1/4} + \nu^y_{j-1/4} \nu^y_{j+1/4} \right) \notag \\
  & - \Gamma \sum_j \xi^x_{j+1/2} ~.
  \label{eq:model_at_dual_two}
\end{align}
By choosing a gauge such that $\xi^z$ is invariant under the symmetries, we can also identify action of the global symmetries as
\begin{align}
  T_x:\ & \nu^{x,y}_{j+1/4} \rightarrow -\nu^{x,y}_{j+5/4} ~, \quad \nu^z_{j+1/4} \rightarrow \nu^z_{j+5/4} ~, \notag \\
  & \nu^{x,y}_{j-1/4} \rightarrow -\nu^{x,y}_{j+3/4} ~, \quad \nu^z_{j-1/4} \rightarrow \nu^z_{j+3/4} ~, \notag \\
  & \xi^x_j \rightarrow -\xi^x_{j+1} ~, \quad \xi^x_{j+1/2} \rightarrow \xi^x_{j+3/2} ~; \notag \\
  g_x:\ & \nu^{x,y}_{j \pm 1/4} \rightarrow -\nu^{x,y}_{j \pm 1/4} ~, \quad \nu^z_{j \pm 1/4} \rightarrow \nu^z_{j \pm 1/4} ~, \notag \\
  & \xi^x_j \rightarrow -\xi^x_j ~, \quad \xi^x_{j+1/2} \rightarrow\xi^x_{j+1/2} ~; \notag \\
  g_z:\ & \nu^x_{j \pm 1/4} \rightarrow \nu^x_{j \pm 1/4} ~, \notag \\
  & \nu^y_{j \pm 1/4} \rightarrow \nu^z_{j \pm 1/4} ~, \quad \nu^z_{j \pm 1/4} \rightarrow -\nu^y_{j \pm 1/4} ~, \notag \\
  & \xi^x_j \rightarrow \xi^x_j ~, \quad \xi^x_{j+1/2} \rightarrow \xi^x_{j+1/2} ~; \notag \\
  \TT:\ & \nu^x_{j \pm 1/4} \rightarrow -\nu^x_{j \pm 1/4} ~, \quad \ii \rightarrow -\ii ~, \notag \\
  & \nu^y_{j \pm 1/4} \rightarrow \nu^z_{j \pm 1/4} ~, \quad \nu^z_{j \pm 1/4} \rightarrow -\nu^y_{j \pm 1/4} ~, \notag\\
  & \xi^x_j \rightarrow -\xi^x_j ~, \quad \xi^x_{j+1/2} \rightarrow \xi^x_{j+1/2} ~.
  \label{eq:sym_at_dual}
\end{align}

Note that $T_x$ is the translation symmetry of the physical spins, which translates the quarter-integer $\nu$-chain by two sites.
When $J = h$ and $\Gamma = 0$, the $\nu$-chain Hamiltonian is invariant under translation by one site.
From the mapping Eq.~(\ref{eq:good_vars}), it is clear that this corresponds to an exact self-duality condition in the domain wall/parton variables~\cite{KohmotoNijsKadanoff, MrossAliceaMotrunich2017, ChewMrossAlicea2018}.
For $\Gamma \neq 0$ and more general interactions, we expect this to be an emergent symmetry at the $z$-FM to VBS transition.

The convenience of the new variables is as follows.
First of all, these variables show that the simple domain wall theory that we wrote in Eq.~(\ref{eq:model_mu_pm}) has an accidental $U(1)$ symmetry of rotations in the $\nu^y - \nu^z$ plane.
This symmetry is, in general, not present; e.g., one can have terms like $\sum_{\sigma = \pm} \mu^x_{\sigma, j-1/2} \mu^x_{\sigma, j+1/2} = 
\xi^z_{j-1/2} \xi^z_{j+1/2} [\nu^y_{j-3/4} \nu^y_{j-1/4} \nu^y_{j+1/4} \nu^y_{j+3/4} + (\nu^y \to \nu^z)]$, etc.
However, both $g_z$ and $\TT$ act like a $Z_4$ symmetry rotating by $\pi/2$ in the $\nu^y - \nu^z$ plane.
Hence, if we ignore the gauge field $\xi$ for a moment, we get a ``$YZ$'' chain with a four-fold anisotropy and with alternating bond strengths.
This is similar to ``good variables'' in the description of the Ashkin-Teller transition, see Appendix 2 in Ref.~\cite{KohmotoNijsKadanoff}.
We emphasize, however, that in the present case starting with our domain wall theory, which can be viewed as a ``gauged'' Ashkin-Teller model, we are keeping track of all global aspects exactly, which is achieved by using the device of the gauge field $\xi$.

Assuming weak such four-fold anisotropy and weak staggering of the bond strengths, we can employ abelian bosonization~\cite{Haldane1981, GiamarchiBook, SachdevBook} to describe this chain in terms of a phase variable $\phigood$ and its conjugate variable $\thtgood$, defined via
\begin{align}
& \frac{1}{2} \left( \nu^z + \ii \nu^y \right)_\text{unif.} \sim e^{\ii \phigood} ~, \notag \\
& \frac{1}{2} \left( \nu^x \right)_\text{unif.} \sim \frac{\partial_x \thtgood}{\pi} ~, \quad
\frac{1}{2} \left( \nu^x \right)_\text{stagg.} \sim D \sin(2\thtgood) ~.
\label{eq:gauged_bosonization}
\end{align}
In the last line, the length units are those of the $\nu$ chain, i.e., $\partial_x$ is the continuum version of the corresponding lattice derivative.
In the above equations, the magnetization components $(\nu^{x,y,z})_\text{unif.}$ and fields $\phigood, \thtgood$ are understood as long-wavelength (slowly-varying) fields; the latter have commutation relations in the continuum $\left[ \partial_x \thtgood(x)/\pi, \phigood(x') \right] = \ii \delta(x-x')$.
We can think of this as a hydrodynamic description of a bosonic system obtained from the $\nu$ spin chain via well-known spin-to-boson mapping adopted to the present case, where on the lattice scale $\nu^x/2 = 1/2 - n$, with $n$ the ``boson number.''
These bosons are at half-filling (which is enforced by the $g_x$ symmetry), and it is well-known that the boson density has an important contribution to its staggered component along the chain, which is also quoted in the last line in Eq.~(\ref{eq:gauged_bosonization}).

With the above conventions, the bond energy density on the $\nu$ chain has a staggered component proportional to $\cos(2\thtgood)$.
Since in the microscopic $\nu$ chain, even and odd bonds have different strengths $J$ and $h$ and also differ due to the $\Gamma$ terms, we expect that the effective continuum action contains a term $\lambda \cos(2\thtgood)$.
Furthermore, since the microscopic $\nu$ chain has only $Z_4$-like symmetry rather than full $U(1)$ symmetry, we expect that the effective action also contains a term $\kappa \cos(4\phigood)$.
Putting these together, the effective action reads,
\begin{align}
S[\phigood, \thtgood] \!=\! & \int\! \dd \tau\, \dd x \left[ \frac{\ii}{\pi} \partial_\tau \phigood \partial_x \thtgood + \frac{\tilde{v}}{2\pi} \left( \frac{1}{\tilde{g}}(\partial_x \thtgood)^2 + \tilde{g} (\partial_x \phigood)^2 \right) \right] ~~~ \notag \\
& \!+\! \!\int\! \dd \tau\, \dd x \left[ \lambda \cos(2\thtgood) + \kappa \cos(4\phigood) \right] + \cdots ~,
  \label{eq:gauged_bosonization_action}
\end{align}
where we have included general Luttinger parameter $\tilde{g}$ (which can be generated, e.g., by allowed $\nu^x \nu^x$ interactions).

We can readily calculate scaling dimensions of the staggered bond coupling and the four-fold spin anisotropy at the Gaussian fixed point ($\lambda = \kappa = 0$):
\begin{align}
 \dim[\cos(2\thtgood)] = \tilde{g} ~, \quad \dim[\cos(4\phigood)] = \frac{4}{\tilde{g}} ~.
\end{align}
When $\tilde{g} < 2$ (which in particular includes our starting point $\tilde{g} \sim 1$), $\cos(2\thtgood)$ is relevant while $\cos(4\phigood)$ is irrelevant.
In this case, we can ignore the four-fold anisotropy at long wavelengths, and the phase transition happens when $\lambda$ changes sign.
In particular, special point with $\lambda = \kappa = 0$ and $\tilde{g} = 1$ corresponds to $J = h$ and $\Gamma = 0$ in the lattice model in Eq.~(\ref{eq:model_at_dual_two}).
Non-zero $\lambda$ corresponds to effective difference between the even and odd bonds of the $\nu$ chain, which comes from both $J \neq h$ and $\Gamma \neq 0$.

Before proceeding further, we should carefully identify Hilbert space for $\thtgood$ and $\phigood$.
In the usual abelian bosonization scheme, $\theta(x) + \pi \sim \theta(x)$, i.e., configurations of the field $\theta(x)$ that differ by a global shift by $\pi$ are identified; similarly, $\phi(x) + 2\pi \sim \phi(x)$.
However, in the present case where the $\nu$ spins actually represent matter field coupled to the dynamical gauge field $\xi$, in order to capture global aspects while using similar hydrodynamic expressions, we require a distinct Hilbert space for $\thtgood$ and $\phigood$.

To see this, let us consider a particular $Z_2$ ``symmetry'' generated by $U = \prod_j \left( \nu^x_{j-1/4} \nu^x_{j+1/4} \right)$, which acts as $\nu^{y,z} \rightarrow -\nu^{y,z},\ \phigood \rightarrow \phigood + \pi$.
However, in the constrained Hilbert space satisfying Eq.~(\ref{eq:at_dual_constraint}), $U$ acts as identity operator.
Hence, $\phigood + \pi$ (i.e., global shift by $\pi$) and $\phigood$ should be identified as the same physical state.

We also point out that Eq.~(\ref{eq:gauged_bosonization}) should be understood as performing bosonization in a fixed gauge field configuration, e.g., $\xi^z = 1$.
However, it is not enough to consider a fixed gauge field configuration.
We should also include instanton operators $\xi^x$ as local physical observables.
Indeed, $\xi^x_{j+1/2}$ is explicitly present in the Hamiltonian, while $(-1)^j \xi^x_j$ corresponds to $\sigma^z_j$ of the physical spin.
To get some intuition how to bosonize $\xi^x$, we observe that we can use constraints, Eq.~(\ref{eq:at_dual_constraint}), to write schematically
\begin{align*}
& \xi^x_j = \dots \nu^x_{j-\frac{5}{4}} \nu^x_{j-\frac{3}{4}} \nu^x_{j-\frac{1}{4}} = e^{\pm \ii \pi (\dots + n_{j-\frac{5}{4}} + n_{j-\frac{3}{4}} + n_{j-\frac{1}{4}})} ~, \\
& \xi^x_{j+1/2} = \dots \nu^x_{j-\frac{3}{4}} \nu^x_{j-\frac{1}{4}} \nu^x_{j+\frac{1}{4}} = e^{\pm \ii \pi (\dots + n_{j-\frac{3}{4}} + n_{j-\frac{1}{4}} + n_{j+\frac{1}{4}}}) ~.
\end{align*}
Since $\nu^x/2 = 1/2 - n \sim \partial_x \thtgood / \pi$, we conclude that the exponents on the right sides give, schematically, $(-1)^j e^{\pm \ii \theta}$, where half-filling for $n$ gave one $(-1)$ per two $n$'s (i.e., per increase of $j$ by $1$).
Thus, vertex operators $e^{\pm \ii \thtgood}$ should also be identified as local observables.
In particular, such vertex operators can appear in the effective action at the lattice scale.
The physics here is that for usual bosons with no gauge fields, vertex operators $e^{\pm \ii 2\theta}$ correspond to allowing vortices, while here instantons of the $Z_2$ gauge field act like half-vortices
\footnote{A more formal demonstration of this can be carried out in Euclidean path integral language along the lines of Appendix A in Ref.~\cite{LaiMotrunich2011}, which asked similar question about $Z_2$ instanton effects but motivated by gapless Majorana spin liquids in 1d.}.
Here we propose that we can capture this physics by requiring that periodicity for $\thtgood$ should be $2\pi$ rather than $\pi$.

The above schematic treatment using string operators in terms of $\nu^x$, while showing the appearance of the $e^{\pm \ii \thtgood}$ vertex operators, does not tell which specific combinations will give us the physical observables.
We can fix this using the following argument.
Consider candidate expressions
\begin{align}
  & \xi^x_j \sim A (-1)^j \sin[\thtgood(j) + \alpha] ~, \notag \\
  & \xi^x_{j+1/2} \sim B (-1)^j \cos[\thtgood(j+1/2) + \beta] ~.
  \label{eq:gauged_bosonization_monopole}
\end{align}
We use these together with the Gauss law constraints, Eq.~(\ref{eq:at_dual_constraint}), to find uniform and staggered components of $\nu^x$ in terms of the long-wavelength field $\thtgood$:
\begin{align*}
(\nu^x)_\text{unif.} & \sim \nu^x_{j+1/4} + \nu^x_{j+3/4} \\
& \approx -\frac{AB}{2} [\cos(\alpha - \beta) + \cos(2\thtgood + \alpha + \beta)] \, \partial_x \thtgood ~, \\
(\nu^x)_\text{stagg.} & \sim \nu^x_{j+1/4} - \nu^x_{j+3/4} \\
& \approx AB [\sin(\alpha - \beta) + \sin(2\thtgood + \alpha + \beta)] ~.
\end{align*}
In order to match with Eq.~(\ref{eq:gauged_bosonization}), we then require $\alpha - \beta = \pi \times \text{int}$ and $\alpha + \beta = \pi \times \text{int}$.
We can absorb $\pi$ shifts in $\alpha$ and $\beta$ into redefinitions of $A$ and $B$; remaining distinct solutions are then $\alpha = \beta = 0$ or $\pi/2$.
In the latter case, we can finally use freedom to change the offset of $\thtgood$ by $\pi/2$:
This does not change the earlier convention where $\sin(2\thtgood)$ appears in $(\nu^x)_\text{stagg.}$ while $\cos(2\thtgood)$ appears in the staggered bond energy in the $\nu$ chain; only the signs of the corresponding amplitudes $D$ and $\lambda$ change, but now these amplitudes are fixed uniquely.
Hence, we can completely fix our convention for the offset of the $\thtgood$ field by choosing $\alpha = \beta = 0$.
To summarize, we can now write bosonized expressions for the $z$-FM and VBS order parameters essentially from the microscopic parton and domain wall perspectives:
\begin{align}
&\!\! M_z^\text{FM} \sim \tau^z_{+, j} \tau^z_{-, j} = (-1)^j \xi^x_j \sim \sin(\thtgood) ~, \label{eq:MzFM} \\
&\!\! \Psi_\text{VBS} \sim \mu^z_{+, j+1/2} \mu^z_{-, j+1/2} = (-1)^j \xi^x_{j+1/2} \sim \cos(\thtgood)\, . \label{eq:PsiVBS}
\end{align}

Let us examine the lattice Hamiltonian Eq.~(\ref{eq:model_at_dual_two}) with these insights.
Due to the oscillating factor $(-1)^j$ in Eq.~(\ref{eq:gauged_bosonization_monopole}), we conclude that terms in the third line in Eq.~(\ref{eq:model_at_dual_two}) are washed out (average to zero) at long wavelengths and can be loosely thought as ``irrelevant'' in the critical theory for small coupling $\Gamma$.
However, note that we are not simply dropping the instanton effects of the gauge field---their physics persists in how precisely we define the continuum theory and physical observables.
(Similar phenomenon was found in Ref.~\cite{LaiMotrunich2011} for physical observables in gapless Majorana spin liquids in 1d.)

Equipped with the above results, we now work out symmetry actions on $\phigood$ and $\thtgood$ that correspond to Eq.(\ref{eq:sym_at_dual}):
\begin{align}
  T_x:\ & \phigood \rightarrow -\phigood ~, \quad \thtgood \rightarrow -\thtgood + \pi ~; \notag \\
  g_x:\ & \phigood \rightarrow -\phigood ~, \quad \thtgood \rightarrow -\thtgood ~; \notag \\
  g_z:\ & \phigood \rightarrow \phigood + \frac{\pi}{2} ~, \quad \thtgood \rightarrow \thtgood ~; \notag \\
  \TT:\ & \phigood \rightarrow \phigood + \frac{\pi}{2} ~, \quad \thtgood \rightarrow -\thtgood ~, \quad \ii \rightarrow -\ii ~.
\end{align}
Note that here we already crucially use that shifting $\thtgood$ by $\pi$ yields a distinct physical state: 
Without this, we would not be able to distinguish actions of $T_x$ and $g_x$.
Of course, any discussion of the physical observables in Eq.~(\ref{eq:gauged_bosonization_monopole}) would not make much sense without requiring $2\pi$ periodicity of $\thtgood$, and we note that their transformation properties are correctly captured in this framework.

Returning to the continuum theory that we wrote earlier in Eq.~(\ref{eq:gauged_bosonization_action}), we can verify that it indeed exhibits the most important symmetry-allowed terms.
Next in importance symmetry-allowed terms are $\partial_x \thtgood \sin(2\thtgood)$ and $\cos(4\thtgood)$ with scaling dimensions $1 + \tilde{g}$ and $4\tilde{g}$ at the Gaussian fixed point, and both are irrelevant for $\tilde{g} > 1$.
The $\partial_x \thtgood \sin(2\thtgood)$ term would actually be prohibited if we also require the spatial inversion symmetry.
The $\cos(4\thtgood)$ term would still be allowed but is irrelevant for $\tilde{g} > 1/2$.
For simplicity, we will assume presence of the inversion symmetry and will assume $1/2 < \tilde{g} < 2$ [guaranteeing also irrelevance of $\cos(4\phigood)$] throughout the discussion of the $z$-FM to VBS transition.

Another consistency check for the theory, in particular the claimed $2\pi$ periodicity of $\thtgood$, is provided by examining neighboring phases for non-zero $\lambda$.
When $\lambda > 0$, the action is minimized by uniform $\thtgood(x) = \pi/2$ or $-\pi/2$.
Thus, there are two degenerate ground states, and it is easy to see that these break the $g_x$ and $\TT$ symmetries but preserve the $T_x$ and $g_z$ symmetries.
So, we obtain the $z$-FM order.

On the other hand, when $\lambda < 0$, the action is minimized by $\thtgood(x) = 0$ or $\pi$.
Thus, again there are two degenerate ground states, which now break the $T_x$ symmetry but preserve all internal symmetries.
So, we obtain the VBS phase.

The transition between the two phases occurs when the effective coupling $\lambda$ for the single relevant operator changes sign.
The correlation length exponent $\nu$ is simply related to the scaling dimension of this term:
\begin{align}
\nu = \frac{1}{2 - \tilde{g}} ~,
\label{eq:nu}
\end{align}
and can vary in the range $\nu \in (2/3, \infty)$ for $\tilde{g} \in (1/2, 2)$.
The $z$-FM and VBS order parameters in Eqs.~(\ref{eq:MzFM}) and (\ref{eq:PsiVBS}) clearly have the same scaling dimension given by
\begin{align}
\text{dim}[M_z^\text{FM}] = \text{dim}[\Psi_\text{VBS}] = \frac{\tilde{g}}{4} ~,
\label{eq:scalingdims_zFM_VBS}
\end{align}
which can vary in the range $(1/8, 1/2)$.
In fact, the critical theory has an emergent continuous symmetry that rotates these parameters into each other, thus ``unifying'' the $z$-FM and VBS orders.

Turning to other observables, we also find that the $x$-FM and $y$-AFM order parameters have equal scaling dimensions.
Indeed, the former can be obtained from the microscopic expression for $\sigma_j^x$ and taking its long-wavelength component:
\begin{align}
M_x^\text{FM} & \sim \tau^x_{+, j} - \tau^x_{-, j} = \xi^z_j \, (\nu^z_{j-1/4}\, \nu^z_{j+1/4} - \nu^y_{j-1/4}\, \nu^y_{j+1/4} ) \notag \\
& \sim \cos(2\phigood) ~. \label{eq:MxFM}
\end{align}
On the other hand, while we do not have a simple microscopic expression for $\sigma_j^y$, we can verify that the following operator has the same transformation properties as $(-1)^j \sigma^y_j$, i.e., the $y$-AFM order parameter:
\begin{align}
M_y^\text{AFM} & \sim \xi^z_j \, (\nu^y_{j-1/4}\, \nu^z_{j+1/4} + \nu^z_{j-1/4}\, \nu^y_{j+1/4} ) \notag \\
& \sim \sin(2\phigood) ~. \label{eq:MyAFM}
\end{align}
The corresponding scaling dimensions are
\begin{align}
\text{dim}[M_x^\text{FM}] = \text{dim}[M_y^\text{AFM}] = \frac{1}{\tilde{g}} ~,
\label{eq:dim_xFM_yAFM}
\end{align}
and can vary between $2$ and $1/2$.
This concludes our discussion of key properties of the $z$-FM to VBS transition.

It is interesting to also examine nearby phases that can be accessed by our theory when the cosine terms that were irrelevant at the $z$-FM to VBS transition become important.
Thus, when $\tilde{g}$ approaches $2$, the term $\kappa \cos(4\phigood)$ becomes important.
It is easy to see that when this term dominates, the system either develops the $x$-FM order when $\kappa < 0$ or the $y$-AFM order when $\kappa > 0$ (the $\pi$ periodicity of the $\phigood$ field ensures that there are two degenerate ground states in each case).
In the present variables, transition to either of these phases either from the $z$-FM phase or the VBS phase is described by a strongly coupled theory where $\cos(2\thtgood)$ and $\cos(4\phigood)$ compete.
In this regard, recall, e.g., that the $z$-FM to $x$-FM transition was actually easy to describe in the direct bosonization variables in Sec.~\ref{sec:bosonization}, where it was difficult to describe the $z$-FM to VBS transition;
thus, the situation is reversed in the present variables.

Let us now consider what happens when $\tilde{g}$ approaches $1/2$ and the term $\lambda' \cos(4\thtgood)$ becomes important (here and below we assume the inversion symmetry to disallow $\partial_x \thtgood \sin(2\thtgood)$ term that would become relevant earlier).
Assuming for a moment that the $\lambda'$ term dominates, we can start by minimizing it; however, it will be important to remember that at the same time we also have the term $\lambda \cos(2\thtgood)$ with generically non-zero $\lambda$.
When $\lambda' < 0$, the corresponding term by itself would have four degenerate ground states: $\thtgood = 0, \pi/2, \pi, 3\pi/2$.
However, the $\lambda$ term will select two of them as ground states of the full action: $\lambda > 0$ will select $\thtgood = \pi/2, 3\pi/2$ corresponding to the $z$-FM phase,
while $\lambda < 0$ will select $\thtgood = 0, \pi$ corresponding to the VBS phase.

On the other hand, when $\lambda' > 0$, the ground states of the corresponding term are $\thtgood = \pi/4, 3\pi/4, 5\pi/4, 7\pi/4$ and are not differentiated by the $\lambda$ term.
More precisely, including the $\lambda$ term will shift the four minima to have the form $\thtgood = \pm (\pi/4 + \delta), \pm (3\pi/4 - \delta)$, and they remain energetically degenerate.
This phase will have coexisting $z$-FM and VBS orders.
It is then natural to guess that our $z$-FM to VBS transition line, upon entering this regime, splits into two lines opening the above phase where the two orders coexist.
Since only one order appears or disappears across each of these lines, we expect that these transitions will be in the Ising universality class.

\subsubsection{Crystalline-SPT-like property of the VBS phase}
We conclude this section by the following interesting observation about a subtle but precise character of the VBS phase in our model, alerted to us by the long-wavelength theory for the $z$-FM to VBS transition.
Our theory implies sharp distinction between the $\sigma^x$ and $\sigma^y$ spin components at the transition:
Indeed, Eq.~(\ref{eq:dim_xFM_yAFM}) shows that there are strong ferromagnetic correlations in the former but antiferromagnetic in the latter.
In our long-wavelength theory, we have looked for and found possible contributions to the ``opposite'' $x$-AFM and $y$-FM order parameters (not shown here), and have concluded that these always have higher scaling dimensions than the $x$-FM and $y$-AFM order parameters discussed earlier.
Furthermore, as we have seen, our theory naturally predicts nearby phases with $x$-FM long range order or $y$-AFM long range order, but not the opposite orders.

At first sight, this is very puzzling since the $z$-FM and VBS phases considered here have only short-range correlations in the $\sigma^x$ and $\sigma^y$ spin components, and naively we did not invoke the $g_x$ and $g_y$ symmetries in specifying these phases.
However, we think that this is too naive and that the resolution of the puzzle is that there is a subtle crystalline-SPT-like property~\cite{YaoKivelson2010, KimLeeJiangWareJianZaletel2016, HaoHuangFuHermele2017} of our VBS phase involving the $g_x$ and $g_y$ symmetries that makes the $\sigma^x$ and $\sigma^y$ spin components inequivalent.

To understand this, we note that in our model with $J_x = J_z$ and $K_{2x} = K_{2z}$, at the exactly solvable Majumdar-Ghosh-like point $K_2/J = 0.5$ inside the VBS phase~\cite{MajumdarGhosh1969, MajumdarGhosh1969II, FurukawaSatoFurusaki2010, FurukawaSatoOnodaFurusaki2012}, the wavefunction for a single dimer has the form
\begin{align*}
& |D_{12} \rangle = \frac{ |\!+\!\hat{y} \rangle_1 |\!-\!\hat{y} \rangle_2 + |\!-\!\hat{y} \rangle_1 |\!+\!\hat{y} \rangle_2}{\sqrt{2}} \\
& = \frac{ |\!+\!\hat{z} \rangle_1 |\!+\!\hat{z} \rangle_2 + |\!-\!\hat{z} \rangle_1 |\!-\!\hat{z} \rangle_2 }{\sqrt{2}} 
 = \frac{ |\!+\!\hat{x} \rangle_1 |\!+\!\hat{x} \rangle_2 + |\!-\!\hat{x} \rangle_1 |\!-\!\hat{x} \rangle_2 }{\sqrt{2}} \, .
\end{align*}
Naturally, it shows ferromagnetic correlations between the $\sigma^z$ spin components of the two spins and also between the $\sigma^x$ components, while the correlations between the $\sigma^y$ components are antiferromagnetic.
While these are, of course, short-range correlations, crucially, the dimer wavefunction is even under the $g_z$ and $g_x$ symmetries but odd under the $g_y$ symmetry.
Then, on a chain of length $L = 4N + 2$ (i.e., with an odd number of dimers), the many-body wavefunction will be similarly even under $g_z$ and $g_x$ but odd under $g_y$.
These ground state quantum numbers will persist also away from the Majumdar-Ghosh point and provide precise additional characterization of the VBS phase, which can be understood as SPT protected by the remaining translation symmetry by two lattice sites and the $g_{x,y,z}$ symmetries.
Near the $z$-FM phase, it is natural to expect that $g_z = +1$, while without further specifications we can have $(g_x, g_y) = (+1, -1)$ or $(-1, +1)$ which are distinct VBS phases.
Our model realizes the first case, while if we had strong ferromagnetic nearest-neighbor $J_y$ interactions instead of $J_x$ interactions, we would expect the second case.

It is interesting to trace why our field theory derivation ``naturally'' produced the first case without us actually specifying such distinction explicitly (since from the point of view of just the symmetries, the two cases are equally likely).
We think the reason is that the derivation was largely guided by the microscopics of the model.
Already in the dual variables, Eq.~(\ref{eq:dw_dual_map}), while we said simply that they describe domain walls in the $z$-FM order, we actually treated the $g_x$ and $g_y$ symmetries in subtly different ways:
The $g_x$ quantum number is encoded in the flux of the gauge field, while the $g_y$ one ``involves'' both the matter and gauge fields, and our analysis further used starting points with ``classical'' (i.e., non-fluctuating) gauge flux.
While we did not emphasize this explicitly, this difference in our treatment of the $g_x$ and $g_y$ symmetries propagated throughout our analysis: 
For example, the parton PSG, while not guided by explicit energetics considerations, was fixed by the precise duality to the domain wall variables which were guided by such considerations.

\subsection{An alternative parton view of the ``good variables''}
\label{subsec:goodvars_altparton}
Having seen the power of the above ``good variables'' for describing the $z$-FM to VBS transition, here we provide another perspective on these variables, which will teach us some interesting lessons.
The $\nu$-chain variables were motivated by looking for an analog of the good variables used to describe the Ashkin-Teller criticality~\cite{KohmotoNijsKadanoff}.
In our case with the additional gauge field and the additional $\Gamma$ term, we were able to carry out all steps exactly (i.e., capturing all global aspects), using the device of the new gauge field $\xi$.
It was particularly convenient to introduce both $\xi_j$ and $\xi_{j+1/2}$ in the derivation, and also to have them to ``unify'' the domain wall and parton variables with the nice structure in Eq.~(\ref{eq:good_vars}), as well as to unify the $z$-FM and VBS order parameters in Eqs.~(\ref{eq:MzFM}) and (\ref{eq:PsiVBS}).
However, note that the $\xi_j^x$ operators, being related to the $\sigma_j^z$ spin operators, cannot be present in the Hamiltonian, while the operators $\xi_{j+1/2}^x$ are present.
This distinction ``disappears'' at criticality, with the latter operators getting ``washed out'' at long wavelengths due to $(-1)^j$ oscillations induced in them by the physics of the $\nu$-chain [see Eq.~(\ref{eq:gauged_bosonization_monopole}) and arguments preceding it].
Still, the gauge structure of the $\nu$-chain theory and the presence of the microscopic instanton operators $\xi^x_{j+1/2}$ in the Hamiltonian do have important consequences for the structure of the critical theory and its observables, as we have already discussed.

The microscopic difference between the $\xi^x_j$ and $\xi^x_{j+1/2}$ variables suggests using the constraints, Eq.~(\ref{eq:at_dual_constraint}), to ``solve'' for $\xi^x_j = \xi^x_{j-1/2} \nu^x_{j-1/4} = \nu^x_{j+1/4} \xi^x_{j+1/2}$.
Such elimination of the $\xi_j$ variables essentially amounts to dropping the $\xi_j^z$ from the $h$ terms in the Hamiltonian, Eq.~(\ref{eq:model_at_dual_two}), obtaining a theory with variables $\nu_{j \pm 1/4}, \xi_{j+1/2}$ satisfying the constraint in the previous sentence.
``Relabeling'' further $\nu_{j \pm 1/4} \to \VV_{\pm, j}$ (for reasons that will become clear below) and $\xi_{j+1/2} \to \XX_{j+1/2}$ , we have an exact reformulation of the problem as
\begin{align}
  \tilde{H} = & -J \sum_j \XX^z_{j+1/2} \left( \VV^z_{+, j} \VV^z_{-, j+1} + \VV^y_{+, j} \VV^y_{-, j+1} \right) \notag \\
  & \! -h \sum_j \left( \VV^z_{-, j} \VV^z_{+, j} + \VV^y_{-, j} \VV^y_{+, j} \right) \!-\! \Gamma \sum_j \XX^x_{j+1/2} \,,
  \label{eq:good_variable_parton_Ham}
\end{align}
with the Hilbert space constraint
\begin{align}
\XX^x_{j-1/2} \XX^x_{j+1/2} = \VV^x_{+, j} \VV^x_{-, j} ~.
\label{eq:good_variable_parton_constraint}
\end{align}
More precisely, we have an exact operator map between the old constrained $\nu_{j \pm 1/4}, \xi_j, \xi_{j+1/2}$ problem and the new constrained $\VV_{\pm, j}, \XX_{j+1/2}$ problem.
The new labels help us to know which setup is being used and to avoid confusions such as that $\VV^z_{-, j} \VV^z_{+, j}$ is a gauge-invariant object in the new setup and corresponds to $\xi^z_j \nu^z_{j-1/4} \nu^z_{j+1/4}$ in the old setup, while $\nu^z_{j-1/4} \nu^z_{j+1/4}$ is not gauge-invariant in the old setup, etc.
The $\VV_{\pm, j}$ variables transform identically to the $\nu_{j \pm 1/4}$; we write the transformations here for readers' convenience and to emphasize concise form in these variables:
\begin{align}
  T_x:\ & \VV^{x,y}_{\pm, j} \to -\VV^{x,y}_{\pm, j+1} ~, \quad \VV^z_{\pm, j} \to \VV^z_{\pm, j+1} ~; \notag \\
  g_x:\ & \VV^{x,y}_{\pm, j} \to -\VV^{x,y}_{\pm, j} ~, \quad \VV^z_{\pm, j} \to \VV^z_{\pm, j} ~; \notag \\
  g_z:\ & \VV^x_{\pm, j} \to \VV^x_{\pm, j} ~, \quad \VV^y_{\pm, j} \to \VV^z_{\pm, j} ~, \quad \VV^z_{\pm, j} \to -\VV^y_{\pm, j} ~; \notag \\
  \TT:\ & \VV^x_{\pm, j} \to -\VV^x_{\pm, j} ~, \quad \VV^y_{\pm, j} \to \VV^z_{\pm, j} ~, \quad \VV^z_{\pm, j} \to -\VV^y_{\pm, j} ~, \notag \\
  & \ii \to -\ii ~.
  \label{eq:good_variable_parton_symmetry}
\end{align}
The gauge field components $\XX^z_{j+1/2}$ and $\XX^x_{j+1/2}$ transform trivially under all symmetries.

We can also express the physical spin operators as:
\begin{align}
\sigma^x_j & \sim \frac{1}{2} (\tau^x_{+, j} - \tau^x_{-, j}) = \frac{1}{2}(\VV^z_{-, j} \VV^z_{+, j} - \VV^y_{-, j} \VV^y_{+, j}) ~, \notag \\
\sigma^z_j & \sim \tau^z_{+, j} \tau^z_{-, j} = (-1)^j \XX^x_{j-1/2} \VV^x_{-, j} = (-1)^j \VV^x_{+, j} \XX^x_{j+1/2} ~.
\label{eq:sigmaz_XXVV}
\end{align}
Since $\XX^x$ fields transform trivially under all symmetries, we can in principle drop the factors of $\XX^x$ in the qualitative contributions to $\sigma^z_j$.
Alternatively, in the $\Gamma \to \infty$ limit, we can replace $\XX^x_{j+1/2}$ by 1 and arrive at the following new parton representation:
\begin{align}
|\sigma^z_j = \pm 1 \rangle\ \leftrightarrow\ |\VV^x_{+, j} = \VV^x_{-, j} = \pm (-1)^j \rangle ~,
\end{align}
\begin{align}
& \sigma^x_j = \frac{1}{2} (\VV^z_{+, j} \VV^z_{-, j} - \VV^y_{+, j} \VV^y_{-, j}) ~, \notag \\
& \sigma^y_j = (-1)^{j+1} \frac{1}{2} (\VV^z_{+, j} \VV^y_{-, j} + \VV^y_{+, j} \VV^z_{-, j}) ~, \notag \\
& \sigma^z_j = (-1)^j \frac{1}{2} (\VV^x_{+, j} + \VV^x_{-, j}) ~.
\label{eq:new_partons}
\end{align}
This is an interesting parton formulation in that it does not try to fractionalize the $\sigma^z$ spin component (or the $z$-FM order parameter); instead, it fractionalizes the $\sigma^{x,y}$ components, in the sense that these are represented as composites of the gauge-charged fields $\VV^{y,z}$.
Notice also rather special form of our Hamiltonian in these variables: e.g., sites $j$ and $j+1$ are coupled only via $\VV_+$ fields at $j$ and $\VV_-$ fields at $j+1$, while symmetries in principle allow either $\VV_+$ or $\VV_-$ at either end of the link.
This special choice of parameters is what lands this parton formulation and the specific ``mean-field'' near the $z$-FM to VBS transition (indeed, the above gauge theory Hamiltonian is an exact rewriting of the setups where we have already established this physics, which is robust to perturbations with generic symmetry-allowed terms).

Everything we did in the $\nu$-chain language readily translates to the new parton language, and here we only emphasize some points that are notable from the perspective of parton approaches.

First, note that the partons are gapped (i.e., ``not condensed'') on either side of the transition:
Recalling the convenience of the quarter-integer lattice of the $\nu$-chain, it is handy to organize the partons into a 1d chain, $\dots, \VV_{-, j}, \VV_{+, j}, \VV_{-, j+1}, \VV_{+, j+1}, \dots$.
The two gapped phases then correspond to different SPT phases of the $\VV$ degrees of freedom, one where they lock into entangled pairs on the even links of this chain, and the other where they lock into entangled pairs on the odd links of this chain.
Thus, for dominant $J$, we have entangled pairs on $(+, j), (-, j+1)$ ``links'' of the form 
\begin{align*}
\frac{1}{\sqrt{2}} (|\VV^x_{+, j} = 1, \VV^x_{-, j+1} = -1 \rangle + |\VV^x_{+, j} = -1, \VV^x_{-, j+1} = 1 \rangle ~.
\end{align*}
This gives $\VV^x_{+, j} \VV^x_{-, j+1} = -1$ and hence $\sigma^z_j \sigma^z_{j+1} = 1$ for each $j$, i.e., the $z$-FM phase.
A careful consideration of Gutzwiller projection into the physical spin space gives the expected two degenerate states, $(1/\sqrt{2}) (|\uparrow, \uparrow, \dots, \uparrow \rangle \pm |\downarrow, \downarrow, \dots, \downarrow \rangle)$, coming from periodic and antiperiodic boundary conditions in the parton Hamiltonian.

On the other hand, for dominant $h$, we have entangled pairs of $\VV_{-, j}$ and $\VV_{+, j}$ with $\VV^x_{-, j} \VV^x_{+, j} = -1$; the Gauss law constraints then give $\XX^x_{j+1/2} \sim (-1)^j$ and hence staggered bond energy in the gauge theory, i.e., the VBS phase.
Here consideration of Gutzwiller projection requires more care since one needs to include effect of the $J$ terms to get non-zero projection; the result is schematically $(1/\sqrt{2}) (|\text{VBS}_\text{even~links} \rangle \pm |\text{VBS}_\text{odd~links} \rangle)$, as expected in the VBS phase.
(A general technique for analyzing phases from the perspective of Gutzwiller-projected wavefunctions is described in Appendix~\ref{app:qn_fermionic_parton}.)

Second, near the critical point, the parton ``mean field'' Hamiltonian is such that there is a strong ``staggered'' component in $\VV^x$ along the $\nu$-chain, i.e., anti-correlation between $\VV^x_+$ and $\VV^x_-$.
Via the Gauss law constraints, this translates to a strong $\sim (-1)^j$ contribution to $\XX^x_{j+1/2}$, meaning that there is an operator in the long-wavelength theory identified to contribute to this physical observable.
Thus, $(\XX^x)_{q = \pi}$, which is precisely the VBS order parameter, has strong contribution, and we have already derived how it is expressed in terms of the long-wavelength fields in the $\nu$-chain section: $(\XX^x)_{q = \pi} \sim \cos(\thtgood)$.

Third, the $z$-FM order parameter necessarily involves also an instanton operator.
Indeed, the parton mean field physics is such that $\VV^\alpha_\pm$ have only $q = 0$ components:
This is clear for the $\VV^{y,z}$ fields (assuming $J, h > 0$).
To avoid confusion with ``staggering'' of $\VV^x$ in the $\nu$-chain, we can explicitly write $(\VV^x_{\pm, j})_{q=0} \sim \partial_x \thtgood/\pi \pm D \sin(2\thtgood)$, i.e., $\VV^x_+$ and $\VV^x_-$ indeed have only $q = 0$ components in the sense of the physical spin chain.
Examination of the transformation properties then immediately shows that it is not possible to construct an object out of such long-wavelength $\VV^\alpha_\pm$ fields that would be even under $T_x$ and odd under $g_x$.
For example, plugging the continuum $\VV$ fields into the expression for $\sigma^z_j$ in Eq.~(\ref{eq:new_partons}) produces only $q = \pi$ component, $(\sigma^z)_{q=\pi} \sim (\VV^x_+ + \VV^x_-)_{q=0} \sim \partial_x \thtgood$.
However, we can combine this or $(\VV^x_+ - \VV^x_-)_{q=0} \sim \sin(2\thtgood)$ with $\XX^x_{q=\pi}$ to obtain contribution to the $z$-FM order parameter.
The second combination is more important (has lower scaling dimension and also transforms correctly under the inversion if such symmetry is present), and the resulting $(\sigma^z)_{q=0} \sim (\VV^x_+ - \VV^x_-)_{q=0} (\XX^x)_{q=\pi}$ will contain $\sin(\thtgood)$, in agreement with the derivation in the $\nu$-chain part.
The $\nu$-chain language had this structure appear more clearly at the microscopic level by keeping the $\xi^x_j$ field, which to some extent is still coded in Eq.~(\ref{eq:sigmaz_XXVV}), while in the new parton formulation we recover it by appealing to symmetry arguments.

\section{Fermionic parton approach} \label{sec:fermionic_parton}
In this section, we use fermionic parton formalism to describe the $z$-FM to VBS transition.
We will see that once a correct PSG is identified that can capture these two phases, this approach directly leads to a convenient field theory description of the transition analogous of the ``good variables'' in the previous section.
This is unlike the initial bosonic parton approach in Sec.~\ref{sec:dw_parton} that required additional two-step duality-like transformation to the good variables.

\subsection{Fermionic parton representation for spins}
We enlarge the local physical spin Hilbert space to a four-dimensional fermionic Hilbert space, generated by two fermionic operators $f_+$ and $f_-$ acting on vacuum.
Our mapping between the physical spin states and fermion states is
\begin{align}
|\sigma^x = \pm 1 \rangle \leftrightarrow |P_+ = \mp 1, P_- = \pm 1 \rangle ~,
\end{align}
where $P_\pm$ are fermion parity operators defined as $P_\pm = (-1)^{f_\pm^\dg f_\pm} = f_\pm f_\pm^\dg - f_\pm^\dg f_\pm$.
Equivalently, we can express the constraint for the physical states as
\begin{align}
P_+ P_- = -1 ~, \quad \text{or} \quad f_+^\dg f_+ + f_-^\dg f_- = 1 ~.
\label{eq:fermionic_parton_constraint}
\end{align}
The physical spin operators are identified as
\begin{align}
& \sigma^x = f_+^\dg f_+ - f_-^\dg f_- ~, \notag \\
& \sigma^y = \ii f^\dg_+ f_- - \ii f_-^\dg f_+ ~, \notag \\
& \sigma^z = f_+^\dg f_- + f_-^\dg f_+ ~.
\label{eq:fermion_parton_to_physical_spin}
\end{align}
The specific parton decomposition and the PSG below were partially motivated by the structure of the bosonic parton approach in Sec.~\ref{sec:dw_parton} but should be considered as an independent formalism.

\subsection{Symmetry analysis and identification of phases}
Similar to the bosonic partons, $f_\pm$ are not local objects but should be viewed as gauge charges for a $Z_2^\zeta$ gauge field.
Consequently, symmetries act projectively on $f_\pm$, and an effective Hamiltonian for the fermionic partons should be invariant under some PSG.
The choice of PSG is not unique, and different choices of PSG describe different sets of phases and transitions.
So, which PSG should we choose in order to describe the $z$-FM and VBS phases, as well as the phase transition between them?
One approach that we tried was to use Jordan-Wigner-like fermionization of the bosonic partons $\tau_\pm$ from Sec.~\ref{sec:dw_parton}.
However, in such attempts, the need to make the $+$ and $-$ species to be mutual fermions led to ``non-local'' structures in important symmetry transformations, which we could not make compatible with the spirit of the parton formalism; on the other hand, naively ignoring these issues and simply using transformed Hamiltonians upon detailed analysis produced incorrect phases.
Instead, with some gained insights, we used trial and error to find an ansatz with desired properties.
In Appendix~\ref{app:fermion_parton_from_good_variable} we tried a particular Jordan-Wigner fermionization of the ``good'' parton variables $\VV_\pm$ from Sec.~\ref{subsec:goodvars_altparton}, which actually worked and produced a different fermionic parton setup which is very closely related to the one guessed here.
In what follows, we will present detailed analysis for our first fermionic parton setup, while in Appendix~\ref{app:fermion_parton_from_good_variable} we will point out key connections between the two setups.

The proposed effective Hamiltonian for the fermionic partons reads
\begin{align}
  H = & \sum_{j, \sigma} \zeta^z_{j+1/2} \left( -t_\sigma f_{j, \sigma}^\dg f_{j+1, \sigma} + \ii \eta_\sigma f_{j, \sigma}^\dg f_{j+1, \sigma}^\dg + \text{H.c.} \right) \notag \\
& -\mu \sum_{j, \sigma} \left( f_{j, \sigma}^\dg f_{j, \sigma} - \frac{1}{2} \right) - \sum_j (-1)^j \Gamma \, \zeta^x_{j+1/2} ~,
\label{eq:model_fermionic_parton}
\end{align}
with gauge constraint
\begin{align}
\zeta^x_{j-1/2} \zeta^x_{j+1/2} = P_{j,+} P_{j,-} ~.
\label{eq:fermparton_gaugeconstraint}
\end{align}
Here the mean field parameters $t_\sigma, \eta_\sigma$ are all real numbers and satisfy $t_- = -t_+$ and $\eta_- = \eta_+$. 

The above Hamiltonian is invariant under the following symmetry actions
\begin{align}
  T_x:\ & f_{j,\pm} \to f_{j+1,\pm} ~; \notag \\
  g_x:\ & f_{j,+} \to f_{j,+} ~, ~~ f_{j,-} \to -f_{j,-} ~; \notag \\
  g_z:\ & f_{j,+} \to (-1)^j \, \ii f_{j,-} ~, ~~ f_{j,-} \to (-1)^j \, \ii f_{j,+} ~; \notag \\
  \TT:\ & f_{j,+} \to (-1)^j f_{j,-} ~, ~~ f_{j,-} \to (-1)^j (-f_{j,+}) ~, \notag \\
  & \ii \to -\ii ~.
  \label{eq:fermionic_parton_sym}
\end{align}
The symmetry transformation rules for $\zeta^x$ are the same as in Eq.~(\ref{eq:bosonic_parton_sym}), since the postulated $\Gamma$ term is similar to that in the bosonic parton theory in Sec.~\ref{sec:dw_parton}.
We point out that the above symmetry actions on $f$ correspond to the following nontrivial PSG equations:
\begin{align}
& T_x g_z \circ f_{j, \sigma} = -g_z T_x \circ f_{j, \sigma} ~, \notag \\
& T_x \TT \circ f_{j, \sigma} = -\TT T_x \circ f_{j, \sigma} ~, \notag \\
& g_x g_z \circ f_{j, \sigma} = -g_z g_x \circ f_{j, \sigma} ~, \notag \\
& g_x \TT \circ f_{j, \sigma} = -\TT g_x \circ f_{j, \sigma} ~, \notag \\
& g_z^2 \circ f_{j, \sigma} = -f_{j, \sigma} ~, \notag \\
& \TT^2 \circ f_{j, \sigma} = -f_{j, \sigma} ~.
\label{eq:fermionic_parton_psg}
\end{align}
We note that this is just one out of 32 PSGs that one finds for this fermionic parton approach with our symmetries.
Different PSGs in general allow accessing different phases in this approach.
While we have not studied all PSGs exhaustively, we will see that the PSG chosen here allows us to realize the $z$-FM and VBS phases of interest to us.

Below, we will identify phases realized by the Hamiltonian Eq.~(\ref{eq:model_fermionic_parton}) in different coupling regimes.
In the infinite $\Gamma$ limit (i.e., when the gauge theory is at strong coupling), we have $P_{j,+} P_{j,-} = \zeta^x_{j-1/2} \zeta^x_{j+1/2} = -1$, which is exactly the constraint for the physical states in Eq.~(\ref{eq:fermionic_parton_constraint}).
Thus, in this limit, we obtain some physical spin Hamiltonian by the operator mapping defined in Eq.~(\ref{eq:fermion_parton_to_physical_spin}).

Here, we assume that a weakly coupled gauge theory, where $\Gamma$ is small compared to $t$ and $\eta$, also captures qualitative physics of the original spin system that can occur for some interaction regimes.
In the corresponding parton mean field theory, Eq.~(\ref{eq:model_fermionic_parton}), we get at least two topologically distinct fermionic phases.
By including gauge fluctuations, these phases lead to different symmetry breaking phases of the original spin system.
Let us now present analysis of these phases.

Due to the $g_x$ symmetry, there are no terms that mix $f_+$ and $f_-$ at the quadratic level.
Hence we can analyze each species separately.
Going to momentum space, $f_{k, \sigma} \equiv (1/\sqrt{L}) \sum_{j=1}^L \ee^{-\ii k j} f_{j, \sigma}$, we have for each species
\begin{align*}
H^\text{mf}_\sigma \!=\! \sum_k \left[\xi_\sigma(k) f_{k, \sigma}^\dg f_{k, \sigma} + \frac{1}{2}\Big(\Delta_\sigma(k) f_{k, \sigma}^\dg f_{-k, \sigma}^\dg + \text{H.c.}\Big) \right] \, ,
\end{align*}
where $\xi_\sigma(k)$ is real by Hermiticity and $\Delta_\sigma(-k) = -\Delta_\sigma(k)$ by convention.
For our nearest-neighbor mean field ansatz, we have
\begin{align*}
\xi_\sigma(k) = -2 t_\sigma \cos(k) - \mu ~, \quad
\Delta_\sigma(k) = -2 \eta_\sigma \sin(k) ~.
\end{align*}
It is easy to check that $\xi_\sigma(k)$ and $\Delta_\sigma(k)$ satisfy
\begin{align}
& \xi_-(k) = \xi_+(k + \pi) ~, \quad \Delta_-(k) = -\Delta_+(k + \pi) ~, \\
& \xi_-(k) = \xi_+^*(-k + \pi) ~, \quad \Delta_-(k) = \Delta_+^*(-k + \pi) ~.
\end{align}
These properties hold also if we include symmetry-allowed further neighbor hopping and pairing terms (the first and second lines follow from the $g_z$ and $\TT$ symmetries respectively).
Since we have $\xi_\sigma(-k) = \xi_\sigma(k)$, for each pair of momenta $\{k, -k\}$ (assuming $k \neq -k$), we have familiar two-fermion pairing problem where we can write the ground state wavefunction as
\begin{align}
\exp\left[ u_\sigma(k) f_{k, \sigma}^\dg f_{-k, \sigma}^\dg \right] |0 \rangle ~,
\end{align}
with the ``pair-function''
\begin{align}
u_\sigma(k) = \frac{-\Delta_\sigma(k)}{\xi_\sigma(k) + \sqrt{\xi_\sigma(k)^2 + |\Delta_\sigma(k)|^2}} ~.
\end{align}
$u_\sigma(k)$ satisfies the same condition as $\Delta_\sigma(k)$
\begin{align}
u_-(k) = -u_+(k + \pi) = u_+^*(-k + \pi) ~,
\label{eq:fp_pair_sym}
\end{align}
maintaining convention $u_\sigma(-k) = -u_\sigma(k)$.
On the other hand, if $k = -k=0\ \text{or}\ \pi$, then the ground state has the mode $f_{k, \sigma}$ occupied or unoccupied depending on whether $\xi_{k, \sigma} < 0$ or $\xi_{k, \sigma} > 0$.

To identify the symmetry breaking pattern in the spin system, we will calculate quantum numbers for $|\Psi_\text{MF} \rangle$ for each of the gauge sectors $\prod_j \zeta^z_{j+1/2} = +1$ or $-1$ separately.
The $+1$ and $-1$ gauge sectors impose correspondingly periodic and antiperiodic boundary conditions on the mean field Hamiltonian for the fermionic partons.
We implement the $+1$ gauge sector by taking $\zeta^z_{j+1/2} = 1$ for all $j = 1, \dots, L$, while for the $-1$ sector we change the sign of a single link variable connecting the $L$th site with the $1$st site: $\zeta_{L+1/2} = -1$.
Note that while the $T_x$ transformation quoted in Eq.~(\ref{eq:fermionic_parton_sym}) is a symmetry of the mean field Hamiltonian in the former case (which we can call $T_x^\text{p.b.c.}$), the precise symmetry of the mean field Hamiltonian in the latter sector is slightly different and reads instead
\begin{align*}
T_x^\text{a.b.c}:\ & f_{j,\pm} \to f_{j+1,\pm}~~\text{for}~~ 1 \leq j < L ~;~~ f_{L,\pm} \to -f_{1,\pm} ~,
\end{align*}
where ``p.b.c.''\ and ``a.b.c.''\ stand for periodic and antiperiodic boundary conditions respectively.
(The precise relation between the translation symmetry $T_x$ in the spin system and the symmetries of the mean field Hamiltonians $T_x^\text{p.b.c./a.b.c.}$ is further explained in Appendix~\ref{app:qn_fermionic_parton}.)
Nevertheless, the preceding momentum space analysis as well as arguments below carry through similarly for both p.b.c.\ and a.b.c., only
the momenta $k$ run over different discrete values for the two sectors: $k^\text{p.b.c.} = 2\pi n/L, n \in \mathbb{Z} \mod L$ vs $k^\text{a.b.c} = \pi (2n + 1)/L, n \in \mathbb{Z} \mod L$.
Throughout, we assume that $L$ is an even integer.
The case with the periodic boundary conditions contains momenta $k = 0$ and $k = \pi$ that satisfy $k = -k$, so the corresponding modes need to be treated separately as described above.
On the other hand, the case with the antiperiodic boundary conditions does not contain such momenta.

As a final preparation for our analysis, we will also need symmetry transformations of the fermion modes $f_{k, \sigma}$:
\begin{align}
  T_x:\ & f_{k, \pm} \to e^{i k} f_{k, \pm} ~; \notag \\
  g_x:\ & f_{k, \pm} \to \pm f_{k, \pm} ~; \notag \\
  g_z:\ & f_{k, \pm} \to \ii f_{k + \pi, \mp} ~; \notag \\
  \TT:\ & f_{k, \pm} \to \pm f_{-k + \pi, \mp} ~, \quad \ii \to -\ii ~.
  \label{eq:fk_sym}
\end{align}

Below, we focus on phases obtained for the nearest-neighbor ansatz.
Let us consider the following two cases:
\begin{itemize}
\item When $|\mu| > 2|t_\sigma|$ and $\eta_\sigma \neq 0$, we obtain trivial $p$-wave superconductor phases for both $f_+$ and $f_-$.
Without loss of generality, we assume $\mu < 0$, so that $\xi_\pm(k=0, \pi) > 0$ and the corresponding modes are unoccupied.
Then the fermionic parton wavefunction is
\begin{align}
|\Psi_\text{MF} \rangle = \prod_{\sigma = \pm} \prod_{0 < k < \pi} \exp \left[ u_\sigma(k) f_{k, \sigma}^\dg  f_{-k, \sigma}^\dg \right] |0 \rangle ~,
\label{eq:wf_fp_trivial}
\end{align}
which holds for both periodic and antiperiodic boundary conditions (with appropriate $k$ in each case).

It is straightforward to see that for both gauge sectors, $|\Psi_\text{MF} \rangle$ in Eq.~(\ref{eq:wf_fp_trivial}) acquires no phase factor under any of the symmetry actions in Eq.~(\ref{eq:fk_sym}).
However, similar to our analysis in the bosonic parton case, $Z_2^\zeta$ instantons carry non-trivial momentum and their condensation would lead to translational symmetry breaking (schematically, $\langle \zeta^x \rangle \neq 0$ breaks $T_x$; more precisely, on a finite chain, the two gauge sectors have exponentially close energies but carry momenta that differ by $\pi$).
Hence, by including gauge fluctuations, the topologically trivial phase of the fermionic partons becomes the VBS ordered phase in the spin variables.

\item When $|\mu| < 2|t_\sigma|$ and $\eta_\sigma \neq 0$, we obtain $p$-wave topological superconductor phases~\cite{Kitaev2001, MotrunichDamleHuse2001} for both $f_+$ and $f_-$.
Without loss of generality, we assume $t_+ > 0$ so that $\xi_+(0) = \xi_-(\pi) < 0$ while $\xi_+(\pi) = \xi_-(0) > 0$.
Then, for the gauge sector $\prod_j \zeta_{j+1/2}^z = 1$, the fermionic parton wavefunction is
\begin{align}
|\Psi_\text{MF}^\text{p.b.c.} \rangle = & f_{k=0, +}^\dg f_{k=\pi, -}^\dg \times \notag \\
& \prod_{\sigma = \pm} \prod_{0 < k < \pi} \exp\left[ u_\sigma(k) f_{k, \sigma}^\dg f_{-k, \sigma}^\dg \right] |0 \rangle ~.
\label{eq:wf_fp_topo}
\end{align}
It is straightforward to check that 
\begin{align}
& g_z |\Psi_\text{MF}^\text{p.b.c.} \rangle =
  |\Psi_\text{MF}^\text{p.b.c.} \rangle ~; \notag \\
& T_x |\Psi_\text{MF}^\text{p.b.c.} \rangle =
  g_x |\Psi_\text{MF}^\text{p.b.c.} \rangle = 
 -|\Psi_\text{MF}^\text{p.b.c.} \rangle ~,
\end{align}
where the phases come entirely from transformation properties of the factor $f_{k=0, +}^\dg f_{k=\pi, -}^\dg$.

For the gauge sector $\prod_j \zeta_{j+1/2}^z = -1$, momenta $k=0$ and $\pi$ cannot be taken.
Hence, the corresponding wavefunction labeled as $|\Psi_\text{MF}^\text{a.b.c.} \rangle$ has the same form as in Eq.~(\ref{eq:wf_fp_trivial}), and $|\Psi_\text{MF}^\text{a.b.c.} \rangle$ is invariant under any of the symmetry actions.

Similarly to the topologically trivial case, the $Z_2^\zeta$ instanton contributes additional momentum difference $\pi$ between the two sectors.
Now we can compare quantum numbers for the two gauge sectors:
They have the same $T_x$ quantum number and the same $g_z$ quantum number, but their $g_x$ quantum numbers are opposite to each other
(the absolute quantum numbers of the corresponding Gutzwiller-projected wavefunctions are calculated in Appendix~\ref{app:qn_fermionic_parton}).
Hence, the resulting phase spontaneously breaks the $g_x$ symmetry but preserves the $T_x$ and $g_z$ symmetries, and it can be identified as the $z$-FM phase.
\end{itemize}

We also mention that in the infinite $\Gamma$ limit, we can use projected wavefunctions as trial wavefunctions for ground states.
In Appendix~\ref{app:qn_fermionic_parton}, we perform a systematic analysis and show how we can extract spontaneous symmetry breaking pattern from such projected wavefunction studies.
Up to a common shift in the $T_x$ quantum number for both sectors, the results are in agreement with the results in this section.

\subsection{Critical theory for fermionic partons}
\label{subsec:fermparton_crit}
Now, let us consider the critical theory for the $z$-FM to VBS transition in this language.
We tune the mean field parameters to a critical point with $\mu = -2 t_+$ at which $\xi_+(k=0) = \xi_-(k=\pi) = 0$; we will allow small deviation of $\mu$ from the critical value.
Then, the low-energy fermionic modes are
\begin{align}
  \psi_+(j a) \sim f_{j, +} ~, \quad \psi_-(j a) \sim (-1)^j f_{j, -} ~,
\end{align}
where $a$ denotes the lattice constant.
We define long-wavelength Majorana modes as $\psi_\pm \sim \gamma_{\pm, 1} + \ii \gamma_{\pm, 2}$.
In continuum, the low-energy theory reads
\begin{align}
  H \approx \int\dd x \Big[ &\ii v \Big(\gamma_{+, 1} \,\partial_x\, \gamma_{+, 1} - \gamma_{+, 2} \,\partial_x\, \gamma_{+, 2} \notag \\
  & \qquad - \gamma_{-, 1} \,\partial_x\, \gamma_{-, 1} + \gamma_{-, 2} \,\partial_x\, \gamma_{-, 2} \Big) \notag \\
  & + 2 \ii m \big(\gamma_{+, 1} \gamma_{+, 2} + \gamma_{-, 1} \gamma_{-, 2} \big) + \cdots \Big] ~,
\end{align}
where $v = 2\eta_+$ is the characteristic velocity, $m = (-2 t_+ - \mu)/a$ measures deviation from the critical point, while $\cdots$ represents terms involving more than two Majorana modes, such as $\gamma_{+, 1} \gamma_{+, 2} \gamma_{-, 1} \gamma_{-, 2}$.

It is convenient to define new complex fermion fields
\begin{align}
& \psi_L = \gamma_{-, 2} + \ii \gamma_{+, 1} ~, \quad
& \psi_R = -\gamma_{-, 1} + \ii \gamma_{+, 2} ~.
\end{align}
In these variables, we have
\begin{align}
  H = \!\int\!\dd x\; \Big[ & \ii v \big(\psi_L^\dg \partial_x \psi_L - \psi_R^\dg \partial_x \psi_R \big) + \ii m( \psi_L^\dg \psi_R - \psi_R^\dg \psi_L) \notag\\
  & + u \psi_L^\dg \psi_L \psi_R^\dg \psi_R + \cdots \Big] ~.
  \label{eq:fermionic_parton_continuum_hamiltonian}
\end{align}

We can deduce symmetry actions on $\psi_{L/R}$ from Eq.~(\ref{eq:fermionic_parton_sym}) as
\begin{align}
  T_x:\ & \psi_L \to \psi_L^\dg ~, \quad \psi_R \to \psi_R^\dg ~; \notag \\
  g_x:\ & \psi_L \to \psi_L^\dg ~, \quad \psi_R \to \psi_R^\dg ~; \notag \\
  g_z:\ & \psi_L \to \ii \psi_L ~, \quad \psi_R \to \ii \psi_R ~; \notag \\
  \TT:\ & \psi_L \to \ii \psi_R^\dg ~, \quad \psi_R \to \ii \psi_L^\dg ~, \quad \ii \to -\ii ~.
  \label{}
\end{align}
Besides, physical states should be invariant under global $Z_2$ gauge action~(i.e., Invariant Gauge Group or $IGG$) $\psi_{L/R} \to -\psi_{L/R}$.
It is easy to verify that the only bilinear terms with no or one derivative and the only quartic term with no derivative are the ones present in the above continuum Hamiltonian.
In particular, the $m$ term is the single allowed fermion mass term, and tuning it across zero corresponds to the $z$-FM to VBS transition.

To connect with the field theory for the transition described in Sec.~\ref{sec:goodvars}, we can apply standard bosonization of the continuum fermions~\cite{Shankar_Acta, GiamarchiBook},
\begin{align}
\psi_L \sim \ee^{\ii (\tilde{\phi} - \tilde{\theta})} ~, \quad \psi_R \sim \ee^{\ii (\tilde{\phi} + \tilde{\theta})} ~.
\end{align}
The fermion kinetic energy plus density-density interaction term give the standard quadratic Hamiltonian for the bosonic field with some effective velocity $\tilde{v}$ and Luttinger parameter $\tilde{g}$.
The mass term becomes $\ii m (\psi_L^\dg \psi_R - \psi_R^\dg \psi_L) \sim m \cos(2 \tilde{\theta})$.
Hence, the continuum action essentially reproduces Eq.~(\ref{eq:gauged_bosonization_action}).
Furthermore, using symmetries we can readily identify the following contributions to the $x$-FM, $y$-AFM, and $z$-AFM order parameters,
\begin{align}
& M_x^\text{FM} \sim \ii \psi_L^\dg \psi_R^\dg - \ii \psi_R \psi_L \sim \cos(2\tilde{\phi}) ~, \notag \\
& M_y^\text{AFM} \sim \psi_L^\dg \psi_R^\dg + \psi_R \psi_L \sim \sin(2\tilde{\phi}) ~, \notag \\
& M_z^\text{AFM} \sim \psi_L^\dg \psi_R + \psi_R^\dg \psi_L \sim \sin(2 \tilde{\theta}) ~. \label{eq:MzAFM}
\end{align}
The first two lines match Eqs.~(\ref{eq:MxFM}) and (\ref{eq:MyAFM}), while the exhibited contribution to the $z$-AFM order parameter will be useful below.

Notably, any object constructed using local fermion fields has identical transformation properties under the $T_x$ and $g_x$ symmetries, so one cannot construct the $z$-FM or VBS order parameters using such objects.
This is where we need to remember that the full theory also contains the $Z_2$ gauge field, and the $z$-FM and VBS order parameters necessarily involve instanton operators of the gauge field (this could be already anticipated from our rigorous lattice analysis of the phases in the previous subsection).
The structure is similar to our good bosonic parton variables, and we should be able to express these order parameters similarly to our analysis in Sec.~\ref{sec:goodvars}.
However, we have to proceed slightly differently here since using the gauge theory constraints in Eq.~(\ref{eq:fermparton_gaugeconstraint}) to extract the long-wavelength VBS order parameter $\xi^x_{q=0}$ does not work immediately.

To this end, we go back to the microscopic hard parton constraint Eq.~(\ref{eq:fermionic_parton_constraint}) and note that it is mathematically equivalent to the condition $\exp[\pm \ii \pi (f_+^\dg f_- + f_-^\dg f_+)] = -1$.
When proposing an effective $Z_2$ gauge theory, which effectively ``softens'' the hard parton constraint, we could instead write a different Gauss law
\begin{align}
\zeta^{\prime\, x}_{j-1/2} \zeta^{\prime\, x}_{j+1/2} = \ee^{\pm \ii \pi (f_{j,+}^\dg f_{j,-} + f_{j,-}^\dg f_{j,+})} ~,
\end{align}
with a term in the Hamiltonian $\Gamma^\prime \sum_j (-1)^j \zeta^{\prime\, x}_{j+1/2}$, which in the $\Gamma^\prime \to \infty$ limit would give the exact hard parton constraint.
Then, we can obtain contributions to the VBS order parameter immediately from
\begin{align}
\zeta^{\prime\, x}_{j+1/2} = \prod_{j' \leq j} \ee^{\pm \ii \pi (-1)^{j'+1} (f_{j',+}^\dg f_{j',-} + f_{j',-}^\dg f_{j',+})} \sim \ee^{\pm \ii \tilde{\theta}} ~,
\end{align}
where we used
\begin{align*}
(-1)^{j+1} (f_{j,+}^\dg f_{j,-} + f_{j,-}^\dg f_{j,+}) = \psi_L^\dg \psi_L + \psi_R^\dg \psi_R = \frac{\partial_x \tilde{\theta}}{\pi} ~.
\end{align*}
We can then fix the offset on $\tilde{\theta}$ so that the VBS order parameter reads identically to Eq.~(\ref{eq:PsiVBS}).
With this in hand, we can combine the VBS order parameter with the contribution to the $z$-AFM order parameter in Eq.~(\ref{eq:MzAFM}) to obtain a contribution to the $z$-FM order parameter whose dominant part matches Eq.~(\ref{eq:MzFM}).
Even though this treatment matches our original fermionic parton model only in the $\Gamma^\prime \to \infty$ and $\Gamma \to \infty$ limits, while at finite $\Gamma^\prime$ and $\Gamma$ these are somewhat different gauge theory models, we believe that the qualitative properties are the same in both models.

One may also wonder how to deduce correct periodicities for the $\tilde{\phi}$ and $\tilde{\theta}$ fields in the fermionic parton language to match the discussion in Sec.~\ref{sec:goodvars}.
One answer to this lies in the precise mapping between the present fermionic parton theory and the fermionic parton theory in Appendix~\ref{app:fermion_parton_from_good_variable}.
The latter in turn maps onto the good bosonic parton theory in Sec.~\ref{sec:goodvars}, which essentially provides precise bosonization of the fermionic parton theory.

Despite these technicalities, which we can basically resolve by connecting with the good bosonic parton theory, the relative ease with which we obtained the critical theory for the $z$-FM to VBS transition in the fermionic language is quite remarkable.
The key point is that the transition is between fully symmetric phases of the fermionic partons but which differ in their SPT indices with respect to the PSG.
The order parameters that develop on one or the other side of the transition in the physical spin model are encoded in the instanton operators of the gauge field.

The fermionic treatment also more readily reveals emergent symmetries at the critical point.
It is straightforward to see that at the critical point, $m = 0$ in Eq.~(\ref{eq:fermionic_parton_continuum_hamiltonian}), $\psi_L$ and $\psi_R$ are separately conserved.
Namely, the critical theory is invariant under the following transformations:
\begin{align}
\psi_L \rightarrow \ee^{\ii \alpha_L} \psi_L ~, \quad \psi_R\rightarrow \ee^{\ii \alpha_R} \psi_R ~.
\end{align}
Correspondingly, the bosonized theory defined in Eq.~(\ref{eq:gauged_bosonization_action}) is invariant under
\begin{align}
\tilde{\phi} \rightarrow \tilde{\phi} + \frac{1}{2}(\alpha_L + \alpha_R) ~, \quad \tilde{\theta} \rightarrow \tilde{\theta} + \frac{1}{2}(-\alpha_L + \alpha_R) ~.
\end{align}

We point out that the $U(1)$ transformation on $\tilde{\phi}$ only, obtained for $\alpha_L = \alpha_R \equiv \alpha$, is actually present in the fermionic quadratic Hamiltonian with the nontrivial PSG defined in Eq.~(\ref{eq:fermionic_parton_psg}) for any mean field parameters (i.e., both inside the phases and at the critical point).
On the fermionic partons, this transformation reads
\begin{align*}
f_{j,+} \to & \cos(\alpha) f_{j,+} - \ii (-1)^j \sin(\alpha) f_{j,-} ~, \\
f_{j,-} \to & -\ii (-1)^j \sin(\alpha) f_{j,+} + \cos(\alpha) f_{j,-} ~.
\end{align*}
This symmetry is more manifest in the bipartite hopping formulation in Appendix~\ref{app:fermion_parton_from_good_variable}, which is how we first noticed it, but is readily checked in the present setup.
Notice that this symmetry is broken down to $Z_4$ subgroup (generated by $\alpha = \pi/2$) if we go beyond the mean field level and add generic symmetric fermion-fermion interactions.
One can actually identify this $Z_4$ as generated by the $g_z$ symmetry action defined in Eq.~(\ref{eq:fermionic_parton_sym}).
In fact, the above $U(1)$ transformation rotates the physical spins in the $x$-$y$ plane as follows:
\begin{align}
\sigma^x_j \to & \cos(2\alpha) \sigma^x_j - (-1)^j \sin(2\alpha) \sigma^y_j ~, \\
\sigma^y_j \to & (-1)^j \sin(2\alpha) \sigma^x_j + \cos(2\alpha) \sigma^y_j ~.
\end{align}
which is not a microscopic symmetry of the spin system.
However, terms that do not obey it appear only as quartic terms with derivatives and are irrelevant at the critical point.
Thus, this action becomes a symmetry operation in the long-wavelength limit at the critical point.
(As a side remark, the above spin transformation corresponds to a uniform rotation of spin variables $S^\prime_j$ defined in Appendix~\ref{app:nonparton_goodvars}, becoming emergent symmetry in the analysis there.)
Finally, note that the $U(1)$ transformation on $\tilde{\theta}$ clearly requires tuning to the massless point.

\section{Conclusions and future directions} \label{sec:conclusion}
In this work, we have studied in detail the phase transition from the $z$-FM order to the VBS order in an anisotropic spin-1/2 system in 1d.
We have provided many different and complementary perspectives on this transition, including direct abelian bosonization, duality to domain walls, as well as parton techniques.
Most notably, we have obtained a particularly nice formulation beyond the natural domain walls and partons, and have found that this phase transition can be captured by a Luttinger-liquid-like theory, with varying critical exponents depending on interaction details.
There is a single relevant cosine perturbation, and the transition is achieved by tuning its coupling through zero.
Furthermore, this formulation unifies the $z$-FM and VBS orders and allows us to easily read off all critical properties.
We have already discussed our main results in the introduction and throughout the main text, including parallels with the DQCP theories in 2d.
[For a lightning recap, the key framework underlying our work is the exact equivalence between the following effective gauge theories: Eq.~(\ref{eq:model_mu_pm}) for the domain wall variables, Eq.~(\ref{eq:model_bosonic_parton}) for the original partons, and Eq.~(\ref{eq:model_at_dual_two}) for the $\nu$-chain ``good variables,'' while further interesting highlights include interpretation of the latter in Eq.~(\ref{eq:good_variable_parton_Ham}) as ``new partons,'' its fermionization in Eq.~(\ref{eq:good_fermionic_parton_Ham}) and connection to the fermionic partons in Eq.~(\ref{eq:model_fermionic_parton})].
We now discuss some lessons and possible future directions.

In this paper, we have provided strong theoretical arguments that this transition is continuous.
It is important to perform unbiased numerical studies of concrete models to test this, as well as to check our predictions for the critical properties.
Some of this work is in progress~\footnote{B.~Roberts, S.~Jiang, and O.~I.~Motrunich, in preparation}.
We would like to look also for additional phenomena in models, and if they can be described by our theory or its generalizations.
For example, when our field theory ceases to describe the continuous transition because another operator becomes relevant, we conjecture that, rather than the transition becoming first-order, a new phase opens up where the $z$-FM and VBS orders coexist.
This conjecture follows by analogy with what happens in the Ashkin-Teller model, but our full theory is distinct from the AT model, and an unbiased study is warranted.
As another example, we can ask if some other phases can appear that are proximate to the phases discussed here, and if they can be understood from some domain wall or parton perspectives.

It will be interesting to also study spin-1/2 models with fewer on-site symmetries than in the present paper.
Thus, to have LSM-like theorem, it is sufficient to have either the $Z_2^x \times Z_2^z$ symmetry or the $\TT$ symmetry.
We required all these symmetries in order to have familiar and fully controlled field theory, but what happens if we relax this while maintaining such an LSM condition?

Another interesting direction is to study systems with more complex symmetries in 1d, e.g., higher-spin chains or $Z_N$ clock systems with additional discrete symmetries, where one guide is to look for systems with LSM-like theorem~\cite{TanizakiSulejmanpasic2018arXiv}.

Thinking about possible applications to the 2d DQCP theories, an important lesson from our study is that we found a new formulation of the transition that is superior to the domain wall/parton descriptions that ``fractionalize'' the VBS order parameter or the $z$-FM order parameter.
We found this formulation by a non-trivial transformation on the original domain wall/parton descriptions, and it would be very interesting to look for similar new formulations in the 2d DQCP problems.
We were also able to interpret this formulation as a new parton approach, where partons are gapped on either side of the transition, while the phases are distinguished as different SPT phases of partons with the given PSG.
No such description is presently known for the 2d DQCPs, and it would be interesting to try our approaches to look for such descriptions, both with bosonic and fermionic partons.

Finally, although the model presented here is fairly simple, we believe the methods we have used can be applied to many other strongly-correlated systems, including in higher dimensions and in fermionic systems.
A general scheme starts by identifying a PSG, either for topological defects or fractionalized particles, and then treats the PSG as a symmetry group for the gauge charges, studying their symmetry-breaking, SPT, or even SET phases, and translating these to phases of the original physical system.
While there are many separate instances of applications of this scheme in the literature, a systematic study---particularly of symmetric distinct phases of gauge charges---has not been attempted and could potentially be used to access many exotic phases and phase transitions.

\begin{acknowledgments}
The authors would like to thank J.~Alicea, Y.-M.~Lu, D.~Mross, Y.~Ran, B.~Roberts, and A.~Vishwanath for useful discussions.
This work was supported by the Institute for Quantum Information and Matter, an NSF Physics Frontiers Center, with support of the Gordon and Betty Moore Foundation, and also by NSF through grant DMR-1619696.
\end{acknowledgments}

\appendix
\section{Review of $Z_2$ gauge theory in 1d} \label{app:z2_gauge_theory}
Here we give a brief review of $Z_2$ gauge theory in 1d.
The gauge fields are two-level systems residing on links of the 1d lattice; we label the corresponding Pauli operators as $\zeta^{x,y,z}_{j+1/2}$.
The Hilbert space is defined by constraints (Gauss law for each $j$)
\begin{align}
\zeta^x_{j-1/2} \, \zeta^x_{j+1/2} = 1 ~,
\end{align}
and the Hamiltonian is
\begin{align}
H = -\sum_{j=1}^L \Gamma_j \zeta^x_{j+1/2} + \dots ~,
\end{align}
where ``$\dots$'' denote any additional ``symmetry-allowed'' terms (which depend on the context, see below).
We assume periodic boundary conditions.

On one hand, the pure gauge theory is very simple, since the full Hilbert space is two-dimensional, with basis states $|X+ \rangle \equiv |\{ \zeta^x_{j+1/2} = 1~, \forall j \} \rangle$ and $|X- \rangle \equiv |\{ \zeta^x_{j+1/2} = -1~, \forall j \} \rangle$.
However, there are still several distinct cases that we need to consider, which arise in different contexts and show different physics.

First, consider the case with $\Gamma_j \equiv \Gamma \neq 0$.
For concreteness, we assume $\Gamma > 0$.
The ground state is simply $\zeta^x_{j+1/2} = 1$ for all $j$, while the excited state is $\zeta^x_{j+1/2} = -1$ with energy $2 \Gamma L$ above the ground state.
This is the most familiar Ising gauge theory in 1d~\cite{KogutRMP}, which has ``proliferated'' instantons [$Z_2$ fluxes in (1+1)D Euclidean path integral language] and is confining for any $\Gamma \neq 0$.

The second case is with $\Gamma_j = (-1)^j \Gamma$ (we also assume that $L$ is even).
This case is equivalent to so-called odd Ising gauge theory~\cite{MoessnerSondhiFradkin2001}.
In this case, the ground state is two-fold degenerate.
Generic local perturbations preserve the two-fold degeneracy as long as one has ``translation'' symmetry $T_x: \zeta^x_{j-1/2} \to -\zeta^x_{j+1/2}$.
Indeed, any diagonal term with such symmetry does not distinguish between the two basis states $|X+ \rangle$ and $|X- \rangle$, while off-diagonal terms necessarily involve all $L$ spins and hence can produce only $\sim \exp(-c L)$ splitting.
The two-fold ground state degeneracy is due to breaking of the translation symmetry.

The final case is with $\Gamma_j \equiv 0$, protected by ``flux conservation'' realized by a unitary $U = \prod_j \zeta^z_{j+1/2}$.
Such flux conservation is natural when the gauge field arises during Ising duality maps, as described in Appendix~\ref{app:ising_dual}.
In the pure gauge theory, the ground state is two-fold degenerate.
Indeed, here also any allowed diagonal term does not distinguish between the two basis states $|X+ \rangle$ and $|X- \rangle$, now because of the flux conservation, while off-diagonal terms are exponentially small in $L$.
The exact eigenstates are of course labeled by $U = 1$ and $U = -1$; the two-fold degeneracy reflects breaking the flux conservation symmetry.
In this last case, when the gauge field is coupled to a matter field that condenses, the condensation effectively introduces a finite energy difference between the two states with $U = 1$ and $U = -1$; the ground state is non-degenerate and the flux conservation is restored.

\section{Ising duality in 1d and interpretation as $Z_2$ matter-gauge theory} \label{app:ising_dual}
In this appendix, we review quantum Ising duality in 1d. 
Let us consider quantum Ising model defined on a chain of length $L$:
\begin{align}
  H = -J \sum_{j=1}^L \sigma^z_j \sigma^z_{j+1} - h_x \sum_{j=1}^L \sigma^x_j - h_z \sum_{j=1}^L \sigma^z_j ~,
  \label{eq:ising_model}
\end{align}
where we impose periodic boundary conditions. 
For concreteness, we assume $J, h_x \geq 0$.
When $h_z = 0$, this model has $Z_2$ global symmetry generated by $g = \prod_j \sigma^x_j$. 

One can introduce dual variables $\mu$ on links to represent domain walls.
The standard definition is
\begin{align}
& \mu^x_{j+1/2} = \sigma^z_j \sigma^z_{j+1} ~, ~~~\text{or}~~~ \sigma^z_j = \prod_{j'<j} \mu^x_{j'+1/2} ~, \notag \\
& \mu^z_{j-1/2} \, \mu^z_{j+1/2} = \sigma^x_j ~, ~~~\text{or}~~~ \mu^z_{j+1/2} = \prod_{j' \leq j} \sigma^x_{j'} ~,
\end{align}
where we have also indicated how one typically ``solves'' for the $\sigma^z$ or $\mu^z$ operators in terms of a string of $\mu^x$ or $\sigma^x$ respectively.
In the absence of the $h_z$ term, the Hamiltonian in Eq.~(\ref{eq:ising_model}) is also local in the dual variables
\begin{align}
\!  H[h_z \!=\! 0] = -J \sum_j \mu^x_{j+1/2} - h_x \sum_j\mu^z_{j-1/2} \mu^z_{j+1/2} \, .
\end{align}
When $J > h_x$, we get Ising ordered phase in the $\sigma$ variables, with two-fold ground state degeneracy\,(GSD), while for $J < h_x$, we get disordered phase with unique ground state.
At $J = h_x$, this model is at a self-dual critical point.

By examining this duality transformation more carefully, one observes that it is actually not a one-to-one mapping (and the ``solutions'' in terms of string operators are not exact inversions):
For example, taking the first equation in the first line as defining $\mu^x$ basis, spin states in the $\sigma^z$ basis $|s \rangle$ and $g|s \rangle$ are mapped to the same state in the $\mu^x$ basis.
Accordingly, the physical states should satisfy constraint $\prod_j \mu^x_{j+1/2} = 1$.
This makes identification of the original spin phases more subtle in the dual variables. 
The Ising ordered phase for $\mu$'s has no GSD after projection to the physical Hilbert space and is thus identified as disordered phase for $\sigma$'s.
On the other hand, the Ising disordered phase for $\mu$'s has two-fold GSD due to the two-to-one mapping, which is interpreted as symmetry breaking phase for $\sigma$'s.

In order to make this duality mapping exact and capture global aspects such as GSD correctly, one can introduce a $Z_2^\rho$ gauge field $\rho_j$ defined on links $(j-1/2, j+1/2)$ of the dual lattice.
We define the exact mapping as
\begin{align}
& \mu^x_{j+1/2} = \sigma^z_j \sigma^z_{j+1} ~, \notag \\ 
& \mu^z_{j-1/2} \, \rho^z_j \, \mu^z_{j+1/2} = \sigma^x_j ~, \notag \\ 
& \rho^x_j = \sigma^z_j ~,
\end{align}
with the gauge constraint (``Gauss law'') 
\begin{align}
\rho^x_j \rho^x_{j+1} = \mu^x_{j+1/2} ~,
\label{eq:appIGTconstraint}
\end{align}
In these variables, the Hamiltonian becomes
\begin{align*}
 H \!=\! -J \sum_j \mu^x_{j+1/2} - h_x \sum_j \mu^z_{j-1/2} \, \rho^z_j \, \mu^z_{j+1/2} - h_z \sum_j \rho^x_j \, ,
\end{align*}
and is understood to act in the constrained Hilbert space.
(The mapping can be easily proved by solving for $\mu^x$ in terms of the $\rho^x$, thus obtaining an unconstrained reformulation of the dual matter-gauge theory, and matching this to the original spin model.)
In this language, domain walls are identified as $Z_2^\rho$ gauge charges.
As reviewed in Appendix~\ref{app:z2_gauge_theory}, such a 1d $Z_2^\rho$ gauge theory is divided into two sectors, labeled by $\prod_j \rho^z_j = \pm 1$, interpreted as sectors with even/odd $Z_2^\rho$ flux.
Operator $\rho^x_j$ anti-commutes with $\prod_j \rho^z_j$, and its action corresponds to an instanton event creating/annihilating $Z_2^\rho$ flux.

When $h_z = 0$, the dual Hamiltonian has a global $Z_2$ symmetry generated by $g = \prod_j \rho_j^z$, which is interpreted as $Z_2^\rho$ flux conservation symmetry; it
corresponds precisely to the global $Z_2$ symmetry of the original Ising model.
With the global $Z_2$ symmetry, there are two phases:
\begin{itemize}
  \item When $J > h_x$, gauge charges are ``gapped,'' which we can write schematically as $\langle \mu^z_j\rangle = 0$. 
  The low-energy sector is a pure $Z_2^\rho$ gauge theory, which is a two-dimensional Hilbert space. 
  Due to the flux conservation symmetry, states with even and odd flux do not mix, and the energy splitting between these two states is exponentially small in the system size, where the splitting is generated in perturbation theory only at $L$-th order.
  This phase is identified as the $Z_2$ spontaneous symmetry breaking phase, as an infinitesimally small $Z_2$ breaking term $\rho^x_j$ would split the degeneracy and choose a classical configuration.
  
  \item When $J < h_x$, gauge charges are ``condensed,'' schematically, $\langle \mu^z \rangle \neq 0$. 
  In this case, one gets $Z_2^\rho$ Higgs phase. 
  To see the nature of this phase, we consider the limit $h_x \to \infty$.
  The ground state is characterized by $\mu^z_{j-1/2} \, \rho^z_j \, \mu^z_{j+1/2} = 1$ for all $j$. 
  Thus, in the ground state manifold, we have $\prod_j \rho^z_j = \prod_j \left( \mu^z_{j-1/2} \, \rho^z_j \, \mu^z_{j+1/2} \right) = 1$. 
  The even flux sector has lower energy than the odd flux sector, and the ground state is non-degenerate. 
  One gets the $Z_2$ symmetric phase.
\end{itemize}

Notice that when $h_z \neq 0$, there is no distinction between the two phases discussed above, and there is only one gapped phase.
All of the above, working with the exact dual Hamiltonian, of course, matches thinking directly about the original Ising spin model.

\section{Equivalence of the domain wall and parton descriptions}
\label{app:dw_parton_equiv}
In this appendix, we will prove equivalence between the lattice models in Eqs.~(\ref{eq:model_mu_pm}) and (\ref{eq:model_bosonic_parton}) via the operator mapping defined in Eq.~(\ref{eq:dw_to_parton}).

By using the gauge constraints $\rho_j^x \rho_{j+1}^x = \mu^x_{+, j+1/2} \mu^x_{-, j+1/2}$, we can replace $\mu^x_{+, j+1/2}$ by $\mu^x_{-, j+1/2} \, \rho^x_j \rho^x_{j+1}$, and effectively drop $\mu^z_{+, j+1/2}$.
Then, the Hamiltonian in Eq.~(\ref{eq:model_mu_pm}) acting within the constrained Hilbert space is equivalent to the following unconstrained Hamiltonian
\begin{align}
  \tilde{H} = & -J \sum_j \left( \rho_j^x \, \mu^x_{-, j+1/2} \, \rho^x_{j+1} + \mu^x_{-, j+1/2} \right) \notag \\
  & -h \sum_j \left( \rho^z_j + \mu^z_{-, j-1/2} \, \rho^z_j \, \mu^z_{-, j+1/2} \right) \notag \\
  & -\Gamma \sum_j (-1)^j \mu^z_{-, j+1/2} ~.
  \label{eq:model_dw_unconstrained}
\end{align}

Similarly, we can replace the parton variables $\left\{ \tau_\pm, \zeta \right\}$ with the constraints by variables $\left\{\tau_+, \zeta \right\}$ without constraints.
In these variables, the parton Hamiltonian Eq.~(\ref{eq:model_bosonic_parton}) becomes
\begin{align}
  \tilde{H} = & -J \sum_j \left( \tau^z_{+, j} \, \zeta^z_{j+1/2} \, \tau^z_{+, j+1} + \zeta^z_{j+1/2} \right) \notag \\
  & -h \sum_j \left( \tau^x_{+, j} + \zeta^x_{j-1/2} \, \tau^x_{+, j} \, \zeta^x_{j+1/2} \right) \notag \\
  & -\Gamma \sum_j (-1)^j \zeta^x_{j+1/2} ~.
  \label{eq:model_parton_unconstrained}
\end{align}

These two unconstrained Hamiltonians are clearly equivalent via the following operator mappings:
\begin{align}
& \mu^x_{-, j+1/2} = \zeta^z_{j+1/2} ~, \quad \mu^z_{-, j+1/2} = \zeta^x_{j+1/2} ~, \notag \\
& \rho^x_j = \tau^z_{+,j} ~, \quad \rho^z_j = \tau^x_{+,j} ~.
  \label{eq:op_map_unconstrained}
\end{align}
Returning to the original domain wall and parton setups with constraints, it is straightforward to see that the above operator mappings correspond precisely to the mappings in Eq.~(\ref{eq:dw_to_parton}) in the main text.

When $\Gamma = 0$, the unconstrained model in Eq.~(\ref{eq:model_parton_unconstrained}) has two $Z_2$ symmetries given by $\prod_j \tau^x_{+, j+1/2}$ and $\prod_j \zeta^z_{j+1/2}$.
Consider a more general model in the $\tau_+, \zeta$ variables with such symmetries:
\begin{align}
  \tilde{H}' = & -\sum_j \left( J_+ \tau^z_{+, j} \, \zeta^z_{j+1/2} \, \tau^z_{+, j+1} + J_- \zeta^z_{j+1/2} \right) \notag \\
  & -\sum_j \left( h_+ \tau^x_{+, j} + h_- \zeta^x_{j-1/2} \, \tau^x_{+, j} \, \zeta^x_{j+1/2} \right) ~.
  \label{eq:model_Z2xZ2_SPT}
\end{align}
For $J_-, h_+ \gg J_+, h_-$, this model is deep in the trivial paramagnetic phase; on the other hand, for $J_-, h_+ \ll J_+, h_-$, it approaches the celebrated cluster model and is deep in the SPT phase for the $Z_2 \times Z_2$ symmetry.\cite{Nielsen2006, ChenLuVishwanath2014}
Varying the parameters along a ray $J_- = h_+$, $J_+ = h_-$, where the model further enjoys a symmetry interchanging the $\tau_+$ and $\zeta$ variables, this model undergoes a direct transition between the trivial and SPT phases when $J_- = h_+ = J_+ = h_-$.
On the other hand, when we view $\tilde{H}'$ as an effective theory for the $z$-FM to VBS transition in the main text, where the irrelevant instanton operators are dropped from the outset, we have physical symmetries that interchange the $\tau_+$ and $\tau_-$ parton species, which here put the parameters on a ray $J_- = J_+$, $h_- = h_+$, and the $z$-FM to VBS transition happens when $J_- = J_+ = h_- = h_+$.
Thus, we conclude that the transition in the effective no-instanton RP$^1$ model (i.e., the gauged Ashkin-Teller model with no instantons, which emerges at the Ising DQCP) corresponds also to the criticality between the trivial and SPT phases in a different system with the $Z_2 \times Z_2$ symmetry~\cite{TsuiJiangLuLee2015, TsuiHuangJiangLee2017, VerresenMoessnerPollmann2017}.
This structure is similar to recent observations in 2d~\cite{GeraedtsMotrunich2017, QinHeYouLuSenSandvikXuMeng2017}, where the transition in the NCCP$^1$ model (which is conjectured to describe the EP DQCP) maps exactly to criticality between trivial and SPT phases~\cite{GroverVishwanath2013, LuLee2014} in a different physical system with a $U(1) \times U(1)$ symmetry, where the SPT phase is an integer quantum Hall state of bosons~\cite{LuVishwanath2012, SenthilLevin2013, GeraedtsMotrunich2013}.

\section{Two-step duality derivation of ``good variables''}
\label{app:deriv_good_vars}
In this appendix, we provide details of the derivation of the ``good variables'' used in Sec.~\ref{sec:goodvars} via a two-step duality transformation starting from the domain wall (equivalently, parton) variables of Sec.~\ref{sec:dw_parton}.
We have already done the first step in Appendix~\ref{app:dw_parton_equiv} when we reformulated the domain wall theory as an unconstrained Hamiltonian in Eq.~(\ref{eq:model_dw_unconstrained}).
We now define a new set of variables:
\begin{align}
  & \eta^x_{j+1/2} \equiv \mu^x_{-, j+1/2} ~, \quad \eta^z_{j+1/2} \equiv (-1)^j \mu^z_{-, j+1/2} ~, \notag \\
  & \eta^x_j \equiv \rho^z_j ~, \quad \eta^z_j \equiv (-1)^j \rho^x_j ~.
  \label{eq:at_dual_map_one}
\end{align}
The Hamiltonian in Eq.~(\ref{eq:model_dw_unconstrained}) becomes
\begin{align}
  \tilde{H} = & -J \sum_j \left( -\eta^z_j \, \eta^x_{j+1/2} \,  \eta^z_{j+1} + \eta^x_{j+1/2} \right) \notag \\
  & -h \sum_j \left( \eta^x_j - \eta^z_{j-1/2} \, \eta^x_j \, \eta^z_{j+1/2} \right) \notag \\
  & -\Gamma \sum_j \eta^z_{j+1/2} ~.
  \label{eq:model_at_dual_one}
\end{align}

We now perform the exact Ising duality defined in Appendix~\ref{app:ising_dual} on the $\eta$ chain.
We label dual Ising variables as $\nu$, which are defined at positions $j \pm 1/4$, and the corresponding gauge fields as $\xi$, which are defined at positions $j$ (original lattice site) or $j + 1/2$ (original lattice link).
The operator mapping between the $\eta$ and $\{\nu, \xi\}$ reads
\begin{align}
  & \eta^x_j = \nu^z_{j-1/4} \, \xi^z_j \, \nu^z_{j+1/4} ~, \quad \eta^z_j = \xi^x_j ~, \notag \\
  & \eta^x_{j+1/2} = \nu^z_{j+1/4} \, \xi^z_{j+1/2} \, \nu^z_{j+3/4} ~, \quad \eta^z_{j+1/2} = \xi^x_{j+1/2} ~,
  \label{eq:at_dual_map_two}
\end{align}
with gauge constraints
\begin{align*}
  \xi^x_j \xi^x_{j+1/2} = \nu^x_{j+1/4} ~, \quad \xi^x_{j+1/2} \xi^x_{j+1} = \nu^x_{j+3/4} ~.
\end{align*}
In these dual variables, the Hamiltonian becomes
\begin{align}
\tilde{H} = & -J \sum_j \xi^z_{j+1/2} \left( \nu^y_{j+1/4} \nu^y_{j+3/4} + \nu^z_{j+1/4} \nu^z_{j+3/4} \right) \notag \\
  & -h \sum_j \xi^z_j \left( \nu^z_{j-1/4} \nu^z_{j+1/4} + \nu^y_{j-1/4} \nu^y_{j+1/4} \right) \notag \\
  & -\Gamma \sum_j \xi^x_{j+1/2} ~,
\end{align}
where for the first term in the first line we used $\eta^z_j \eta^z_{j+1} = \xi^x_j \xi^x_{j+1} = \nu^x_{j+1/4} \, \nu^x_{j+3/4}$, and similarly for the second term in the second line.
This is precisely the claimed $\nu$-chain Hamiltonian in Eq.~(\ref{eq:model_at_dual_two}) in the main text.
It is now straightforward to return to the constrained domain wall variables $\{\mu_\pm, \rho\}$ and obtain operator mappings between these and the new constrained variables $\{\nu, \xi\}$.
The explicit mappings are precisely the ones claimed in Eq.(\ref{eq:good_vars}) in the main text, where we have also included mappings to the constrained parton variables $\{\tau_\pm, \zeta\}$.

\section{Non-parton view of ``good variables''} \label{app:nonparton_goodvars}
The available tools and knowledge for 1d correlated systems are so powerful and extensive that we can provide an alternative derivation of the ``good variables'' for the $z$-FM to VBS transition that does not involve any parton fields; this is the main goal in this appendix.
Starting with our model with ferromagnetic nearest-neighbor (nnb) $S^z S^z$ and $S^x S^x$ interactions, consider adding also comparable antiferromagnetic nnb $S^y S^y$ interactions.
We then perform $\pi$ rotation around the $S^x$ axis on every other site, defining new ``primed'' spin variables
\begin{align}
& S^{\prime\, x}_j = S^x_j ~, \\
& S^{\prime\, y,z}_j = (-1)^j S^{y,z}_j ~. 
\end{align}
The new spins have ferromagnetic nnb $S^{\prime\, x} S^{\prime\, x}$ and $S^{\prime\, y} S^{\prime\, y}$ interactions and antiferromagnetic nnb $S^{\prime\, z} S^{\prime\, z}$ interactions.
Under this change of variables, antiferromagnetic second-neighbor interactions of the original spins become antiferromagnetic interactions of the new spins.
Interestingly, such a translationally invariant Hamiltonian with nearest-neighbor and second-nearest-neighbor exchanges in terms of the $S_j$ spins becomes a translationally invariant Hamiltonian in terms of the $S^\prime_j$ spins.
In fact, this holds generally as long as the original Hamiltonian has the $g_x$ and $T_x$ symmetries, which one can prove using the following identity for a unitary $U_\text{odd} = \prod_{j \in \text{odd~integers}} \sigma^x_j$ that ``performs'' the above change of variables:
\begin{align}
U_\text{odd} T_x U_\text{odd}^{-1} = g_x T_x ~.
\end{align}
Note also that the on-site symmetries $g_x$, $g_z$, and $\TT$ of the original spins become similar symmetries in terms of the $S^\prime$ spins; these symmetries are crucial to constrain the form of the field theory below and are assumed throughout.

Let us now consider the case where the ferromagnetic nnb $S^{\prime\, x} S^{\prime\, x}$ and $S^{\prime\, y} S^{\prime\, y}$ interactions are dominant and equal, and the model has $U(1)$ symmetry of continuous rotations in the $S^{\prime\, x}$-$S^{\prime\, y}$ plane (we will later break this symmetry down to only discrete $\pi$ rotation corresponding to the $g_z$ symmetry).
We can now perform standard bosonization in the spirit of Sec.~\ref{sec:bosonization} but for this primed spin system, with a phase variable $\phi^\prime$ describing spin components in the $S^{\prime\, x}$-$S^{\prime\, y}$ plane:
\begin{align}
& S^{\prime\, x}_j \sim \cos(\phi^\prime) ~, \quad S^{\prime\, y}_j \sim \sin(\phi^\prime) ~, \\
& S^{\prime\, z}_j \sim \frac{\partial_x \theta^\prime}{\pi} + A' (-1)^j \sin(2\theta^\prime) ~, \\
& B_{j+1/2} \sim C' (-1)^j \cos(2\theta^\prime) ~.
\end{align}
Here, we have already assumed that the phase field $\phi^\prime$ and the conjugate field $\theta^\prime$ are long-wavelength fields, and we have also written out important contributions to the staggered part of the spin component $S^{\prime\, z}_j$ and bond energy $B_{j+1/2}$.
From these expressions, one can infer how the symmetries act on the continuum fields and write down a continuum description of the system as 
\begin{align}
S = & \!\int\! \dd \tau\, \dd x \left[ \frac{\ii}{\pi} \partial_\tau \phi^\prime \partial_x \theta^\prime + \frac{v^\prime}{2\pi} \left( \frac{1}{g^\prime}(\partial_x \theta^\prime)^2 + g^\prime (\partial_x \phi^\prime)^2 \right) \right] \notag \\
& \!+\! \!\int\! \dd \tau\, \dd x \left[ \lambda^\prime \cos(4\theta^\prime) + \kappa^\prime \cos(2\phi^\prime) \right] + \cdots ~. \label{eq:S_nonparton_primevars}
\end{align}
In the last line, we have already included the leading perturbation allowed when we have only the $\pi$ rotation symmetry in the $S^{\prime\, x}$-$S^{\prime\, y}$ plane, but let us ignore this for a moment and set $\kappa^\prime = 0$.
With the $U(1)$ symmetry and dominant in-plane spin interactions, we have quasi-long-range-ordered phase when the Luttinger parameter $g' > 1/2$ so that the $\lambda^\prime$ term is irrelevant.
As we increase either the second-neighbor antiferromagnetic interactions or the nearest-neighbor antiferromagnetic $S^{\prime\, z} S^{\prime\, z}$ interactions, $g'$ decreases, and for $g' < 1/2$ the $\lambda^\prime$ term becomes relevant and drives the system into either a VBS phase (for $\lambda' > 0$) or a $z$-AFM phase (for $\lambda' < 0$) of the $S^\prime$ spins.
In the regime when $g^\prime < 1/2$, we can then induce a transition between the VBS and $z$-AFM phases by varying $\lambda^\prime$ through $0$.
In the $U(1)$-symmetric system, the transition is described by a Gaussian fixed point (essentially the Gaussian part of the above action with renormalized $g'$), where the $\lambda^\prime$ term is the only relevant perturbation.
When we break the $U(1)$ symmetry by including the $\kappa'$ term, this term is actually irrelevant for such $g^\prime < 1/2$.
Hence, even when the $U(1)$ is replaced by the $\pi$ rotation symmetry, the VBS to $z$-AFM transition of the $S^\prime$ spins is still described by the same field theory.

We can now go back to the physical spins and see that the $z$-AFM phase of the $S^\prime$ spins becomes the $z$-FM phase of the physical spins.
The $z$-FM order parameter becomes $M_z^\text{FM} \sim \sin(2\theta')$, while the VBS order parameter is $\Psi_\text{VBS} \sim \cos(2\theta^\prime)$.
This can be directly compared with our expressions in the main text, Eqs.~(\ref{eq:MzFM}) and (\ref{eq:PsiVBS}), which agree upon identification $\thtgood = 2\theta^\prime$ (remember also from Sec.~\ref{sec:goodvars} that $\thtgood$ has $2\pi$-periodicity, which agrees with the $\pi$ periodicity of $\theta^\prime$ in the standard bosonization).
Furthermore, we see that the $S^x$ correlations in this setting are naturally ferromagnetic, while the $S^y$ correlations are antiferromagnetic, with $M_x^\text{FM} \sim \cos(\phi^\prime)$ and $M_y^\text{AFM} \sim \sin(\phi^\prime)$.
This can be compared with expressions Eqs.~(\ref{eq:MxFM}) and (\ref{eq:MyAFM}) in the main text, with the identification $\phigood = \phi^\prime/2$ (with corresponding agreement between periodicities of these variables).
This identification shows that the $\phigood$ variables in the main text can be thought of as ``fractionalizing'' the $x$-FM and $y$-AFM order parameters.
The theory for the transition in Eq.~(\ref{eq:S_nonparton_primevars}) is thus equivalent to the theory in Eq.~(\ref{eq:gauged_bosonization_action}) in the main text (using also the fact that the transformations of the fields under the action of the symmetries of the original spin chain are completely fixed by the expressions for the spin components and the bond energy).

We conclude with some remarks.
First, we note that from the point of view of studying the $z$-FM to VBS  transition, we invoked a highly non-obvious starting point with dominant and comparable $x$-FM and $y$-AFM nnb interactions, in order to use abelian bosonization near the $S^{\prime\, x}$-$S^{\prime\, y}$ easy-plane limit; we then employed bosonization tools to treat sufficiently strong interactions that can produce the desired phases and the transition between them.
It is natural to ask if there may be variants of this approach that would work for the 2d DQCP theories; even if such approaches do not lead to easily tractable field theories, perhaps they could provide new arguments or tests for some non-trivial conjectures about the 2d DQCP theories~\cite{SeibergSenthilWangWitten2016, WangNahumMetlitskiXuSenthil2017, MetlitskiThorngren2017}.

Second, we remark that this more direct non-parton approach to the $z$-FM to VBS transition does not start with the corresponding order parameters and hence does not by itself try to unify the two order parameters.
The unification does happen, but the two order parameters are encoded as ``vortex instantons'' in the physical spin phase variables in this setting.
When we do try to ``unify'' the two order parameters in an $O(2)$ vector, $(\Psi_\text{VBS}, M_z^\text{FM}) \sim (\cos\alpha, \sin\alpha)$, it appears that we should identify the corresponding ``angle'' as $\alpha = 2\theta^\prime = \thtgood$, i.e., as the conjugate (or dual) variable to the phase variable $\tilde{\phi}$ that is fractionalizing the physical spin variable.

\section{Derivation of the fermionic parton ansatz from the ``good'' bosonic partons}
\label{app:fermion_parton_from_good_variable}
In this part, we will derive the fermionic parton description starting from the ``good'' variable bosonic partons.
First, we perform Jordan-Wigner transformation as
\begin{align}
& \FF_{-,j} = \left( \prod_{1 \le j' < j} \VV^x_{-,j'} \VV^x_{+,j'} \right) \cdot \frac{1}{2} (\VV^z_{-,j} - \ii \VV^y_{-,j} ) ~, \notag \\
& \FF_{+,j} = \left( \prod_{1 \le j' < j} \VV^x_{-,j'} \VV^x_{+,j'} \right) \VV^x_{-,j} \cdot \frac{1}{2} (\VV^z_{+,j} - \ii \VV^y_{+,j}) ~,
\label{eq:good_fermion_JW}
\end{align}
where $\FF_{\pm,j}$ are annihilation operators for fermions.

Under this mapping, the effective Hamiltonian Eq.~(\ref{eq:good_variable_parton_Ham}) becomes
\begin{align}
\tilde{H} = & -J \sum_j \XX^z_{j+1/2} \left( \FF_{+,j}^\dg \FF_{-,j+1} + \text{H.c.} \right) \notag \\
& -h \sum_j \left( \FF_{-,j}^\dg \FF_{+,j} + \text{H.c.} \right) - \Gamma \sum_j \XX^x_{j+1/2} ~.
\label{eq:good_fermionic_parton_Ham}
\end{align}
The Hilbert constraint is
\begin{align}
\XX^x_{j-1/2} \XX^x_{j+1/2} = P_{+,j} P_{-,j}
\label{eq:good_fermionic_parton_constraint}
\end{align}
where $P_{\pm,j} \equiv (-1)^{\FF_{\pm,j}^\dg \FF_{\pm,j}} = \FF_{\pm,j} \FF_{\pm,j}^\dg - \FF_{\pm,j}^\dg \FF_{\pm,j}$ is the parity operator for each fermion.
We point out that the total fermion parity is fixed due to the gauge constraints, which imply $\prod_j P_{+,j} P_{-,j} = 1$.
Under the formal Jordan-Wigner transformation, this means that the boundary conditions for the fermions $\FF$ are also fixed and are opposite in Eq.~(\ref{eq:good_fermionic_parton_Ham}) compared to that for the bosonic partons $\VV$ in
Eq.~(\ref{eq:good_variable_parton_Ham}).
However, in our gauge theory, the boundary conditions can be absorbed in redefinition of the gauge field on one link.

The spin operators in terms of the Jordan-Wigner fermions read:
\begin{align}
\sigma^x_j & = \FF_{+,j} \FF_{-,j} + \FF{-,j}^\dg \FF_{+,j}^\dg ~, \notag \\
\sigma^y_j & = (-1)^j (-\ii \FF_{+,j} \FF_{-,j} + \ii \FF{-,j}^\dg \FF_{+,j}^\dg) ~, \notag \\
\sigma^z_j & = (-1)^j (1 - \FF_{+,j}^\dg \FF_{+,j} - \FF_{-,j}^\dg \FF_{-,j}) ~.
\label{eq:sigmaz_FF}
\end{align}
We can also derive how the symmetries act on $\FF_{\pm,j}$ from Eq.~(\ref{eq:good_variable_parton_symmetry}:
\begin{align}
& T_x:\ \FF_{+,j} \to \FF_{+,j+1}^\dg ~, \quad \FF_{-,j} \to -\FF_{-,j+1}^\dg ~; \notag \\
& g_x:\ \FF_{+,j} \to \FF_{+,j}^\dg ~, \quad \FF_{-,j} \to -\FF_{-,j}^\dg ~; \notag \\
& g_z:\ \FF_{\pm,j} \to \ii \FF_{\pm,j} ~; \notag \\
& \TT:\ \FF_{+,j} \to \ii \FF_{+,j}^\dg ~, \quad \FF_{-,j} \to -\ii \FF_{-,j}^\dg ~, \quad \ii \to -\ii ~.
\label{eq:good_fermionic_parton_symmetry}
\end{align}
There is some subtlety when extracting the $T_x$ transformation properties of the $\FF$ fields in Eq.~(\ref{eq:good_fermion_JW}) because of the string operator: 
The translated string is essentially a new string but with a missing operator at the origin, which we did not write out explicitly.
Instead of dealing with this, we take a perspective where we consider the above writing of the spin operators, the symmetry actions, and the mean field Hamiltonian as a proposal for a fermionic parton approach, where we can then independently verify that it produces the desired $z$-FM and VBS phases and the phase transition between them.

Surprisingly, although the symmetry transformation rules on $\FF_{\pm,j}$ in Eq.~(\ref{eq:good_fermionic_parton_symmetry}) look differently from those on $f_{\pm,j}$ in Eq.(\ref{eq:fermionic_parton_sym}), they actually have the same PSG equations defined in Eq.~(\ref{eq:fermionic_parton_psg}).
Hence, we expect the fermionic theory derived here to be the same theory as presented in Sec.~\ref{sec:fermionic_parton} up to some basis change.

We can indeed find such an explicit basis change as:
\begin{align*}
& f_{+,j} = \frac{e^{\ii 3\pi/4} (-1)^j}{2} \left( \FF_{+,j}^\dg - \FF_{+,j} + \FF_{-,j}^\dg + \FF_{-,j} \right) ~, \\
& f_{-,j} = \frac{e^{\ii 3\pi/4}}{2} \left( \FF_{+,j}^\dg + \FF_{+,j} + \FF_{-,j}^\dg - \FF_{-,j} \right) ~.
\end{align*}
One can easily check that the spin representations Eqs.~(\ref{eq:sigmaz_FF}) and (\ref{eq:fermion_parton_to_physical_spin}) match, and also the symmetry transformation rules Eq.~(\ref{eq:good_fermionic_parton_symmetry}) and Eq.~(\ref{eq:fermionic_parton_sym}).
The $h$ term in Eq.~(\ref{eq:good_fermionic_parton_Ham}) becomes the ``chemical potential'' term in Eq.~(\ref{eq:model_fermionic_parton}) with $\mu = -h$, while the $J$ term becomes the ``pairing'' plus ``hopping'' term with specific $t_+ = -t_- = \eta_+ = \eta_- = J/2$.
(General $t$ and $\eta$ correspond to general symmetry-allowed $\FF$ fermion hopping with real-valued amplitudes and with ``bipartite'' structure, i.e., only hopping between the $+$ and $-$ ``sublattices.'')

Let us finally consider the Gauss law constraints and the gauge field terms.
It is easy to check that
\begin{align}
\FF_{+,j}^\dg \FF_{+,j} + \FF_{-,j}^\dg \FF_{-,j} = 1 - (-1)^j (f_{+,j}^\dg f_{-,j} + f_{-,j}^\dg f_{+,j}) ~.
\end{align}
Hence, by defining $\zeta^{\prime\, x}_{j+1/2} = (-1)^j \XX^x_{j+1/2}$, the constraints in Eq.~(\ref{eq:good_fermionic_parton_constraint}) become
\begin{align}
\zeta^{\prime\, x}_{j-1/2} \zeta^{\prime\, x}_{j+1/2} = \ee^{\ii \pi (f_{+,j}^\dg f_{-,j} + f_{-,j}^\dg f_{+,j})} ~,
\end{align}
while the $\Gamma$ term in Eq.~(\ref{eq:good_fermionic_parton_Ham}) becomes $\Gamma \sum_j (-1)^j \zeta^{\prime\, x}_{j+1/2}$.
The structure of the $\Gamma$ term matches that in Eq.~(\ref{eq:model_fermionic_parton}) for the fermionic parton theory in Sec.~\ref{sec:fermionic_parton}.
On the other hand, the Gauss law constraints actually differ from the ``natural'' constraints used in Sec.~\ref{sec:fermionic_parton}, Eq.~(\ref{eq:fermionic_parton_constraint}).
Thus, strictly speaking, the parton-gauge model introduced in this Appendix is different from the one in Sec.~\ref{sec:fermionic_parton}.
However, in the strong coupling limit, $\Gamma \to \infty$, the two models agree, since the Hilbert space space constraint obtained in this limit, $\exp\left[\ii \pi (f_{+,j}^\dg f_{-,j} + f_{-,j}^\dg f_{+,j}) \right]  = -1$, is equivalent to the single-occupancy constraint Eq.~(\ref{eq:fermionic_parton_constraint}) in  Sec.~\ref{sec:fermionic_parton}.
As discussed in the main text, we believe that this difference is only quantitative and not qualitative, as long as one is accessing the same phases and transitions.
The gauge theory constraints in this Appendix are more convenient for discussing key observables at the $z$-FM to VBS transition, and in fact we essentially used this insight in Sec.~\ref{sec:fermionic_parton}.

\section{Extracting quantum numbers from Gutzwiller-projected fermionic wavefunctions}
\label{app:qn_fermionic_parton}
In this Appendix, we will discuss a generic algorithm to extract quantum numbers from Gutzwiller-projected wavefunctions.
We will use the fermionic parton setup of Sec.~\ref{sec:fermionic_parton} as an example.

We start from a mean field Hamiltonian for fermionic partons,
\begin{align}
H_\text{MF} \!=\!\! \sum_{j\sigma, j'\sigma'} \!\! \left[-t_{j\sigma, j'\sigma'} f_{j\sigma}^\dg f_{j'\sigma'} + (\Delta_{j\sigma, j'\sigma'} f_{j\sigma}^\dg f_{j'\sigma'}^\dg + \text{H.c.} ) \right] \,,
\label{eq:fp_mf_ham}
\end{align}
where we ignore the gauge field at the mean field level. 
From this, we construct a physical spin wavefunction using Gutzwiller projection,
\begin{align}
|\Psi \rangle = P_\text{Gutzw} |\Psi_\text{MF} \rangle ~,
\end{align}
where $|\Psi_\text{MF} \rangle$ is the ground state of the mean field Hamiltonian, and $P_\text{Gutzw}$ projects out configurations with zero or double occupancy.
Formally, we identify physical spin states with the fermionic states with precisely one fermion per site as follows:
\begin{align}
|\sigma_1, \sigma_2, \dots, \sigma_L \rangle = f_{1, \sigma_1}^\dagger f_{2, \sigma_2}^\dagger \dots f_{L, \sigma_L}^\dagger |0 \rangle ~,
\label{eq:physical_states}
\end{align}
where the spin labels $\sigma_j$ refer to the $\sigma^x_j$ basis per Eq.~(\ref{eq:fermion_parton_to_physical_spin}), and $|0 \rangle$ is the fermion vacuum.

Let $SG$ be the symmetry group for the original spin model.
By definition, the quantum number for a symmetry $g \in SG$ can be obtained as
\begin{align}
q_g = \frac{\langle s|g|\Psi \rangle}{\langle s|\Psi\rangle} = \frac{\langle s|g|\Psi_\text{MF} \rangle}{\langle s|\Psi_\text{MF} \rangle} ~,
\end{align}
where $|s \rangle$ is an arbitrary spin configuration with nonzero overlap with $|\Psi \rangle$.
In the last equation, we used $\langle s|\Psi \rangle = \langle s|\Psi_\text{MF} \rangle$ and $\langle s|g|\Psi \rangle = \langle s|g|\Psi_\text{MF} \rangle$, since any physical state is invariant under the projection $P_\text{Gutzw}$.
Here, we applied this argument to states $|s \rangle$ and $|g^\dagger s \rangle$, and, strictly speaking, the numerator should still be written as $\langle g^\dagger s| \Psi_\text{MF} \rangle$; however, we will define an action of $g$ on the whole fermion Fock space that is consistent with its action on the physical states, so indeed $\langle g^\dagger s| \Psi_\text{MF} \rangle = \langle s|g|\Psi_\text{MF} \rangle$. 
In what follows, we will develop an algorithm to obtain $q_g$.

In general, $H_\text{MF}$ is not invariant under $g$.
Instead, one can find a gauge transformation $W_g$ such that
\begin{align}
\left[H_\text{MF}, W_g g \right] = 0 ~.
\label{eq:HMF_commute_Wgg}
\end{align}
Note that in the main text in Sec.~\ref{sec:fermionic_parton}, e.g., in Eq.~(\ref{eq:fermionic_parton_sym}), we quoted the combined transformations $W_g g$, while here for more precise arguments we find it convenient to separate the action of symmetries of the mean field Hamiltonians into such two parts, both of which will be defined for our specific example below.
The gauge transformation $W_g \equiv \prod_j W_g(j)$ acts non-trivially on the whole fermion Fock space but leaves all physical states invariant up to a global phase.
In other words, we have
\begin{align}
P_\text{Gutzw} W_g = \ee^{\ii \alpha(g)} P_\text{Gutzw} ~.
\label{eq:gauge_transf_phase}
\end{align}
(If the reader is not comfortable with this general statement, this property can be considered as a postulate and is easily verified for each exhibited gauge transformation below.)

Due to nontrivial gauge transformations, the symmetry group for the mean field Hamiltonian, denoted as $PSG$, would be different from the $SG$.
Note that even for a fixed $SG$, the choice of $PSG$ is far from unique.
Different $PSG$s correspond to different gauge theories, which in general would describe different phases of the spin system.

We consider the case where $|\Psi_\text{MF} \rangle$ is a trivial or one-dimensional representation under the $PSG$.
For a given $PSG$, there may exist more than one fully symmetric phases, depending on the mean field parameters.
These phases are what one would call symmetry protected topological~(SPT) phases in the formal fermionic problem defined in the whole Fock space with the $PSG$ as the symmetry group.
In general, ground states for different SPT phases would have different quantum numbers under the $PSG$.
In particular, under $W_g g \in PSG$, 
\begin{align}
W_g g |\Psi_\text{MF} \rangle = \ee^{\ii \beta(g)} |\Psi_\text{MF} \rangle ~,
\label{eq:PSG_phase}
\end{align}
where the number $\ee^{\ii \beta(g)}$ depends on both the $PSG$ and SPT classes.

We can now deduce the quantum numbers of the physical states as follows:
\begin{align*}
& g |\Psi \rangle = g P_\text{Gutzw} |\Psi_\text{MF} \rangle = P_\text{Gutzw} g |\Psi_\text{MF} \rangle \\
& = P_\text{Gutzw} W_g^{-1} W_g g |\Psi_\text{MF} \rangle = P_\text{Gutzw} e^{-\ii \alpha(g)} e^{\ii \beta(g)} |\Psi_\text{MF} \rangle ~.
\end{align*}
In the first line, we used $g P_\text{Gutzw} = P_\text{Gutzw} g$, which follows from how we extend the action of $g$ on the full fermion Fock space as already mentioned earlier.
In the second line, we used $P_\text{Gutzw} W_g^{-1} = \ee^{-\ii \alpha(g)} P_\text{Gutzw}$, which is a simple corollary of Eq.~(\ref{eq:gauge_transf_phase}), and also Eq.~(\ref{eq:PSG_phase}).
Hence, the physical symmetry quantum number is
\begin{align}
q_g = e^{-\ii \alpha(g)} e^{\ii \beta(g)} ~.
\end{align}

In the following, we will use this method to identify phases for the Hamiltonian in Eq.~(\ref{eq:model_fermionic_parton}). 
The mean field Hamiltonian is obtained by simply setting $\zeta^z_{j+1/2} = 1$ on all links, which corresponds to the sector with no gauge flux and gives periodic boundary conditions for the fermions.
By changing the sign of $\zeta^z$ on one link, we obtain the sector with non-trivial gauge flux, which gives antiperiodic boundary conditions for the fermions.

We first identify the $PSG$ for this mean field Hamiltonian. 
We consider a spin chain with $L$ sites, where $L$ is an even integer; the sites are labeled $j = 1, 2, \dots, L$.
We also impose periodic boundary conditions in the spin system.
We can readily deduce desired extensions of actions of the global symmetries to the fermionic operators from their actions on the spins. 
We get
\begin{align}
& T_x \cdot f_{j,\pm} \cdot T_x^{-1}=
\begin{cases}
f_{j+1,\pm} & \text{if}\ j<L ~, \\
-f_{1,\pm} & \text{if}\ j=L ~;
\end{cases} \notag \\
& g_x \cdot f_{j,\pm} \cdot g_x^{-1} = \pm f_{j,\pm} ~; \notag \\
& g_z \cdot f_{j,\pm} \cdot g_z^{-1} = f_{j,\mp} ~.
\end{align}
The extra minus for the translation symmetry action on the last fermion comes from the anti-commutation relation for the fermionic operators (e.g., for bosonic partons, we would not have this extra minus sign).
Note that when defining the action of the symmetries on the fermion operators, we require that the physical states in Eq.~(\ref{eq:physical_states}) are transformed correctly, and we have also postulated that the fermion vacuum is transformed trivially.
Other choices are possible that would meet this requirement, and we have just picked one.
Note that this part can be used for any $PSG$.
The $PSG$ associated with a given ansatz is encoded in additional gauge transformations that need to be performed on top of the above action of symmetries to make the mean field Hamiltonian invariant; that is, combination $W_g g$ is a formal symmetry of the mean field Hamiltonian defined in the whole fermion Fock space, cf.\ Eq.~(\ref{eq:HMF_commute_Wgg}).

The $PSG$ for the specific ansatz in Eq.~(\ref{eq:model_fermionic_parton}) can be read from the symmetry actions in Eq.~(\ref{eq:fermionic_parton_sym})
(remember that in this equation in the main text, we quoted the combined action $W_g g$ on the fermion field).
Here, we emphasize a subtle point for the gauge transformation associated with the translation symmetry. 
As we have already mentioned, to identify phases it is necessary to consider the effect of gauge fluctuations.
For the $Z_2^\zeta$ gauge field, there are two gauge-inequivalent sectors, labeled by $\prod_j \zeta^z_{j+1/2} = \pm 1$, which correspond to periodic and antiperiodic boundary conditions for $f_{j,\pm}$.
The gauge transformation associated with the translation symmetry actually depends on the sector, and we label the corresponding gauge transformations as $W^\text{p.b.c}_{T_x}$ and $W^\text{a.b.c}_{T_x}$.
For the on-site symmetries, the gauge transformations do not depend on the sector, and we simply omit the sector label.

We can now list the associated gauge transformations:
\begin{align}
& W^\text{p.b.c}_{T_x} \cdot f_{j,\pm} \cdot (W^\text{p.b.c}_{T_x})^{-1} =
\begin{cases} 
f_{j,\pm} & \text{if}\ j > 1 ~, \notag\\
-f_{1,\pm} & \text{if}\ j = 1 ~;
\end{cases} \notag \\
& W^\text{a.b.c}_{T_x} \cdot f_{j,\pm} \cdot (W^\text{a.b.c}_{T_x})^{-1} = f_{j,\pm} ~; \notag\\
& W_{g_x} \cdot f_{j,\pm} \cdot W_{g_x}^{-1} = f_{j,\pm} ~; \notag \\
& W_{g_z} \cdot f_{j,\pm} \cdot W_{g_z}^{-1} = (-1)^j \, \ii f_{j,\pm} ~.
\end{align}
The corresponding $\alpha(g)$ defined in Eq.~(\ref{eq:gauge_transf_phase}) are readily obtained as
\begin{align}
\alpha(T_x^\text{p.b.c}) = \pi ~, ~~ 
\alpha(T_x^\text{a.b.c}) = \alpha(g_x) = \alpha(g_z) = 0 ~.
\label{eq:psg_phase_example}
\end{align}
Note that these depend only on the $PSG$ and not on which phase the fermions are in.
We need to remember, however, that for a fixed $PSG$ viewed as a formal symmetry in the fermion Fock space, the fermions can be in distinct fully symmetric (under the $PSG$) phases that differ by their SPT index.

We now consider the action of $W_g g$ on the mean field ground state and the corresponding quantum number $e^{\ii \beta(g)}$ in Eq.~(\ref{eq:PSG_phase}).
As we will see below, this depends on the SPT index of the fermion state, and it is this difference that will give different symmetry-breaking phases of the physical spins.
The detailed analysis of the mean field Hamiltonian was already performed in the main text.
In the nearest-neighbor ansatz, we find two different phases of fermions: for $|\mu| > 2 t_+$, we get the topologically trivial phase, while for $|\mu| < 2 t_+$, we get the topological (Kitaev) phase for both $+$ and $-$ fermion species separately~\cite{Kitaev2001, MotrunichDamleHuse2001}.
In each case, we need to analyze the action of the symmetries in the two flux sectors of the gauge field, i.e., for the periodic and antiperiodic boundary conditions for fermions.
This analysis was essentially performed in the main text, giving us the following quantum numbers of $|\Psi_\text{MF} \rangle$ for the elements of the $PSG$:
\begin{itemize}
\item Topologically trivial phase:
\begin{align}
\beta(T_x^\text{p.b.c./a.b.c}) = \beta(g_x^\text{p.b.c./a.b.c}) = \beta(g_z^\text{p.b.c./a.b.c}) = 0 ~.  
\end{align}
Combined with Eq.~(\ref{eq:psg_phase_example}), we conclude that the wavefunctions for the two gauge sectors differ by momentum $\pi$ but have identical $g_x$ quantum numbers and identical $g_z$ quantum numbers.
Hence, we get the VBS phase of the spin system.

\item Topological superconductor phase:
\begin{align}
& \beta(T_x^\text{p.b.c.}) = \pi ~, \quad \beta(T_x^\text{a.b.c.}) = 0 ~; \\ 
& \beta(g_x^\text{p.b.c.}) = \pi ~, \quad \beta(g_x^\text{a.b.c.}) = 0 ~; \\ 
& \beta(g_z^\text{p.b.c.}) = \beta(g_z^\text{a.b.c.}) = 0 ~.
\label{eq:beta_g}
\end{align}
Combined with Eq.~(\ref{eq:psg_phase_example}), we conclude that the wavefunctions for these two gauge sectors have opposite $g_x$ quantum numbers but have identical $T_x$ quantum numbers and identical $g_z$ quantum numbers. 
Hence, we get the $z$-FM phase of the spin system. 
\end{itemize}

Comparing with the main text, there we simply said that the condensation of visons by itself introduces momentum $\pi$ difference between the two sectors; this is the $\alpha(T_x)$ part here, computed in Eq.~(\ref{eq:psg_phase_example}).
This is added to the contribution from the fermion mean field in the corresponding sectors, computed in the main text and summarized in the $\beta$ part in Eq.~(\ref{eq:beta_g}).
We note that from the analysis here, we can actually assign absolute quantum numbers to the wavefunctions and not just the differences in the quantum numbers.
We also note the subtlety in assigning the translation quantum numbers to flux sectors from simply invoking the vison condensation without regards to other aspects of the parton setup: for bosonic partons, $\alpha(T_x^\text{p.b.c})$ and $\alpha(T_x^\text{a.b.c})$ would actually be interchanged compared to fermionic partons here.

\bibliography{bib1Ddqcp} 

\begin{thebibliography}{100}%
\makeatletter
\providecommand \@ifxundefined [1]{%
 \@ifx{#1\undefined}
}%
\providecommand \@ifnum [1]{%
 \ifnum #1\expandafter \@firstoftwo
 \else \expandafter \@secondoftwo
 \fi
}%
\providecommand \@ifx [1]{%
 \ifx #1\expandafter \@firstoftwo
 \else \expandafter \@secondoftwo
 \fi
}%
\providecommand \natexlab [1]{#1}%
\providecommand \enquote  [1]{``#1''}%
\providecommand \bibnamefont  [1]{#1}%
\providecommand \bibfnamefont [1]{#1}%
\providecommand \citenamefont [1]{#1}%
\providecommand \href@noop [0]{\@secondoftwo}%
\providecommand \href [0]{\begingroup \@sanitize@url \@href}%
\providecommand \@href[1]{\@@startlink{#1}\@@href}%
\providecommand \@@href[1]{\endgroup#1\@@endlink}%
\providecommand \@sanitize@url [0]{\catcode `\\12\catcode `\$12\catcode
  `\&12\catcode `\#12\catcode `\^12\catcode `\_12\catcode `\%12\relax}%
\providecommand \@@startlink[1]{}%
\providecommand \@@endlink[0]{}%
\providecommand \url  [0]{\begingroup\@sanitize@url \@url }%
\providecommand \@url [1]{\endgroup\@href {#1}{\urlprefix }}%
\providecommand \urlprefix  [0]{URL }%
\providecommand \Eprint [0]{\href }%
\providecommand \doibase [0]{http://dx.doi.org/}%
\providecommand \selectlanguage [0]{\@gobble}%
\providecommand \bibinfo  [0]{\@secondoftwo}%
\providecommand \bibfield  [0]{\@secondoftwo}%
\providecommand \translation [1]{[#1]}%
\providecommand \BibitemOpen [0]{}%
\providecommand \bibitemStop [0]{}%
\providecommand \bibitemNoStop [0]{.\EOS\space}%
\providecommand \EOS [0]{\spacefactor3000\relax}%
\providecommand \BibitemShut  [1]{\csname bibitem#1\endcsname}%
\let\auto@bib@innerbib\@empty
\bibitem [{\citenamefont {Senthil}\ \emph
  {et~al.}(2004{\natexlab{a}})\citenamefont {Senthil}, \citenamefont
  {Vishwanath}, \citenamefont {Balents}, \citenamefont {Sachdev},\ and\
  \citenamefont {Fisher}}]{DQCP_science}%
  \BibitemOpen
  \bibfield  {author} {\bibinfo {author} {\bibfnamefont {T.}~\bibnamefont
  {Senthil}}, \bibinfo {author} {\bibfnamefont {A.}~\bibnamefont {Vishwanath}},
  \bibinfo {author} {\bibfnamefont {L.}~\bibnamefont {Balents}}, \bibinfo
  {author} {\bibfnamefont {S.}~\bibnamefont {Sachdev}}, \ and\ \bibinfo
  {author} {\bibfnamefont {M.~P.~A.}\ \bibnamefont {Fisher}},\ }\href@noop {}
  {\bibfield  {journal} {\bibinfo  {journal} {Science}\ }\textbf {\bibinfo
  {volume} {303}},\ \bibinfo {pages} {1490} (\bibinfo {year}
  {2004}{\natexlab{a}})}\BibitemShut {NoStop}%
\bibitem [{\citenamefont {Senthil}\ \emph
  {et~al.}(2004{\natexlab{b}})\citenamefont {Senthil}, \citenamefont {Balents},
  \citenamefont {Sachdev}, \citenamefont {Vishwanath},\ and\ \citenamefont
  {Fisher}}]{DQCP_prb}%
  \BibitemOpen
  \bibfield  {author} {\bibinfo {author} {\bibfnamefont {T.}~\bibnamefont
  {Senthil}}, \bibinfo {author} {\bibfnamefont {L.}~\bibnamefont {Balents}},
  \bibinfo {author} {\bibfnamefont {S.}~\bibnamefont {Sachdev}}, \bibinfo
  {author} {\bibfnamefont {A.}~\bibnamefont {Vishwanath}}, \ and\ \bibinfo
  {author} {\bibfnamefont {M.~P.~A.}\ \bibnamefont {Fisher}},\ }\href@noop {}
  {\bibfield  {journal} {\bibinfo  {journal} {Phys. Rev. B}\ }\textbf {\bibinfo
  {volume} {70}},\ \bibinfo {pages} {144407} (\bibinfo {year}
  {2004}{\natexlab{b}})}\BibitemShut {NoStop}%
\bibitem [{\citenamefont {Sandvik}(2007)}]{Sandvik2007}%
  \BibitemOpen
  \bibfield  {author} {\bibinfo {author} {\bibfnamefont {A.~W.}\ \bibnamefont
  {Sandvik}},\ }\href {\doibase 10.1103/PhysRevLett.98.227202} {\bibfield
  {journal} {\bibinfo  {journal} {Phys. Rev. Lett.}\ }\textbf {\bibinfo
  {volume} {98}},\ \bibinfo {pages} {227202} (\bibinfo {year}
  {2007})}\BibitemShut {NoStop}%
\bibitem [{\citenamefont {Melko}\ and\ \citenamefont
  {Kaul}(2008)}]{MelkoKaul2008}%
  \BibitemOpen
  \bibfield  {author} {\bibinfo {author} {\bibfnamefont {R.~G.}\ \bibnamefont
  {Melko}}\ and\ \bibinfo {author} {\bibfnamefont {R.~K.}\ \bibnamefont
  {Kaul}},\ }\href {\doibase 10.1103/PhysRevLett.100.017203} {\bibfield
  {journal} {\bibinfo  {journal} {Phys. Rev. Lett.}\ }\textbf {\bibinfo
  {volume} {100}},\ \bibinfo {pages} {017203} (\bibinfo {year}
  {2008})}\BibitemShut {NoStop}%
\bibitem [{\citenamefont {Lou}\ \emph {et~al.}(2009)\citenamefont {Lou},
  \citenamefont {Sandvik},\ and\ \citenamefont
  {Kawashima}}]{LuoSandvikKawashima2009}%
  \BibitemOpen
  \bibfield  {author} {\bibinfo {author} {\bibfnamefont {J.}~\bibnamefont
  {Lou}}, \bibinfo {author} {\bibfnamefont {A.~W.}\ \bibnamefont {Sandvik}}, \
  and\ \bibinfo {author} {\bibfnamefont {N.}~\bibnamefont {Kawashima}},\ }\href
  {\doibase 10.1103/PhysRevB.80.180414} {\bibfield  {journal} {\bibinfo
  {journal} {Phys. Rev. B}\ }\textbf {\bibinfo {volume} {80}},\ \bibinfo
  {pages} {180414} (\bibinfo {year} {2009})}\BibitemShut {NoStop}%
\bibitem [{\citenamefont {Banerjee}\ \emph {et~al.}(2010)\citenamefont
  {Banerjee}, \citenamefont {Damle},\ and\ \citenamefont
  {Alet}}]{BanerjeeDamleAlet2010}%
  \BibitemOpen
  \bibfield  {author} {\bibinfo {author} {\bibfnamefont {A.}~\bibnamefont
  {Banerjee}}, \bibinfo {author} {\bibfnamefont {K.}~\bibnamefont {Damle}}, \
  and\ \bibinfo {author} {\bibfnamefont {F.}~\bibnamefont {Alet}},\ }\href
  {\doibase 10.1103/PhysRevB.82.155139} {\bibfield  {journal} {\bibinfo
  {journal} {Phys. Rev. B}\ }\textbf {\bibinfo {volume} {82}},\ \bibinfo
  {pages} {155139} (\bibinfo {year} {2010})}\BibitemShut {NoStop}%
\bibitem [{\citenamefont {Sandvik}(2010)}]{Sandvik2010}%
  \BibitemOpen
  \bibfield  {author} {\bibinfo {author} {\bibfnamefont {A.~W.}\ \bibnamefont
  {Sandvik}},\ }\href {\doibase 10.1103/PhysRevLett.104.177201} {\bibfield
  {journal} {\bibinfo  {journal} {Phys. Rev. Lett.}\ }\textbf {\bibinfo
  {volume} {104}},\ \bibinfo {pages} {177201} (\bibinfo {year}
  {2010})}\BibitemShut {NoStop}%
\bibitem [{\citenamefont {Harada}\ \emph {et~al.}(2013)\citenamefont {Harada},
  \citenamefont {Suzuki}, \citenamefont {Okubo}, \citenamefont {Matsuo},
  \citenamefont {Lou}, \citenamefont {Watanabe}, \citenamefont {Todo},\ and\
  \citenamefont
  {Kawashima}}]{HaradaSuzukiOkuboMatsuoLuoWatanabeTodoKawashima2013}%
  \BibitemOpen
  \bibfield  {author} {\bibinfo {author} {\bibfnamefont {K.}~\bibnamefont
  {Harada}}, \bibinfo {author} {\bibfnamefont {T.}~\bibnamefont {Suzuki}},
  \bibinfo {author} {\bibfnamefont {T.}~\bibnamefont {Okubo}}, \bibinfo
  {author} {\bibfnamefont {H.}~\bibnamefont {Matsuo}}, \bibinfo {author}
  {\bibfnamefont {J.}~\bibnamefont {Lou}}, \bibinfo {author} {\bibfnamefont
  {H.}~\bibnamefont {Watanabe}}, \bibinfo {author} {\bibfnamefont
  {S.}~\bibnamefont {Todo}}, \ and\ \bibinfo {author} {\bibfnamefont
  {N.}~\bibnamefont {Kawashima}},\ }\href {\doibase 10.1103/PhysRevB.88.220408}
  {\bibfield  {journal} {\bibinfo  {journal} {Phys. Rev. B}\ }\textbf {\bibinfo
  {volume} {88}},\ \bibinfo {pages} {220408} (\bibinfo {year}
  {2013})}\BibitemShut {NoStop}%
\bibitem [{\citenamefont {Jiang}\ \emph {et~al.}(2008)\citenamefont {Jiang},
  \citenamefont {Nyfeler}, \citenamefont {Chandrasekharan},\ and\ \citenamefont
  {Wiese}}]{JiangNyfelerChandrasekharanWises2008}%
  \BibitemOpen
  \bibfield  {author} {\bibinfo {author} {\bibfnamefont {F.-J.}\ \bibnamefont
  {Jiang}}, \bibinfo {author} {\bibfnamefont {M.}~\bibnamefont {Nyfeler}},
  \bibinfo {author} {\bibfnamefont {S.}~\bibnamefont {Chandrasekharan}}, \ and\
  \bibinfo {author} {\bibfnamefont {U.-J.}\ \bibnamefont {Wiese}},\ }\href@noop
  {} {\bibfield  {journal} {\bibinfo  {journal} {Journal of Statistical
  Mechanics: Theory and Experiment}\ }\textbf {\bibinfo {volume} {2008}},\
  \bibinfo {pages} {P02009} (\bibinfo {year} {2008})}\BibitemShut {NoStop}%
\bibitem [{\citenamefont {Chen}\ \emph {et~al.}(2013)\citenamefont {Chen},
  \citenamefont {Huang}, \citenamefont {Deng}, \citenamefont {Kuklov},
  \citenamefont {Prokof'ev},\ and\ \citenamefont
  {Svistunov}}]{ChenHuangDengKuklovProkofevSvistunov2013}%
  \BibitemOpen
  \bibfield  {author} {\bibinfo {author} {\bibfnamefont {K.}~\bibnamefont
  {Chen}}, \bibinfo {author} {\bibfnamefont {Y.}~\bibnamefont {Huang}},
  \bibinfo {author} {\bibfnamefont {Y.}~\bibnamefont {Deng}}, \bibinfo {author}
  {\bibfnamefont {A.~B.}\ \bibnamefont {Kuklov}}, \bibinfo {author}
  {\bibfnamefont {N.~V.}\ \bibnamefont {Prokof'ev}}, \ and\ \bibinfo {author}
  {\bibfnamefont {B.~V.}\ \bibnamefont {Svistunov}},\ }\href {\doibase
  10.1103/PhysRevLett.110.185701} {\bibfield  {journal} {\bibinfo  {journal}
  {Phys. Rev. Lett.}\ }\textbf {\bibinfo {volume} {110}},\ \bibinfo {pages}
  {185701} (\bibinfo {year} {2013})}\BibitemShut {NoStop}%
\bibitem [{\citenamefont {Nahum}\ \emph
  {et~al.}(2015{\natexlab{a}})\citenamefont {Nahum}, \citenamefont {Chalker},
  \citenamefont {Serna}, \citenamefont {Ortu\~no},\ and\ \citenamefont
  {Somoza}}]{NahumChalkerSernaOrtunoSomoza2015}%
  \BibitemOpen
  \bibfield  {author} {\bibinfo {author} {\bibfnamefont {A.}~\bibnamefont
  {Nahum}}, \bibinfo {author} {\bibfnamefont {J.~T.}\ \bibnamefont {Chalker}},
  \bibinfo {author} {\bibfnamefont {P.}~\bibnamefont {Serna}}, \bibinfo
  {author} {\bibfnamefont {M.}~\bibnamefont {Ortu\~no}}, \ and\ \bibinfo
  {author} {\bibfnamefont {A.~M.}\ \bibnamefont {Somoza}},\ }\href {\doibase
  10.1103/PhysRevX.5.041048} {\bibfield  {journal} {\bibinfo  {journal} {Phys.
  Rev. X}\ }\textbf {\bibinfo {volume} {5}},\ \bibinfo {pages} {041048}
  (\bibinfo {year} {2015}{\natexlab{a}})}\BibitemShut {NoStop}%
\bibitem [{\citenamefont {Nahum}\ \emph
  {et~al.}(2015{\natexlab{b}})\citenamefont {Nahum}, \citenamefont {Serna},
  \citenamefont {Chalker}, \citenamefont {Ortu\~no},\ and\ \citenamefont
  {Somoza}}]{NahumSernaChalkerOrtunoSomoza2015II}%
  \BibitemOpen
  \bibfield  {author} {\bibinfo {author} {\bibfnamefont {A.}~\bibnamefont
  {Nahum}}, \bibinfo {author} {\bibfnamefont {P.}~\bibnamefont {Serna}},
  \bibinfo {author} {\bibfnamefont {J.~T.}\ \bibnamefont {Chalker}}, \bibinfo
  {author} {\bibfnamefont {M.}~\bibnamefont {Ortu\~no}}, \ and\ \bibinfo
  {author} {\bibfnamefont {A.~M.}\ \bibnamefont {Somoza}},\ }\href {\doibase
  10.1103/PhysRevLett.115.267203} {\bibfield  {journal} {\bibinfo  {journal}
  {Phys. Rev. Lett.}\ }\textbf {\bibinfo {volume} {115}},\ \bibinfo {pages}
  {267203} (\bibinfo {year} {2015}{\natexlab{b}})}\BibitemShut {NoStop}%
\bibitem [{\citenamefont {Motrunich}\ and\ \citenamefont
  {Vishwanath}(1494)}]{MotrunichVishwanath2008}%
  \BibitemOpen
  \bibfield  {author} {\bibinfo {author} {\bibfnamefont {O.~I.}\ \bibnamefont
  {Motrunich}}\ and\ \bibinfo {author} {\bibfnamefont {A.}~\bibnamefont
  {Vishwanath}},\ }\href@noop {} {\  (\bibinfo {year}
  {arXiv:0805.1494})}\BibitemShut {NoStop}%
\bibitem [{\citenamefont {Kuklov}\ \emph {et~al.}(2008)\citenamefont {Kuklov},
  \citenamefont {Matsumoto}, \citenamefont {Prokof'ev}, \citenamefont
  {Svistunov},\ and\ \citenamefont
  {Troyer}}]{KuklovMatsumotoProkofevSvistunovTroyer2008}%
  \BibitemOpen
  \bibfield  {author} {\bibinfo {author} {\bibfnamefont {A.~B.}\ \bibnamefont
  {Kuklov}}, \bibinfo {author} {\bibfnamefont {M.}~\bibnamefont {Matsumoto}},
  \bibinfo {author} {\bibfnamefont {N.~V.}\ \bibnamefont {Prokof'ev}}, \bibinfo
  {author} {\bibfnamefont {B.~V.}\ \bibnamefont {Svistunov}}, \ and\ \bibinfo
  {author} {\bibfnamefont {M.}~\bibnamefont {Troyer}},\ }\href {\doibase
  10.1103/PhysRevLett.101.050405} {\bibfield  {journal} {\bibinfo  {journal}
  {Phys. Rev. Lett.}\ }\textbf {\bibinfo {volume} {101}},\ \bibinfo {pages}
  {050405} (\bibinfo {year} {2008})}\BibitemShut {NoStop}%
\bibitem [{\citenamefont {Bartosch}(2013)}]{Bartosch2013}%
  \BibitemOpen
  \bibfield  {author} {\bibinfo {author} {\bibfnamefont {L.}~\bibnamefont
  {Bartosch}},\ }\href {\doibase 10.1103/PhysRevB.88.195140} {\bibfield
  {journal} {\bibinfo  {journal} {Phys. Rev. B}\ }\textbf {\bibinfo {volume}
  {88}},\ \bibinfo {pages} {195140} (\bibinfo {year} {2013})}\BibitemShut
  {NoStop}%
\bibitem [{\citenamefont {Charrier}\ \emph {et~al.}(2008)\citenamefont
  {Charrier}, \citenamefont {Alet},\ and\ \citenamefont
  {Pujol}}]{CharrierAletPujol2008}%
  \BibitemOpen
  \bibfield  {author} {\bibinfo {author} {\bibfnamefont {D.}~\bibnamefont
  {Charrier}}, \bibinfo {author} {\bibfnamefont {F.}~\bibnamefont {Alet}}, \
  and\ \bibinfo {author} {\bibfnamefont {P.}~\bibnamefont {Pujol}},\ }\href
  {\doibase 10.1103/PhysRevLett.101.167205} {\bibfield  {journal} {\bibinfo
  {journal} {Phys. Rev. Lett.}\ }\textbf {\bibinfo {volume} {101}},\ \bibinfo
  {pages} {167205} (\bibinfo {year} {2008})}\BibitemShut {NoStop}%
\bibitem [{\citenamefont {Chen}\ \emph {et~al.}(2009)\citenamefont {Chen},
  \citenamefont {Gukelberger}, \citenamefont {Trebst}, \citenamefont {Alet},\
  and\ \citenamefont {Balents}}]{ChenGukelbergerTrebstFabienBalents2009}%
  \BibitemOpen
  \bibfield  {author} {\bibinfo {author} {\bibfnamefont {G.}~\bibnamefont
  {Chen}}, \bibinfo {author} {\bibfnamefont {J.}~\bibnamefont {Gukelberger}},
  \bibinfo {author} {\bibfnamefont {S.}~\bibnamefont {Trebst}}, \bibinfo
  {author} {\bibfnamefont {F.}~\bibnamefont {Alet}}, \ and\ \bibinfo {author}
  {\bibfnamefont {L.}~\bibnamefont {Balents}},\ }\href {\doibase
  10.1103/PhysRevB.80.045112} {\bibfield  {journal} {\bibinfo  {journal} {Phys.
  Rev. B}\ }\textbf {\bibinfo {volume} {80}},\ \bibinfo {pages} {045112}
  (\bibinfo {year} {2009})}\BibitemShut {NoStop}%
\bibitem [{\citenamefont {Charrier}\ and\ \citenamefont
  {Alet}(2010)}]{CharrierAlet2010}%
  \BibitemOpen
  \bibfield  {author} {\bibinfo {author} {\bibfnamefont {D.}~\bibnamefont
  {Charrier}}\ and\ \bibinfo {author} {\bibfnamefont {F.}~\bibnamefont
  {Alet}},\ }\href {\doibase 10.1103/PhysRevB.82.014429} {\bibfield  {journal}
  {\bibinfo  {journal} {Phys. Rev. B}\ }\textbf {\bibinfo {volume} {82}},\
  \bibinfo {pages} {014429} (\bibinfo {year} {2010})}\BibitemShut {NoStop}%
\bibitem [{\citenamefont {Sreejith}\ and\ \citenamefont
  {Powell}(2015)}]{SreejithPowell2015}%
  \BibitemOpen
  \bibfield  {author} {\bibinfo {author} {\bibfnamefont {G.~J.}\ \bibnamefont
  {Sreejith}}\ and\ \bibinfo {author} {\bibfnamefont {S.}~\bibnamefont
  {Powell}},\ }\href {\doibase 10.1103/PhysRevB.92.184413} {\bibfield
  {journal} {\bibinfo  {journal} {Phys. Rev. B}\ }\textbf {\bibinfo {volume}
  {92}},\ \bibinfo {pages} {184413} (\bibinfo {year} {2015})}\BibitemShut
  {NoStop}%
\bibitem [{\citenamefont {Shao}\ \emph {et~al.}(2016)\citenamefont {Shao},
  \citenamefont {Guo},\ and\ \citenamefont {Sandvik}}]{ShaoGuoSandvik2016}%
  \BibitemOpen
  \bibfield  {author} {\bibinfo {author} {\bibfnamefont {H.}~\bibnamefont
  {Shao}}, \bibinfo {author} {\bibfnamefont {W.}~\bibnamefont {Guo}}, \ and\
  \bibinfo {author} {\bibfnamefont {A.~W.}\ \bibnamefont {Sandvik}},\
  }\href@noop {} {\bibfield  {journal} {\bibinfo  {journal} {Science}\ }\textbf
  {\bibinfo {volume} {352}},\ \bibinfo {pages} {213} (\bibinfo {year}
  {2016})}\BibitemShut {NoStop}%
\bibitem [{\citenamefont {{Serna}}\ and\ \citenamefont
  {{Nahum}}(2018)}]{SernaNahum2018}%
  \BibitemOpen
  \bibfield  {author} {\bibinfo {author} {\bibfnamefont {P.}~\bibnamefont
  {{Serna}}}\ and\ \bibinfo {author} {\bibfnamefont {A.}~\bibnamefont
  {{Nahum}}},\ }\href@noop {} {\bibfield  {journal} {\bibinfo  {journal} {ArXiv
  e-prints}\ } (\bibinfo {year} {2018})},\ \Eprint
  {http://arxiv.org/abs/1805.03759} {arXiv:1805.03759 [cond-mat.str-el]}
  \BibitemShut {NoStop}%
\bibitem [{\citenamefont {Lieb}\ \emph {et~al.}(1961)\citenamefont {Lieb},
  \citenamefont {Schultz},\ and\ \citenamefont
  {Mattis}}]{LiebSchultzMattis1961}%
  \BibitemOpen
  \bibfield  {author} {\bibinfo {author} {\bibfnamefont {E.}~\bibnamefont
  {Lieb}}, \bibinfo {author} {\bibfnamefont {T.}~\bibnamefont {Schultz}}, \
  and\ \bibinfo {author} {\bibfnamefont {D.}~\bibnamefont {Mattis}},\
  }\href@noop {} {\bibfield  {journal} {\bibinfo  {journal} {Annals of
  Physics}\ }\textbf {\bibinfo {volume} {16}},\ \bibinfo {pages} {407}
  (\bibinfo {year} {1961})}\BibitemShut {NoStop}%
\bibitem [{\citenamefont {Oshikawa}(2000)}]{Oshikawa2000}%
  \BibitemOpen
  \bibfield  {author} {\bibinfo {author} {\bibfnamefont {M.}~\bibnamefont
  {Oshikawa}},\ }\href {\doibase 10.1103/PhysRevLett.84.1535} {\bibfield
  {journal} {\bibinfo  {journal} {Phys. Rev. Lett.}\ }\textbf {\bibinfo
  {volume} {84}},\ \bibinfo {pages} {1535} (\bibinfo {year}
  {2000})}\BibitemShut {NoStop}%
\bibitem [{\citenamefont {Hastings}(2004)}]{Hastings2004}%
  \BibitemOpen
  \bibfield  {author} {\bibinfo {author} {\bibfnamefont {M.~B.}\ \bibnamefont
  {Hastings}},\ }\href@noop {} {\bibfield  {journal} {\bibinfo  {journal}
  {Phys. Rev. B}\ }\textbf {\bibinfo {volume} {69}},\ \bibinfo {pages} {104431}
  (\bibinfo {year} {2004})}\BibitemShut {NoStop}%
\bibitem [{\citenamefont {Watanabe}\ \emph {et~al.}(2015)\citenamefont
  {Watanabe}, \citenamefont {Po}, \citenamefont {Vishwanath},\ and\
  \citenamefont {Zaletel}}]{WatanabePoVishwanathZaletel2015}%
  \BibitemOpen
  \bibfield  {author} {\bibinfo {author} {\bibfnamefont {H.}~\bibnamefont
  {Watanabe}}, \bibinfo {author} {\bibfnamefont {H.~C.}\ \bibnamefont {Po}},
  \bibinfo {author} {\bibfnamefont {A.}~\bibnamefont {Vishwanath}}, \ and\
  \bibinfo {author} {\bibfnamefont {M.}~\bibnamefont {Zaletel}},\ }\href@noop
  {} {\bibfield  {journal} {\bibinfo  {journal} {Proceedings of the National
  Academy of Sciences}\ }\textbf {\bibinfo {volume} {112}},\ \bibinfo {pages}
  {14551} (\bibinfo {year} {2015})}\BibitemShut {NoStop}%
\bibitem [{\citenamefont {Cheng}\ \emph {et~al.}(2016)\citenamefont {Cheng},
  \citenamefont {Zaletel}, \citenamefont {Barkeshli}, \citenamefont
  {Vishwanath},\ and\ \citenamefont
  {Bonderson}}]{ChengZaletelBarkeshliVishwanathBonderson2016}%
  \BibitemOpen
  \bibfield  {author} {\bibinfo {author} {\bibfnamefont {M.}~\bibnamefont
  {Cheng}}, \bibinfo {author} {\bibfnamefont {M.}~\bibnamefont {Zaletel}},
  \bibinfo {author} {\bibfnamefont {M.}~\bibnamefont {Barkeshli}}, \bibinfo
  {author} {\bibfnamefont {A.}~\bibnamefont {Vishwanath}}, \ and\ \bibinfo
  {author} {\bibfnamefont {P.}~\bibnamefont {Bonderson}},\ }\href {\doibase
  10.1103/PhysRevX.6.041068} {\bibfield  {journal} {\bibinfo  {journal} {Phys.
  Rev. X}\ }\textbf {\bibinfo {volume} {6}},\ \bibinfo {pages} {041068}
  (\bibinfo {year} {2016})}\BibitemShut {NoStop}%
\bibitem [{\citenamefont {Po}\ \emph {et~al.}(2017)\citenamefont {Po},
  \citenamefont {Watanabe}, \citenamefont {Jian},\ and\ \citenamefont
  {Zaletel}}]{PoWatanbeJianZaletel2017}%
  \BibitemOpen
  \bibfield  {author} {\bibinfo {author} {\bibfnamefont {H.~C.}\ \bibnamefont
  {Po}}, \bibinfo {author} {\bibfnamefont {H.}~\bibnamefont {Watanabe}},
  \bibinfo {author} {\bibfnamefont {C.-M.}\ \bibnamefont {Jian}}, \ and\
  \bibinfo {author} {\bibfnamefont {M.~P.}\ \bibnamefont {Zaletel}},\ }\href
  {\doibase 10.1103/PhysRevLett.119.127202} {\bibfield  {journal} {\bibinfo
  {journal} {Phys. Rev. Lett.}\ }\textbf {\bibinfo {volume} {119}},\ \bibinfo
  {pages} {127202} (\bibinfo {year} {2017})}\BibitemShut {NoStop}%
\bibitem [{\citenamefont {Qi}\ \emph {et~al.}(2017)\citenamefont {Qi},
  \citenamefont {Fang},\ and\ \citenamefont {Fu}}]{QiFangFu2017}%
  \BibitemOpen
  \bibfield  {author} {\bibinfo {author} {\bibfnamefont {Y.}~\bibnamefont
  {Qi}}, \bibinfo {author} {\bibfnamefont {C.}~\bibnamefont {Fang}}, \ and\
  \bibinfo {author} {\bibfnamefont {L.}~\bibnamefont {Fu}},\ }\href@noop {}
  {\bibfield  {journal} {\bibinfo  {journal} {arXiv preprint arXiv:1705.09190}\
  } (\bibinfo {year} {2017})}\BibitemShut {NoStop}%
\bibitem [{\citenamefont {Huang}\ \emph {et~al.}(2017)\citenamefont {Huang},
  \citenamefont {Song}, \citenamefont {Huang},\ and\ \citenamefont
  {Hermele}}]{Huang2017}%
  \BibitemOpen
  \bibfield  {author} {\bibinfo {author} {\bibfnamefont {S.-J.}\ \bibnamefont
  {Huang}}, \bibinfo {author} {\bibfnamefont {H.}~\bibnamefont {Song}},
  \bibinfo {author} {\bibfnamefont {Y.-P.}\ \bibnamefont {Huang}}, \ and\
  \bibinfo {author} {\bibfnamefont {M.}~\bibnamefont {Hermele}},\ }\href@noop
  {} {\bibfield  {journal} {\bibinfo  {journal} {Physical Review B}\ }\textbf
  {\bibinfo {volume} {96}},\ \bibinfo {pages} {205106} (\bibinfo {year}
  {2017})}\BibitemShut {NoStop}%
\bibitem [{\citenamefont {Lu}\ \emph {et~al.}(2017)\citenamefont {Lu},
  \citenamefont {Ran},\ and\ \citenamefont {Oshikawa}}]{LuRanOshikawa2017}%
  \BibitemOpen
  \bibfield  {author} {\bibinfo {author} {\bibfnamefont {Y.-M.}\ \bibnamefont
  {Lu}}, \bibinfo {author} {\bibfnamefont {Y.}~\bibnamefont {Ran}}, \ and\
  \bibinfo {author} {\bibfnamefont {M.}~\bibnamefont {Oshikawa}},\ }\href@noop
  {} {\bibfield  {journal} {\bibinfo  {journal} {arXiv preprint
  arXiv:1705.09298}\ } (\bibinfo {year} {2017})}\BibitemShut {NoStop}%
\bibitem [{\citenamefont {Lu}(2017)}]{Lu2017}%
  \BibitemOpen
  \bibfield  {author} {\bibinfo {author} {\bibfnamefont {Y.-M.}\ \bibnamefont
  {Lu}},\ }\href@noop {} {\bibfield  {journal} {\bibinfo  {journal} {arXiv
  preprint arXiv:1705.04691}\ } (\bibinfo {year} {2017})}\BibitemShut {NoStop}%
\bibitem [{\citenamefont {Yang}\ \emph {et~al.}(2017)\citenamefont {Yang},
  \citenamefont {Jiang}, \citenamefont {Vishwanath},\ and\ \citenamefont
  {Ran}}]{YangJiangVishwanathRan2017}%
  \BibitemOpen
  \bibfield  {author} {\bibinfo {author} {\bibfnamefont {X.}~\bibnamefont
  {Yang}}, \bibinfo {author} {\bibfnamefont {S.}~\bibnamefont {Jiang}},
  \bibinfo {author} {\bibfnamefont {A.}~\bibnamefont {Vishwanath}}, \ and\
  \bibinfo {author} {\bibfnamefont {Y.}~\bibnamefont {Ran}},\ }\href@noop {}
  {\bibfield  {journal} {\bibinfo  {journal} {arXiv preprint arXiv:1705.05421}\
  } (\bibinfo {year} {2017})}\BibitemShut {NoStop}%
\bibitem [{\citenamefont {{Komargodski}}\ \emph {et~al.}(2017)\citenamefont
  {{Komargodski}}, \citenamefont {{Sharon}}, \citenamefont {{Thorngren}},\ and\
  \citenamefont {{Zhou}}}]{KomargodskiSharonThorngrenZhou2017}%
  \BibitemOpen
  \bibfield  {author} {\bibinfo {author} {\bibfnamefont {Z.}~\bibnamefont
  {{Komargodski}}}, \bibinfo {author} {\bibfnamefont {A.}~\bibnamefont
  {{Sharon}}}, \bibinfo {author} {\bibfnamefont {R.}~\bibnamefont
  {{Thorngren}}}, \ and\ \bibinfo {author} {\bibfnamefont {X.}~\bibnamefont
  {{Zhou}}},\ }\href@noop {} {\bibfield  {journal} {\bibinfo  {journal} {ArXiv
  e-prints}\ } (\bibinfo {year} {2017})},\ \Eprint
  {http://arxiv.org/abs/1705.04786} {arXiv:1705.04786 [hep-th]} \BibitemShut
  {NoStop}%
\bibitem [{\citenamefont {Komargodski}\ \emph {et~al.}(2018)\citenamefont
  {Komargodski}, \citenamefont {Sulejmanpasic},\ and\ \citenamefont
  {\"Unsal}}]{KomargodskiSulejmanpasicUnsal2018}%
  \BibitemOpen
  \bibfield  {author} {\bibinfo {author} {\bibfnamefont {Z.}~\bibnamefont
  {Komargodski}}, \bibinfo {author} {\bibfnamefont {T.}~\bibnamefont
  {Sulejmanpasic}}, \ and\ \bibinfo {author} {\bibfnamefont {M.}~\bibnamefont
  {\"Unsal}},\ }\href {\doibase 10.1103/PhysRevB.97.054418} {\bibfield
  {journal} {\bibinfo  {journal} {Phys. Rev. B}\ }\textbf {\bibinfo {volume}
  {97}},\ \bibinfo {pages} {054418} (\bibinfo {year} {2018})}\BibitemShut
  {NoStop}%
\bibitem [{\citenamefont {{Metlitski}}\ and\ \citenamefont
  {{Thorngren}}(2017)}]{MetlitskiThorngren2017}%
  \BibitemOpen
  \bibfield  {author} {\bibinfo {author} {\bibfnamefont {M.~A.}\ \bibnamefont
  {{Metlitski}}}\ and\ \bibinfo {author} {\bibfnamefont {R.}~\bibnamefont
  {{Thorngren}}},\ }\href@noop {} {\bibfield  {journal} {\bibinfo  {journal}
  {ArXiv e-prints}\ } (\bibinfo {year} {2017})},\ \Eprint
  {http://arxiv.org/abs/1707.07686} {arXiv:1707.07686 [cond-mat.str-el]}
  \BibitemShut {NoStop}%
\bibitem [{\citenamefont {Sulejmanpasic}\ and\ \citenamefont
  {Tanizaki}(2018)}]{SulejmanpasicTanizaki2018}%
  \BibitemOpen
  \bibfield  {author} {\bibinfo {author} {\bibfnamefont {T.}~\bibnamefont
  {Sulejmanpasic}}\ and\ \bibinfo {author} {\bibfnamefont {Y.}~\bibnamefont
  {Tanizaki}},\ }\href {\doibase 10.1103/PhysRevB.97.144201} {\bibfield
  {journal} {\bibinfo  {journal} {Phys. Rev. B}\ }\textbf {\bibinfo {volume}
  {97}},\ \bibinfo {pages} {144201} (\bibinfo {year} {2018})}\BibitemShut
  {NoStop}%
\bibitem [{\citenamefont {{Tanizaki}}\ and\ \citenamefont
  {{Sulejmanpasic}}(2018)}]{TanizakiSulejmanpasic2018arXiv}%
  \BibitemOpen
  \bibfield  {author} {\bibinfo {author} {\bibfnamefont {Y.}~\bibnamefont
  {{Tanizaki}}}\ and\ \bibinfo {author} {\bibfnamefont {T.}~\bibnamefont
  {{Sulejmanpasic}}},\ }\href@noop {} {\bibfield  {journal} {\bibinfo
  {journal} {ArXiv e-prints}\ } (\bibinfo {year} {2018})},\ \Eprint
  {http://arxiv.org/abs/1805.11423} {arXiv:1805.11423 [cond-mat.str-el]}
  \BibitemShut {NoStop}%
\bibitem [{\citenamefont {Majumdar}\ and\ \citenamefont
  {Ghosh}(1969{\natexlab{a}})}]{MajumdarGhosh1969}%
  \BibitemOpen
  \bibfield  {author} {\bibinfo {author} {\bibfnamefont {C.~K.}\ \bibnamefont
  {Majumdar}}\ and\ \bibinfo {author} {\bibfnamefont {D.~K.}\ \bibnamefont
  {Ghosh}},\ }\href@noop {} {\bibfield  {journal} {\bibinfo  {journal} {Journal
  of Mathematical Physics}\ }\textbf {\bibinfo {volume} {10}},\ \bibinfo
  {pages} {1388} (\bibinfo {year} {1969}{\natexlab{a}})}\BibitemShut {NoStop}%
\bibitem [{\citenamefont {Majumdar}\ and\ \citenamefont
  {Ghosh}(1969{\natexlab{b}})}]{MajumdarGhosh1969II}%
  \BibitemOpen
  \bibfield  {author} {\bibinfo {author} {\bibfnamefont {C.~K.}\ \bibnamefont
  {Majumdar}}\ and\ \bibinfo {author} {\bibfnamefont {D.~K.}\ \bibnamefont
  {Ghosh}},\ }\href@noop {} {\bibfield  {journal} {\bibinfo  {journal} {Journal
  of Mathematical Physics}\ }\textbf {\bibinfo {volume} {10}},\ \bibinfo
  {pages} {1399} (\bibinfo {year} {1969}{\natexlab{b}})}\BibitemShut {NoStop}%
\bibitem [{\citenamefont {Haldane}(1982)}]{Haldane1982}%
  \BibitemOpen
  \bibfield  {author} {\bibinfo {author} {\bibfnamefont {F.~D.~M.}\
  \bibnamefont {Haldane}},\ }\href {\doibase 10.1103/PhysRevB.25.4925}
  {\bibfield  {journal} {\bibinfo  {journal} {Phys. Rev. B}\ }\textbf {\bibinfo
  {volume} {25}},\ \bibinfo {pages} {4925} (\bibinfo {year}
  {1982})}\BibitemShut {NoStop}%
\bibitem [{\citenamefont {{Kosterlitz}}\ and\ \citenamefont
  {{Thouless}}(1973)}]{KosterlitzThouless1973}%
  \BibitemOpen
  \bibfield  {author} {\bibinfo {author} {\bibfnamefont {J.~M.}\ \bibnamefont
  {{Kosterlitz}}}\ and\ \bibinfo {author} {\bibfnamefont {D.~J.}\ \bibnamefont
  {{Thouless}}},\ }\href {\doibase 10.1088/0022-3719/6/7/010} {\bibfield
  {journal} {\bibinfo  {journal} {Journal of Physics C Solid State Physics}\
  }\textbf {\bibinfo {volume} {6}},\ \bibinfo {pages} {1181} (\bibinfo {year}
  {1973})}\BibitemShut {NoStop}%
\bibitem [{\citenamefont {{Kosterlitz}}(1974)}]{Kosterlitz1974}%
  \BibitemOpen
  \bibfield  {author} {\bibinfo {author} {\bibfnamefont {J.~M.}\ \bibnamefont
  {{Kosterlitz}}},\ }\href {\doibase 10.1088/0022-3719/7/6/005} {\bibfield
  {journal} {\bibinfo  {journal} {Journal of Physics C Solid State Physics}\
  }\textbf {\bibinfo {volume} {7}},\ \bibinfo {pages} {1046} (\bibinfo {year}
  {1974})}\BibitemShut {NoStop}%
\bibitem [{\citenamefont {Jos\'e}\ \emph {et~al.}(1977)\citenamefont {Jos\'e},
  \citenamefont {Kadanoff}, \citenamefont {Kirkpatrick},\ and\ \citenamefont
  {Nelson}}]{JoseKadanoffKirkpatrickNelson1977}%
  \BibitemOpen
  \bibfield  {author} {\bibinfo {author} {\bibfnamefont {J.~V.}\ \bibnamefont
  {Jos\'e}}, \bibinfo {author} {\bibfnamefont {L.~P.}\ \bibnamefont
  {Kadanoff}}, \bibinfo {author} {\bibfnamefont {S.}~\bibnamefont
  {Kirkpatrick}}, \ and\ \bibinfo {author} {\bibfnamefont {D.~R.}\ \bibnamefont
  {Nelson}},\ }\href {\doibase 10.1103/PhysRevB.16.1217} {\bibfield  {journal}
  {\bibinfo  {journal} {Phys. Rev. B}\ }\textbf {\bibinfo {volume} {16}},\
  \bibinfo {pages} {1217} (\bibinfo {year} {1977})}\BibitemShut {NoStop}%
\bibitem [{\citenamefont {Lecheminant}\ \emph {et~al.}(2002)\citenamefont
  {Lecheminant}, \citenamefont {Gogolin},\ and\ \citenamefont
  {Nersesyan}}]{LecheminantGogolinNersesyan2002}%
  \BibitemOpen
  \bibfield  {author} {\bibinfo {author} {\bibfnamefont {P.}~\bibnamefont
  {Lecheminant}}, \bibinfo {author} {\bibfnamefont {A.~O.}\ \bibnamefont
  {Gogolin}}, \ and\ \bibinfo {author} {\bibfnamefont {A.~A.}\ \bibnamefont
  {Nersesyan}},\ }\href {\doibase
  https://doi.org/10.1016/S0550-3213(02)00474-1} {\bibfield  {journal}
  {\bibinfo  {journal} {Nuclear Physics B}\ }\textbf {\bibinfo {volume}
  {639}},\ \bibinfo {pages} {502 } (\bibinfo {year} {2002})}\BibitemShut
  {NoStop}%
\bibitem [{\citenamefont {Peskin}(1978)}]{Peskin1978}%
  \BibitemOpen
  \bibfield  {author} {\bibinfo {author} {\bibfnamefont {M.~E.}\ \bibnamefont
  {Peskin}},\ }\href@noop {} {\bibfield  {journal} {\bibinfo  {journal} {Annals
  of Physics}\ }\textbf {\bibinfo {volume} {113}},\ \bibinfo {pages} {122}
  (\bibinfo {year} {1978})}\BibitemShut {NoStop}%
\bibitem [{\citenamefont {Dasgupta}\ and\ \citenamefont
  {Halperin}(1981)}]{DasguptaHalperin1981}%
  \BibitemOpen
  \bibfield  {author} {\bibinfo {author} {\bibfnamefont {C.}~\bibnamefont
  {Dasgupta}}\ and\ \bibinfo {author} {\bibfnamefont {B.~I.}\ \bibnamefont
  {Halperin}},\ }\href {\doibase 10.1103/PhysRevLett.47.1556} {\bibfield
  {journal} {\bibinfo  {journal} {Phys. Rev. Lett.}\ }\textbf {\bibinfo
  {volume} {47}},\ \bibinfo {pages} {1556} (\bibinfo {year}
  {1981})}\BibitemShut {NoStop}%
\bibitem [{\citenamefont {Fisher}\ and\ \citenamefont
  {Lee}(1989)}]{FisherLee1989}%
  \BibitemOpen
  \bibfield  {author} {\bibinfo {author} {\bibfnamefont {M.~P.~A.}\
  \bibnamefont {Fisher}}\ and\ \bibinfo {author} {\bibfnamefont {D.~H.}\
  \bibnamefont {Lee}},\ }\href {\doibase 10.1103/PhysRevB.39.2756} {\bibfield
  {journal} {\bibinfo  {journal} {Phys. Rev. B}\ }\textbf {\bibinfo {volume}
  {39}},\ \bibinfo {pages} {2756} (\bibinfo {year} {1989})}\BibitemShut
  {NoStop}%
\bibitem [{\citenamefont {Lannert}\ \emph {et~al.}(2001)\citenamefont
  {Lannert}, \citenamefont {Fisher},\ and\ \citenamefont
  {Senthil}}]{LannertFisherSenthil2001}%
  \BibitemOpen
  \bibfield  {author} {\bibinfo {author} {\bibfnamefont {C.}~\bibnamefont
  {Lannert}}, \bibinfo {author} {\bibfnamefont {M.~P.~A.}\ \bibnamefont
  {Fisher}}, \ and\ \bibinfo {author} {\bibfnamefont {T.}~\bibnamefont
  {Senthil}},\ }\href {\doibase 10.1103/PhysRevB.63.134510} {\bibfield
  {journal} {\bibinfo  {journal} {Phys. Rev. B}\ }\textbf {\bibinfo {volume}
  {63}},\ \bibinfo {pages} {134510} (\bibinfo {year} {2001})}\BibitemShut
  {NoStop}%
\bibitem [{\citenamefont {Ashkin}\ and\ \citenamefont
  {Teller}(1943)}]{AshkinTeller1943}%
  \BibitemOpen
  \bibfield  {author} {\bibinfo {author} {\bibfnamefont {J.}~\bibnamefont
  {Ashkin}}\ and\ \bibinfo {author} {\bibfnamefont {E.}~\bibnamefont
  {Teller}},\ }\href {\doibase 10.1103/PhysRev.64.178} {\bibfield  {journal}
  {\bibinfo  {journal} {Phys. Rev.}\ }\textbf {\bibinfo {volume} {64}},\
  \bibinfo {pages} {178} (\bibinfo {year} {1943})}\BibitemShut {NoStop}%
\bibitem [{\citenamefont {Kadanoff}\ and\ \citenamefont
  {Wegner}(1971)}]{KadanoffWegner1971}%
  \BibitemOpen
  \bibfield  {author} {\bibinfo {author} {\bibfnamefont {L.~P.}\ \bibnamefont
  {Kadanoff}}\ and\ \bibinfo {author} {\bibfnamefont {F.~J.}\ \bibnamefont
  {Wegner}},\ }\href {\doibase 10.1103/PhysRevB.4.3989} {\bibfield  {journal}
  {\bibinfo  {journal} {Phys. Rev. B}\ }\textbf {\bibinfo {volume} {4}},\
  \bibinfo {pages} {3989} (\bibinfo {year} {1971})}\BibitemShut {NoStop}%
\bibitem [{\citenamefont {Kadanoff}(1979)}]{Kadanoff1979}%
  \BibitemOpen
  \bibfield  {author} {\bibinfo {author} {\bibfnamefont {L.~P.}\ \bibnamefont
  {Kadanoff}},\ }\href {\doibase https://doi.org/10.1016/0003-4916(79)90280-X}
  {\bibfield  {journal} {\bibinfo  {journal} {Annals of Physics}\ }\textbf
  {\bibinfo {volume} {120}},\ \bibinfo {pages} {39 } (\bibinfo {year}
  {1979})}\BibitemShut {NoStop}%
\bibitem [{\citenamefont {Kohmoto}\ \emph {et~al.}(1981)\citenamefont
  {Kohmoto}, \citenamefont {den Nijs},\ and\ \citenamefont
  {Kadanoff}}]{KohmotoNijsKadanoff}%
  \BibitemOpen
  \bibfield  {author} {\bibinfo {author} {\bibfnamefont {M.}~\bibnamefont
  {Kohmoto}}, \bibinfo {author} {\bibfnamefont {M.}~\bibnamefont {den Nijs}}, \
  and\ \bibinfo {author} {\bibfnamefont {L.~P.}\ \bibnamefont {Kadanoff}},\
  }\href {\doibase 10.1103/PhysRevB.24.5229} {\bibfield  {journal} {\bibinfo
  {journal} {Phys. Rev. B}\ }\textbf {\bibinfo {volume} {24}},\ \bibinfo
  {pages} {5229} (\bibinfo {year} {1981})}\BibitemShut {NoStop}%
\bibitem [{\citenamefont {Delfino}(1999)}]{Delfino1999}%
  \BibitemOpen
  \bibfield  {author} {\bibinfo {author} {\bibfnamefont {G.}~\bibnamefont
  {Delfino}},\ }\href {\doibase https://doi.org/10.1016/S0370-2693(99)00123-9}
  {\bibfield  {journal} {\bibinfo  {journal} {Physics Letters B}\ }\textbf
  {\bibinfo {volume} {450}},\ \bibinfo {pages} {196 } (\bibinfo {year}
  {1999})}\BibitemShut {NoStop}%
\bibitem [{\citenamefont {Ramola}\ \emph {et~al.}(2015)\citenamefont {Ramola},
  \citenamefont {Damle},\ and\ \citenamefont {Dhar}}]{RamolaDamleDhar2015}%
  \BibitemOpen
  \bibfield  {author} {\bibinfo {author} {\bibfnamefont {K.}~\bibnamefont
  {Ramola}}, \bibinfo {author} {\bibfnamefont {K.}~\bibnamefont {Damle}}, \
  and\ \bibinfo {author} {\bibfnamefont {D.}~\bibnamefont {Dhar}},\ }\href
  {\doibase 10.1103/PhysRevLett.114.190601} {\bibfield  {journal} {\bibinfo
  {journal} {Phys. Rev. Lett.}\ }\textbf {\bibinfo {volume} {114}},\ \bibinfo
  {pages} {190601} (\bibinfo {year} {2015})}\BibitemShut {NoStop}%
\bibitem [{\citenamefont {{Chew}}\ \emph {et~al.}(2018)\citenamefont {{Chew}},
  \citenamefont {{Mross}},\ and\ \citenamefont
  {{Alicea}}}]{ChewMrossAlicea2018}%
  \BibitemOpen
  \bibfield  {author} {\bibinfo {author} {\bibfnamefont {A.}~\bibnamefont
  {{Chew}}}, \bibinfo {author} {\bibfnamefont {D.~F.}\ \bibnamefont {{Mross}}},
  \ and\ \bibinfo {author} {\bibfnamefont {J.}~\bibnamefont {{Alicea}}},\
  }\href@noop {} {\bibfield  {journal} {\bibinfo  {journal} {ArXiv e-prints}\ }
  (\bibinfo {year} {2018})},\ \Eprint {http://arxiv.org/abs/1802.04809}
  {arXiv:1802.04809 [cond-mat.str-el]} \BibitemShut {NoStop}%
\bibitem [{\citenamefont {Motrunich}\ and\ \citenamefont
  {Vishwanath}(2004)}]{MotrunichVishwanath2004}%
  \BibitemOpen
  \bibfield  {author} {\bibinfo {author} {\bibfnamefont {O.~I.}\ \bibnamefont
  {Motrunich}}\ and\ \bibinfo {author} {\bibfnamefont {A.}~\bibnamefont
  {Vishwanath}},\ }\href@noop {} {\bibfield  {journal} {\bibinfo  {journal}
  {Phys. Rev. B}\ }\textbf {\bibinfo {volume} {70}},\ \bibinfo {pages} {075104}
  (\bibinfo {year} {2004})}\BibitemShut {NoStop}%
\bibitem [{\citenamefont {Tanaka}\ and\ \citenamefont
  {Hu}(2005)}]{TanakaHu2005}%
  \BibitemOpen
  \bibfield  {author} {\bibinfo {author} {\bibfnamefont {A.}~\bibnamefont
  {Tanaka}}\ and\ \bibinfo {author} {\bibfnamefont {X.}~\bibnamefont {Hu}},\
  }\href {\doibase 10.1103/PhysRevLett.95.036402} {\bibfield  {journal}
  {\bibinfo  {journal} {Phys. Rev. Lett.}\ }\textbf {\bibinfo {volume} {95}},\
  \bibinfo {pages} {036402} (\bibinfo {year} {2005})}\BibitemShut {NoStop}%
\bibitem [{\citenamefont {Senthil}\ and\ \citenamefont
  {Fisher}(2006)}]{SenthilFisher2006}%
  \BibitemOpen
  \bibfield  {author} {\bibinfo {author} {\bibfnamefont {T.}~\bibnamefont
  {Senthil}}\ and\ \bibinfo {author} {\bibfnamefont {M.~P.~A.}\ \bibnamefont
  {Fisher}},\ }\href@noop {} {\bibfield  {journal} {\bibinfo  {journal} {Phys.
  Rev. B}\ }\textbf {\bibinfo {volume} {74}},\ \bibinfo {pages} {064405}
  (\bibinfo {year} {2006})}\BibitemShut {NoStop}%
\bibitem [{\citenamefont {Sachdev}(2011)}]{SachdevBook}%
  \BibitemOpen
  \bibfield  {author} {\bibinfo {author} {\bibfnamefont {S.}~\bibnamefont
  {Sachdev}},\ }\href@noop {} {\emph {\bibinfo {title} {Quantum Phase
  Transitions}}},\ \bibinfo {edition} {second edition}\ ed.\ (\bibinfo
  {publisher} {Cambridge University Press},\ \bibinfo {year}
  {2011})\BibitemShut {NoStop}%
\bibitem [{\citenamefont {Furukawa}\ \emph {et~al.}(2010)\citenamefont
  {Furukawa}, \citenamefont {Sato},\ and\ \citenamefont
  {Furusaki}}]{FurukawaSatoFurusaki2010}%
  \BibitemOpen
  \bibfield  {author} {\bibinfo {author} {\bibfnamefont {S.}~\bibnamefont
  {Furukawa}}, \bibinfo {author} {\bibfnamefont {M.}~\bibnamefont {Sato}}, \
  and\ \bibinfo {author} {\bibfnamefont {A.}~\bibnamefont {Furusaki}},\ }\href
  {\doibase 10.1103/PhysRevB.81.094430} {\bibfield  {journal} {\bibinfo
  {journal} {Phys. Rev. B}\ }\textbf {\bibinfo {volume} {81}},\ \bibinfo
  {pages} {094430} (\bibinfo {year} {2010})}\BibitemShut {NoStop}%
\bibitem [{\citenamefont {Furukawa}\ \emph {et~al.}(2012)\citenamefont
  {Furukawa}, \citenamefont {Sato}, \citenamefont {Onoda},\ and\ \citenamefont
  {Furusaki}}]{FurukawaSatoOnodaFurusaki2012}%
  \BibitemOpen
  \bibfield  {author} {\bibinfo {author} {\bibfnamefont {S.}~\bibnamefont
  {Furukawa}}, \bibinfo {author} {\bibfnamefont {M.}~\bibnamefont {Sato}},
  \bibinfo {author} {\bibfnamefont {S.}~\bibnamefont {Onoda}}, \ and\ \bibinfo
  {author} {\bibfnamefont {A.}~\bibnamefont {Furusaki}},\ }\href {\doibase
  10.1103/PhysRevB.86.094417} {\bibfield  {journal} {\bibinfo  {journal} {Phys.
  Rev. B}\ }\textbf {\bibinfo {volume} {86}},\ \bibinfo {pages} {094417}
  (\bibinfo {year} {2012})}\BibitemShut {NoStop}%
\bibitem [{\citenamefont {Xu}\ and\ \citenamefont {You}(2015)}]{XuYou2015}%
  \BibitemOpen
  \bibfield  {author} {\bibinfo {author} {\bibfnamefont {C.}~\bibnamefont
  {Xu}}\ and\ \bibinfo {author} {\bibfnamefont {Y.-Z.}\ \bibnamefont {You}},\
  }\href {\doibase 10.1103/PhysRevB.92.220416} {\bibfield  {journal} {\bibinfo
  {journal} {Phys. Rev. B}\ }\textbf {\bibinfo {volume} {92}},\ \bibinfo
  {pages} {220416} (\bibinfo {year} {2015})}\BibitemShut {NoStop}%
\bibitem [{\citenamefont {Karch}\ and\ \citenamefont
  {Tong}(2016)}]{KarchTong2016}%
  \BibitemOpen
  \bibfield  {author} {\bibinfo {author} {\bibfnamefont {A.}~\bibnamefont
  {Karch}}\ and\ \bibinfo {author} {\bibfnamefont {D.}~\bibnamefont {Tong}},\
  }\href {\doibase 10.1103/PhysRevX.6.031043} {\bibfield  {journal} {\bibinfo
  {journal} {Phys. Rev. X}\ }\textbf {\bibinfo {volume} {6}},\ \bibinfo {pages}
  {031043} (\bibinfo {year} {2016})}\BibitemShut {NoStop}%
\bibitem [{\citenamefont {Seiberg}\ \emph {et~al.}(2016)\citenamefont
  {Seiberg}, \citenamefont {Senthil}, \citenamefont {Wang},\ and\ \citenamefont
  {Witten}}]{SeibergSenthilWangWitten2016}%
  \BibitemOpen
  \bibfield  {author} {\bibinfo {author} {\bibfnamefont {N.}~\bibnamefont
  {Seiberg}}, \bibinfo {author} {\bibfnamefont {T.}~\bibnamefont {Senthil}},
  \bibinfo {author} {\bibfnamefont {C.}~\bibnamefont {Wang}}, \ and\ \bibinfo
  {author} {\bibfnamefont {E.}~\bibnamefont {Witten}},\ }\href {\doibase
  https://doi.org/10.1016/j.aop.2016.08.007} {\bibfield  {journal} {\bibinfo
  {journal} {Annals of Physics}\ }\textbf {\bibinfo {volume} {374}},\ \bibinfo
  {pages} {395 } (\bibinfo {year} {2016})}\BibitemShut {NoStop}%
\bibitem [{\citenamefont {Wang}\ \emph {et~al.}(2017)\citenamefont {Wang},
  \citenamefont {Nahum}, \citenamefont {Metlitski}, \citenamefont {Xu},\ and\
  \citenamefont {Senthil}}]{WangNahumMetlitskiXuSenthil2017}%
  \BibitemOpen
  \bibfield  {author} {\bibinfo {author} {\bibfnamefont {C.}~\bibnamefont
  {Wang}}, \bibinfo {author} {\bibfnamefont {A.}~\bibnamefont {Nahum}},
  \bibinfo {author} {\bibfnamefont {M.~A.}\ \bibnamefont {Metlitski}}, \bibinfo
  {author} {\bibfnamefont {C.}~\bibnamefont {Xu}}, \ and\ \bibinfo {author}
  {\bibfnamefont {T.}~\bibnamefont {Senthil}},\ }\href {\doibase
  10.1103/PhysRevX.7.031051} {\bibfield  {journal} {\bibinfo  {journal} {Phys.
  Rev. X}\ }\textbf {\bibinfo {volume} {7}},\ \bibinfo {pages} {031051}
  (\bibinfo {year} {2017})}\BibitemShut {NoStop}%
\bibitem [{\citenamefont {Mross}\ \emph {et~al.}(2017)\citenamefont {Mross},
  \citenamefont {Alicea},\ and\ \citenamefont
  {Motrunich}}]{MrossAliceaMotrunich2017}%
  \BibitemOpen
  \bibfield  {author} {\bibinfo {author} {\bibfnamefont {D.~F.}\ \bibnamefont
  {Mross}}, \bibinfo {author} {\bibfnamefont {J.}~\bibnamefont {Alicea}}, \
  and\ \bibinfo {author} {\bibfnamefont {O.~I.}\ \bibnamefont {Motrunich}},\
  }\href {\doibase 10.1103/PhysRevX.7.041016} {\bibfield  {journal} {\bibinfo
  {journal} {Phys. Rev. X}\ }\textbf {\bibinfo {volume} {7}},\ \bibinfo {pages}
  {041016} (\bibinfo {year} {2017})}\BibitemShut {NoStop}%
\bibitem [{\citenamefont {Haldane}(1981)}]{Haldane1981}%
  \BibitemOpen
  \bibfield  {author} {\bibinfo {author} {\bibfnamefont {F.~D.~M.}\
  \bibnamefont {Haldane}},\ }\href {\doibase 10.1103/PhysRevLett.47.1840}
  {\bibfield  {journal} {\bibinfo  {journal} {Phys. Rev. Lett.}\ }\textbf
  {\bibinfo {volume} {47}},\ \bibinfo {pages} {1840} (\bibinfo {year}
  {1981})}\BibitemShut {NoStop}%
\bibitem [{\citenamefont {Shankar}(1995)}]{Shankar_Acta}%
  \BibitemOpen
  \bibfield  {author} {\bibinfo {author} {\bibfnamefont {R.}~\bibnamefont
  {Shankar}},\ }\href@noop {} {\bibfield  {journal} {\bibinfo  {journal} {Acta
  Phys. Pol. B}\ }\textbf {\bibinfo {volume} {26}},\ \bibinfo {pages} {1835}
  (\bibinfo {year} {1995})}\BibitemShut {NoStop}%
\bibitem [{\citenamefont {Giamarchi}(2004)}]{GiamarchiBook}%
  \BibitemOpen
  \bibfield  {author} {\bibinfo {author} {\bibfnamefont {T.}~\bibnamefont
  {Giamarchi}},\ }\href@noop {} {\emph {\bibinfo {title} {Quantum Physics in
  One Dimension}}}\ (\bibinfo  {publisher} {Clarendon Press, Oxford},\ \bibinfo
  {year} {2004})\BibitemShut {NoStop}%
\bibitem [{Note1()}]{Note1}%
  \BibitemOpen
  \bibinfo {note} {We point out that $\DOTSB \sum@ \slimits@ _j (-1)^j \protect
  \qopname \relax o{cos}(2\theta _{j+1/2})$ is allowed on the lattice. However,
  due to the staggered phase, this sum has rapid oscillations (assuming small
  coupling and hence slowly-varying field $\theta $) and hence disappears in
  the continuum.}\BibitemShut {Stop}%
\bibitem [{Note2()}]{Note2}%
  \BibitemOpen
  \bibinfo {note} {For finite-length chains, there are two gauge-inequivalent
  sectors: $\DOTSB \prod@ \slimits@ _j \rho ^z_j = 1$ and $-1$. For the
  nontrivial gauge flux sector, it is impossible to find a uniform $\rho ^z_j$
  configuration, and thus $T_x$ should be defined differently from Eq.~(\ref
  {eq:dw_sym}).}\BibitemShut {Stop}%
\bibitem [{\citenamefont {Wen}(7071)}]{WenPSG}%
  \BibitemOpen
  \bibfield  {author} {\bibinfo {author} {\bibfnamefont {X.-G.}\ \bibnamefont
  {Wen}},\ }\href {\doibase 10.1103/PhysRevB.65.165113} {\bibfield  {journal}
  {\bibinfo  {journal} {Phys. Rev. B}\ }\textbf {\bibinfo {volume} {65}},\
  \bibinfo {pages} {165113} (\bibinfo {year} {2002,
  cond-mat/0107071})}\BibitemShut {NoStop}%
\bibitem [{\citenamefont {Motrunich}\ and\ \citenamefont
  {Senthil}(2005)}]{MotrunichSenthil2005}%
  \BibitemOpen
  \bibfield  {author} {\bibinfo {author} {\bibfnamefont {O.~I.}\ \bibnamefont
  {Motrunich}}\ and\ \bibinfo {author} {\bibfnamefont {T.}~\bibnamefont
  {Senthil}},\ }\href {\doibase 10.1103/PhysRevB.71.125102} {\bibfield
  {journal} {\bibinfo  {journal} {Phys. Rev. B}\ }\textbf {\bibinfo {volume}
  {71}},\ \bibinfo {pages} {125102} (\bibinfo {year} {2005})}\BibitemShut
  {NoStop}%
\bibitem [{\citenamefont {Senthil}\ and\ \citenamefont
  {Fisher}(2000)}]{SenthilFisher2000}%
  \BibitemOpen
  \bibfield  {author} {\bibinfo {author} {\bibfnamefont {T.}~\bibnamefont
  {Senthil}}\ and\ \bibinfo {author} {\bibfnamefont {M.~P.~A.}\ \bibnamefont
  {Fisher}},\ }\href {\doibase 10.1103/PhysRevB.62.7850} {\bibfield  {journal}
  {\bibinfo  {journal} {Phys. Rev. B}\ }\textbf {\bibinfo {volume} {62}},\
  \bibinfo {pages} {7850} (\bibinfo {year} {2000})}\BibitemShut {NoStop}%
\bibitem [{Note3()}]{Note3}%
  \BibitemOpen
  \bibinfo {note} {Strictly speaking, it is possible to have a $U(1)$ gauge
  group rather than $Z_2^\zeta $. However, for our purposes here, we always
  allow Higgs terms to break the $U(1)$ gauge group to $Z_2^\zeta
  $.}\BibitemShut {Stop}%
\bibitem [{Note4()}]{Note4}%
  \BibitemOpen
  \bibinfo {note} {A perturbative scheme where we treat the $\Gamma $ and $h$
  terms in Eq.~(\ref {eq:model_bosonic_parton}) as an unperturbed Hamiltonian
  and the $J$ term as a perturbation, assuming $\Gamma > 2h$ [so that the
  unperturbed ground state satisfies the bare parton constraints in Eq.~(\ref
  {eq:bosonic_parton_constraint})] and $\Gamma - 2h \gg J$, gives the following
  spin Hamiltonian at second order in $J$: \begin {align} H_\protect \text
  {spin} = \DOTSB \sum@ \slimits@ _j \protect \frac {-J^2 \Gamma }{\Gamma ^2 -
  2 h^2 (1 + \sigma ^x_j \sigma ^x_{j+1})} (1 + \sigma ^z_j \sigma ^z_{j+1}) ~.
  \end {align} This Hamiltonian wants to have ferromagnetic nearest-neighbor
  $\sigma ^z \sigma ^z$ and $\sigma ^x \sigma ^x$ correlations, akin to the
  effect of the $J_z$ and $J_x$ terms in Eq.~(\ref
  {eq:spin_model}).}\BibitemShut {Stop}%
\bibitem [{\citenamefont {Kamal}\ and\ \citenamefont
  {Murthy}(1993)}]{KamalMurthy1993}%
  \BibitemOpen
  \bibfield  {author} {\bibinfo {author} {\bibfnamefont {M.}~\bibnamefont
  {Kamal}}\ and\ \bibinfo {author} {\bibfnamefont {G.}~\bibnamefont {Murthy}},\
  }\href {\doibase 10.1103/PhysRevLett.71.1911} {\bibfield  {journal} {\bibinfo
   {journal} {Phys. Rev. Lett.}\ }\textbf {\bibinfo {volume} {71}},\ \bibinfo
  {pages} {1911} (\bibinfo {year} {1993})}\BibitemShut {NoStop}%
\bibitem [{\citenamefont {Sachdev}\ and\ \citenamefont
  {Park}(2002)}]{SachdevPark2002}%
  \BibitemOpen
  \bibfield  {author} {\bibinfo {author} {\bibfnamefont {S.}~\bibnamefont
  {Sachdev}}\ and\ \bibinfo {author} {\bibfnamefont {K.}~\bibnamefont {Park}},\
  }\href {\doibase https://doi.org/10.1006/aphy.2002.6232} {\bibfield
  {journal} {\bibinfo  {journal} {Annals of Physics}\ }\textbf {\bibinfo
  {volume} {298}},\ \bibinfo {pages} {58 } (\bibinfo {year}
  {2002})}\BibitemShut {NoStop}%
\bibitem [{Note5()}]{Note5}%
  \BibitemOpen
  \bibinfo {note} {A more formal demonstration of this can be carried out in
  Euclidean path integral language along the lines of Appendix A in Ref.~\cite
  {LaiMotrunich2011}, which asked similar question about $Z_2$ instanton
  effects but motivated by gapless Majorana spin liquids in 1d.}\BibitemShut
  {Stop}%
\bibitem [{\citenamefont {Lai}\ and\ \citenamefont
  {Motrunich}(2011)}]{LaiMotrunich2011}%
  \BibitemOpen
  \bibfield  {author} {\bibinfo {author} {\bibfnamefont {H.-H.}\ \bibnamefont
  {Lai}}\ and\ \bibinfo {author} {\bibfnamefont {O.~I.}\ \bibnamefont
  {Motrunich}},\ }\href {\doibase 10.1103/PhysRevB.84.235148} {\bibfield
  {journal} {\bibinfo  {journal} {Phys. Rev. B}\ }\textbf {\bibinfo {volume}
  {84}},\ \bibinfo {pages} {235148} (\bibinfo {year} {2011})}\BibitemShut
  {NoStop}%
\bibitem [{\citenamefont {Yao}\ and\ \citenamefont
  {Kivelson}(2010)}]{YaoKivelson2010}%
  \BibitemOpen
  \bibfield  {author} {\bibinfo {author} {\bibfnamefont {H.}~\bibnamefont
  {Yao}}\ and\ \bibinfo {author} {\bibfnamefont {S.~A.}\ \bibnamefont
  {Kivelson}},\ }\href {\doibase 10.1103/PhysRevLett.105.166402} {\bibfield
  {journal} {\bibinfo  {journal} {Phys. Rev. Lett.}\ }\textbf {\bibinfo
  {volume} {105}},\ \bibinfo {pages} {166402} (\bibinfo {year}
  {2010})}\BibitemShut {NoStop}%
\bibitem [{\citenamefont {Kim}\ \emph {et~al.}(2016)\citenamefont {Kim},
  \citenamefont {Lee}, \citenamefont {Jiang}, \citenamefont {Ware},
  \citenamefont {Jian}, \citenamefont {Zaletel}, \citenamefont {Han},\ and\
  \citenamefont {Ran}}]{KimLeeJiangWareJianZaletel2016}%
  \BibitemOpen
  \bibfield  {author} {\bibinfo {author} {\bibfnamefont {P.}~\bibnamefont
  {Kim}}, \bibinfo {author} {\bibfnamefont {H.}~\bibnamefont {Lee}}, \bibinfo
  {author} {\bibfnamefont {S.}~\bibnamefont {Jiang}}, \bibinfo {author}
  {\bibfnamefont {B.}~\bibnamefont {Ware}}, \bibinfo {author} {\bibfnamefont
  {C.-M.}\ \bibnamefont {Jian}}, \bibinfo {author} {\bibfnamefont
  {M.}~\bibnamefont {Zaletel}}, \bibinfo {author} {\bibfnamefont {J.~H.}\
  \bibnamefont {Han}}, \ and\ \bibinfo {author} {\bibfnamefont
  {Y.}~\bibnamefont {Ran}},\ }\href {\doibase 10.1103/PhysRevB.94.064432}
  {\bibfield  {journal} {\bibinfo  {journal} {Phys. Rev. B}\ }\textbf {\bibinfo
  {volume} {94}},\ \bibinfo {pages} {064432} (\bibinfo {year}
  {2016})}\BibitemShut {NoStop}%
\bibitem [{\citenamefont {Song}\ \emph {et~al.}(2017)\citenamefont {Song},
  \citenamefont {Huang}, \citenamefont {Fu},\ and\ \citenamefont
  {Hermele}}]{HaoHuangFuHermele2017}%
  \BibitemOpen
  \bibfield  {author} {\bibinfo {author} {\bibfnamefont {H.}~\bibnamefont
  {Song}}, \bibinfo {author} {\bibfnamefont {S.-J.}\ \bibnamefont {Huang}},
  \bibinfo {author} {\bibfnamefont {L.}~\bibnamefont {Fu}}, \ and\ \bibinfo
  {author} {\bibfnamefont {M.}~\bibnamefont {Hermele}},\ }\href {\doibase
  10.1103/PhysRevX.7.011020} {\bibfield  {journal} {\bibinfo  {journal} {Phys.
  Rev. X}\ }\textbf {\bibinfo {volume} {7}},\ \bibinfo {pages} {011020}
  (\bibinfo {year} {2017})}\BibitemShut {NoStop}%
\bibitem [{\citenamefont {{Kitaev}}(2001)}]{Kitaev2001}%
  \BibitemOpen
  \bibfield  {author} {\bibinfo {author} {\bibfnamefont {A.~Y.}\ \bibnamefont
  {{Kitaev}}},\ }\href {\doibase 10.1070/1063-7869/44/10S/S29} {\bibfield
  {journal} {\bibinfo  {journal} {Physics Uspekhi}\ }\textbf {\bibinfo {volume}
  {44}},\ \bibinfo {pages} {131} (\bibinfo {year} {2001})},\ \Eprint
  {http://arxiv.org/abs/cond-mat/0010440} {cond-mat/0010440} \BibitemShut
  {NoStop}%
\bibitem [{\citenamefont {Motrunich}\ \emph {et~al.}(2001)\citenamefont
  {Motrunich}, \citenamefont {Damle},\ and\ \citenamefont
  {Huse}}]{MotrunichDamleHuse2001}%
  \BibitemOpen
  \bibfield  {author} {\bibinfo {author} {\bibfnamefont {O.}~\bibnamefont
  {Motrunich}}, \bibinfo {author} {\bibfnamefont {K.}~\bibnamefont {Damle}}, \
  and\ \bibinfo {author} {\bibfnamefont {D.~A.}\ \bibnamefont {Huse}},\ }\href
  {\doibase 10.1103/PhysRevB.63.224204} {\bibfield  {journal} {\bibinfo
  {journal} {Phys. Rev. B}\ }\textbf {\bibinfo {volume} {63}},\ \bibinfo
  {pages} {224204} (\bibinfo {year} {2001})}\BibitemShut {NoStop}%
\bibitem [{Note6()}]{Note6}%
  \BibitemOpen
  \bibinfo {note} {B.~Roberts, S.~Jiang, and O.~I.~Motrunich, in
  preparation}\BibitemShut {NoStop}%
\bibitem [{\citenamefont {Kogut}(1979)}]{KogutRMP}%
  \BibitemOpen
  \bibfield  {author} {\bibinfo {author} {\bibfnamefont {J.~B.}\ \bibnamefont
  {Kogut}},\ }\href {\doibase 10.1103/RevModPhys.51.659} {\bibfield  {journal}
  {\bibinfo  {journal} {Rev. Mod. Phys.}\ }\textbf {\bibinfo {volume} {51}},\
  \bibinfo {pages} {659} (\bibinfo {year} {1979})}\BibitemShut {NoStop}%
\bibitem [{\citenamefont {Moessner}\ \emph {et~al.}(2001)\citenamefont
  {Moessner}, \citenamefont {Sondhi},\ and\ \citenamefont
  {Fradkin}}]{MoessnerSondhiFradkin2001}%
  \BibitemOpen
  \bibfield  {author} {\bibinfo {author} {\bibfnamefont {R.}~\bibnamefont
  {Moessner}}, \bibinfo {author} {\bibfnamefont {S.~L.}\ \bibnamefont
  {Sondhi}}, \ and\ \bibinfo {author} {\bibfnamefont {E.}~\bibnamefont
  {Fradkin}},\ }\href {\doibase 10.1103/PhysRevB.65.024504} {\bibfield
  {journal} {\bibinfo  {journal} {Phys. Rev. B}\ }\textbf {\bibinfo {volume}
  {65}},\ \bibinfo {pages} {024504} (\bibinfo {year} {2001})}\BibitemShut
  {NoStop}%
\bibitem [{\citenamefont {Nielsen}(2006)}]{Nielsen2006}%
  \BibitemOpen
  \bibfield  {author} {\bibinfo {author} {\bibfnamefont {M.~A.}\ \bibnamefont
  {Nielsen}},\ }\href {\doibase https://doi.org/10.1016/S0034-4877(06)80014-5}
  {\bibfield  {journal} {\bibinfo  {journal} {Reports on Mathematical Physics}\
  }\textbf {\bibinfo {volume} {57}},\ \bibinfo {pages} {147 } (\bibinfo {year}
  {2006})}\BibitemShut {NoStop}%
\bibitem [{\citenamefont {Chen}\ \emph {et~al.}(2014)\citenamefont {Chen},
  \citenamefont {Lu},\ and\ \citenamefont {Vishwanath}}]{ChenLuVishwanath2014}%
  \BibitemOpen
  \bibfield  {author} {\bibinfo {author} {\bibfnamefont {X.}~\bibnamefont
  {Chen}}, \bibinfo {author} {\bibfnamefont {Y.-M.}\ \bibnamefont {Lu}}, \ and\
  \bibinfo {author} {\bibfnamefont {A.}~\bibnamefont {Vishwanath}},\
  }\href@noop {} {\bibfield  {journal} {\bibinfo  {journal} {Nature
  communications}\ }\textbf {\bibinfo {volume} {5}},\ \bibinfo {pages} {3507}
  (\bibinfo {year} {2014})}\BibitemShut {NoStop}%
\bibitem [{\citenamefont {Tsui}\ \emph {et~al.}(2015)\citenamefont {Tsui},
  \citenamefont {Jiang}, \citenamefont {Lu},\ and\ \citenamefont
  {Lee}}]{TsuiJiangLuLee2015}%
  \BibitemOpen
  \bibfield  {author} {\bibinfo {author} {\bibfnamefont {L.}~\bibnamefont
  {Tsui}}, \bibinfo {author} {\bibfnamefont {H.-C.}\ \bibnamefont {Jiang}},
  \bibinfo {author} {\bibfnamefont {Y.-M.}\ \bibnamefont {Lu}}, \ and\ \bibinfo
  {author} {\bibfnamefont {D.-H.}\ \bibnamefont {Lee}},\ }\href {\doibase
  https://doi.org/10.1016/j.nuclphysb.2015.04.020} {\bibfield  {journal}
  {\bibinfo  {journal} {Nuclear Physics B}\ }\textbf {\bibinfo {volume}
  {896}},\ \bibinfo {pages} {330 } (\bibinfo {year} {2015})}\BibitemShut
  {NoStop}%
\bibitem [{\citenamefont {{Tsui}}\ \emph {et~al.}(2017)\citenamefont {{Tsui}},
  \citenamefont {{Huang}}, \citenamefont {{Jiang}},\ and\ \citenamefont
  {{Lee}}}]{TsuiHuangJiangLee2017}%
  \BibitemOpen
  \bibfield  {author} {\bibinfo {author} {\bibfnamefont {L.}~\bibnamefont
  {{Tsui}}}, \bibinfo {author} {\bibfnamefont {Y.-T.}\ \bibnamefont {{Huang}}},
  \bibinfo {author} {\bibfnamefont {H.-C.}\ \bibnamefont {{Jiang}}}, \ and\
  \bibinfo {author} {\bibfnamefont {D.-H.}\ \bibnamefont {{Lee}}},\ }\href
  {\doibase 10.1016/j.nuclphysb.2017.03.021} {\bibfield  {journal} {\bibinfo
  {journal} {Nuclear Physics B}\ }\textbf {\bibinfo {volume} {919}},\ \bibinfo
  {pages} {470} (\bibinfo {year} {2017})},\ \Eprint
  {http://arxiv.org/abs/1701.00834} {arXiv:1701.00834 [cond-mat.str-el]}
  \BibitemShut {NoStop}%
\bibitem [{\citenamefont {Verresen}\ \emph {et~al.}(2017)\citenamefont
  {Verresen}, \citenamefont {Moessner},\ and\ \citenamefont
  {Pollmann}}]{VerresenMoessnerPollmann2017}%
  \BibitemOpen
  \bibfield  {author} {\bibinfo {author} {\bibfnamefont {R.}~\bibnamefont
  {Verresen}}, \bibinfo {author} {\bibfnamefont {R.}~\bibnamefont {Moessner}},
  \ and\ \bibinfo {author} {\bibfnamefont {F.}~\bibnamefont {Pollmann}},\
  }\href {\doibase 10.1103/PhysRevB.96.165124} {\bibfield  {journal} {\bibinfo
  {journal} {Phys. Rev. B}\ }\textbf {\bibinfo {volume} {96}},\ \bibinfo
  {pages} {165124} (\bibinfo {year} {2017})}\BibitemShut {NoStop}%
\bibitem [{\citenamefont {Geraedts}\ and\ \citenamefont
  {Motrunich}(2017)}]{GeraedtsMotrunich2017}%
  \BibitemOpen
  \bibfield  {author} {\bibinfo {author} {\bibfnamefont {S.}~\bibnamefont
  {Geraedts}}\ and\ \bibinfo {author} {\bibfnamefont {O.~I.}\ \bibnamefont
  {Motrunich}},\ }\href {\doibase 10.1103/PhysRevB.96.115137} {\bibfield
  {journal} {\bibinfo  {journal} {Phys. Rev. B}\ }\textbf {\bibinfo {volume}
  {96}},\ \bibinfo {pages} {115137} (\bibinfo {year} {2017})}\BibitemShut
  {NoStop}%
\bibitem [{\citenamefont {Qin}\ \emph {et~al.}(2017)\citenamefont {Qin},
  \citenamefont {He}, \citenamefont {You}, \citenamefont {Lu}, \citenamefont
  {Sen}, \citenamefont {Sandvik}, \citenamefont {Xu},\ and\ \citenamefont
  {Meng}}]{QinHeYouLuSenSandvikXuMeng2017}%
  \BibitemOpen
  \bibfield  {author} {\bibinfo {author} {\bibfnamefont {Y.~Q.}\ \bibnamefont
  {Qin}}, \bibinfo {author} {\bibfnamefont {Y.-Y.}\ \bibnamefont {He}},
  \bibinfo {author} {\bibfnamefont {Y.-Z.}\ \bibnamefont {You}}, \bibinfo
  {author} {\bibfnamefont {Z.-Y.}\ \bibnamefont {Lu}}, \bibinfo {author}
  {\bibfnamefont {A.}~\bibnamefont {Sen}}, \bibinfo {author} {\bibfnamefont
  {A.~W.}\ \bibnamefont {Sandvik}}, \bibinfo {author} {\bibfnamefont
  {C.}~\bibnamefont {Xu}}, \ and\ \bibinfo {author} {\bibfnamefont {Z.~Y.}\
  \bibnamefont {Meng}},\ }\href {\doibase 10.1103/PhysRevX.7.031052} {\bibfield
   {journal} {\bibinfo  {journal} {Phys. Rev. X}\ }\textbf {\bibinfo {volume}
  {7}},\ \bibinfo {pages} {031052} (\bibinfo {year} {2017})}\BibitemShut
  {NoStop}%
\bibitem [{\citenamefont {Grover}\ and\ \citenamefont
  {Vishwanath}(2013)}]{GroverVishwanath2013}%
  \BibitemOpen
  \bibfield  {author} {\bibinfo {author} {\bibfnamefont {T.}~\bibnamefont
  {Grover}}\ and\ \bibinfo {author} {\bibfnamefont {A.}~\bibnamefont
  {Vishwanath}},\ }\href {\doibase 10.1103/PhysRevB.87.045129} {\bibfield
  {journal} {\bibinfo  {journal} {Phys. Rev. B}\ }\textbf {\bibinfo {volume}
  {87}},\ \bibinfo {pages} {045129} (\bibinfo {year} {2013})}\BibitemShut
  {NoStop}%
\bibitem [{\citenamefont {Lu}\ and\ \citenamefont {Lee}(2014)}]{LuLee2014}%
  \BibitemOpen
  \bibfield  {author} {\bibinfo {author} {\bibfnamefont {Y.-M.}\ \bibnamefont
  {Lu}}\ and\ \bibinfo {author} {\bibfnamefont {D.-H.}\ \bibnamefont {Lee}},\
  }\href {\doibase 10.1103/PhysRevB.89.195143} {\bibfield  {journal} {\bibinfo
  {journal} {Phys. Rev. B}\ }\textbf {\bibinfo {volume} {89}},\ \bibinfo
  {pages} {195143} (\bibinfo {year} {2014})}\BibitemShut {NoStop}%
\bibitem [{\citenamefont {Lu}\ and\ \citenamefont
  {Vishwanath}(2012)}]{LuVishwanath2012}%
  \BibitemOpen
  \bibfield  {author} {\bibinfo {author} {\bibfnamefont {Y.-M.}\ \bibnamefont
  {Lu}}\ and\ \bibinfo {author} {\bibfnamefont {A.}~\bibnamefont
  {Vishwanath}},\ }\href {\doibase 10.1103/PhysRevB.86.125119} {\bibfield
  {journal} {\bibinfo  {journal} {Phys. Rev. B}\ }\textbf {\bibinfo {volume}
  {86}},\ \bibinfo {pages} {125119} (\bibinfo {year} {2012})}\BibitemShut
  {NoStop}%
\bibitem [{\citenamefont {Senthil}\ and\ \citenamefont
  {Levin}(2013)}]{SenthilLevin2013}%
  \BibitemOpen
  \bibfield  {author} {\bibinfo {author} {\bibfnamefont {T.}~\bibnamefont
  {Senthil}}\ and\ \bibinfo {author} {\bibfnamefont {M.}~\bibnamefont
  {Levin}},\ }\href {\doibase 10.1103/PhysRevLett.110.046801} {\bibfield
  {journal} {\bibinfo  {journal} {Phys. Rev. Lett.}\ }\textbf {\bibinfo
  {volume} {110}},\ \bibinfo {pages} {046801} (\bibinfo {year}
  {2013})}\BibitemShut {NoStop}%
\bibitem [{\citenamefont {Geraedts}\ and\ \citenamefont
  {Motrunich}(2013)}]{GeraedtsMotrunich2013}%
  \BibitemOpen
  \bibfield  {author} {\bibinfo {author} {\bibfnamefont {S.~D.}\ \bibnamefont
  {Geraedts}}\ and\ \bibinfo {author} {\bibfnamefont {O.~I.}\ \bibnamefont
  {Motrunich}},\ }\href {\doibase https://doi.org/10.1016/j.aop.2013.03.017}
  {\bibfield  {journal} {\bibinfo  {journal} {Annals of Physics}\ }\textbf
  {\bibinfo {volume} {334}},\ \bibinfo {pages} {288 } (\bibinfo {year}
  {2013})}\BibitemShut {NoStop}%
\end{thebibliography}%

\end{document}